\documentclass[12pt]{article}
\pdfoutput=1
\usepackage{amsmath}
\usepackage{amssymb}
\usepackage{epsfig}
\usepackage{mcite}
\usepackage{graphicx}

\def\L{\mathcal L}
\def\e{\varepsilon}

\newcommand{\wt}{\widetilde}

\begin{document}

\def\a{\alpha}
\def\b{\beta}
\def\c{\chi}
\def\d{\delta}
\def\e{\epsilon}
\def\f{\phi}
\def\g{\gamma}
\def\h{\eta}
\def\i{\iota}
\def\j{\psi}
\def\k{\kappa}
\def\l{\lambda}
\def\m{\mu}
\def\n{\nu}
\def\o{\omega}
\def\p{\pi}
\def\q{\theta}
\def\r{\rho}
\def\s{\sigma}
\def\t{\tau}
\def\u{\upsilon}
\def\x{\xi}
\def\z{\zeta}
\def\D{\Delta}
\def\F{\Phi}
\def\G{\Gamma}
\def\J{\Psi}
\def\L{\Lambda}
\def\O{\Omega}
\def\P{\Pi}
\def\Q{\Theta}
\def\S{\Sigma}
\def\U{\Upsilon}
\def\X{\Xi}

%Varletters
\def\ve{\varepsilon}
\def\vf{\varphi}
\def\vr{\varrho}
\def\vs{\varsigma}
\def\vq{\vartheta}

\def\dg{\dagger}                                     % hermitian conjugate
\def\ddg{\ddagger}                                   % double dagger
\def\wt#1{\widetilde{#1}}                    % big tilde
\def\mt{\widetilde{m}_1}
\def\mti{\widetilde{m}_i}
\def\rt{\widetilde{r}_1}
\def\mtt{\widetilde{m}_2}
\def\mttt{\widetilde{m}_3}
\def\rtt{\widetilde{r}_2}
\def\mb{\overline{m}}
\def\VEV#1{\left\langle #1\right\rangle}        % < >
\def\be{\begin{equation}}
\def\ee{\end{equation}}
\def\ds{\displaystyle}
\def\ra{\rightarrow}

\def\bea{\begin{eqnarray}}
\def\eea{\end{eqnarray}}
\def\NO{\nonumber}
\def\Bar#1{\overline{#1}}

% Journal abbreviations (preprints)

\def\pl#1#2#3{Phys.~Lett.~{\bf B {#1}} ({#2}) #3}
\def\np#1#2#3{Nucl.~Phys.~{\bf B {#1}} ({#2}) #3}
\def\prl#1#2#3{Phys.~Rev.~Lett.~{\bf #1} ({#2}) #3}
\def\pr#1#2#3{Phys.~Rev.~{\bf D {#1}} ({#2}) #3}
\def\zp#1#2#3{Z.~Phys.~{\bf C {#1}} ({#2}) #3}
\def\cqg#1#2#3{Class.~and Quantum Grav.~{\bf {#1}} ({#2}) #3}
\def\cmp#1#2#3{Commun.~Math.~Phys.~{\bf {#1}} ({#2}) #3}
\def\jmp#1#2#3{J.~Math.~Phys.~{\bf {#1}} ({#2}) #3}
\def\ap#1#2#3{Ann.~of Phys.~{\bf {#1}} ({#2}) #3}
\def\prep#1#2#3{Phys.~Rep.~{\bf {#1}C} ({#2}) #3}
\def\ptp#1#2#3{Progr.~Theor.~Phys.~{\bf {#1}} ({#2}) #3}
\def\ijmp#1#2#3{Int.~J.~Mod.~Phys.~{\bf A {#1}} ({#2}) #3}
\def\mpl#1#2#3{Mod.~Phys.~Lett.~{\bf A {#1}} ({#2}) #3}
\def\nc#1#2#3{Nuovo Cim.~{\bf {#1}} ({#2}) #3}
\def\ibid#1#2#3{{\it ibid.}~{\bf {#1}} ({#2}) #3}

\title{
%{\normalsize \mbox{ }\hfill
%\begin{minipage}{3cm}
%MPP-2005-118
%\end{minipage}}\\
\vspace*{4mm}
\bf Decrypting $SO(10)$-inspired leptogenesis}
\author{
{\Large Pasquale Di Bari$^1$, Luca Marzola$^2$, Michele Re Fiorentin$^1$}
\\
{\it $^1$ Physics and Astronomy}, 
{\it University of Southampton,} \\
{\it  Southampton, SO17 1BJ, U.K.}
\\
{\it $^2$ Institute of Physics, University of Tartu}, \\
{\it Ravila 14c, 50411 Tartu, Estonia.}
}

\maketitle \thispagestyle{empty}

\vspace{-8mm}
%\centerline{\date{\today}}

\begin{abstract}
Encouraged by the recent results from neutrino oscillation experiments, we perform an analytical study of $SO(10)$-inspired models and leptogenesis with hierarchical right-handed (RH) neutrino spectrum.  
Under the approximation of negligible misalignment  between the neutrino Yukawa basis and the charged lepton basis, we find an analytical expression for the final asymmetry 
directly in terms of the low energy neutrino parameters that fully
reproduces previous numerical results.  This expression  also shows that it is   
possible to identify an effective leptogenesis phase for these models.
When we also impose  the wash-out of a large pre-existing asymmetry $N^{\rm p,i}_{B-L}$, 
the strong thermal (ST) condition,
we derive analytically all those constraints on the low energy neutrino parameters that characterise 
the  {\rm ST}-$SO(10)$-inspired leptogenesis solution, confirming previous numerical results.  
In particular we show why, though neutrino masses have to be necessarily normally ordered,
the solution implies an analytical lower bound on the effective neutrino-less double beta decay neutrino mass, 
$m_{ee} \gtrsim 8\,{\rm meV}$, for $N^{\rm p,i}_{B-L}=10^{-3}$, testable with next generation experiments.
This, in combination with an upper bound on the atmospheric mixing angle, 
necessarily in the first octant, forces the lightest neutrino mass 
within a narrow range $m_1 \simeq (10$--$30)\,{\rm meV}$
(corresponding to $\sum_i \, m_i \simeq (75$--$125)\,{\rm meV}$).
We also show why the solution  could correctly predict
a non-vanishing reactor neutrino mixing angle
and  requires the Dirac phase to be in the fourth quadrant,
implying $\sin\d $ (and $J_{CP}$) negative
as hinted by current global analyses. 
Many of the analytical results presented 
(expressions for the orthogonal matrix, RH neutrino mixing matrix, masses and phases)  
can  have applications beyond leptogenesis. 
\end{abstract}
\newpage
\section{Introduction}
There is no observational evidence of primordial antimatter in our observable Universe
that, therefore, is  matter-antimatter asymmetric. 
The matter-antimatter asymmetry can be expressed in terms of the baryon-to-photon number ratio, currently very precisely and accurately measured by the {\em Planck} satellite from CMB temperature anisotropies,  
finding \cite{planck2013}
\be\label{etaBCMB}
\eta_B^{\rm CMB} = (6.1 \pm 0.1) \times 10^{-10} \,  .
\ee
This value is too high to be explained within the Standard Model (SM) and, therefore, it can  be regarded as an evidence of new physics.  Leptogenesis \cite{fy} provides an attractive explanation since it 
is based on a minimal extension of the SM, 
the see-saw mechanism \cite{minkowski77,*yanagida79,*grs79,*glashow,*barbieri,*rabigoran}, able to address neutrino masses and mixing, 
soundly established by neutrino oscillation experiments. In this way leptogenesis realises
a very interesting connection between two, apparently unrelated, fundamental physical observations:
the cosmological matter-antimatter asymmetry and neutrino masses and mixing. 

Recent important experimental findings  support a traditional thermal picture and 
encourage further investigation in this direction. 
The confirmation of the BEH mechanism with the discovery of the
Higgs boson at the LHC, an important ingredient of the see-saw mechanism, 
certainly corroborates the overall picture of leptogenesis.
In addition, the non-discovery of new physics at LHC so far, quite strongly 
constrains models of baryogenesis at the ${\rm TeV}$ (or lower) energy scale. 
Moreover if the recent claim of a B-mode polarization signal in the CMB will be (even partially) confirmed 
(see \cite{seljak,*spergel,*planckdust} for critical analyses of BICEP2 results), 
it would support high inflationary scale $V^{1/4} \sim  10^{16}\,{\rm GeV}$ \cite{bicep2}. 
In this way high values of the reheat temperature $T_{RH}$,   
%at least as high as those allowed by QCD scatterings, $T_{\rm RH} \lesssim 3\times 10^{14}\,{\rm GeV}$ \cite{enqvist}, 
though not necessarily implied, would be not only possible but even quite natural. 
This phenomenological picture is then well compatible  with 
the original idea of a high energy scale thermal leptogenesis scenario, much above the
{\rm TeV} scale \cite{fy}. 

Values of $T_{\rm RH}$  greater than those required by successful leptogenesis, $T_{\rm RH} \gtrsim 10^9\,{\rm GeV}$ \cite{di,*cmb,predictions}, would be not only possible but even, as already mentioned, quite natural if the BICEP2  signal, even just a small fraction of it, will be confirmed as primordial.
In any case it is legitimate to conceive that other mechanisms of baryogenesis might 
have generated an asymmetry much larger than the observed one  prior to the onset of leptogenesis,  one of the three main working assumptions in our study. 
For example, considering that we will work within the context of
$SO(10)$-inspired conditions (another of our main working assumptions), 
the decays of GUT gauge bosons might have acted as the source of 
such a large pre-existing asymmetry \cite{yoshimura}.  
It is then quite legitimate to impose in addition to the successful leptogenesis condition, 
requiring that the asymmetry produced from right-handed (RH) neutrinos decays 
reproduces the observed asymmetry,  that at the same time the RH neutrino inverse processes 
wash-out a possible pre-existing (too) large asymmetry 
generated by some external mechanism: we will refer to this condition as the ST (equilibrium) condition.
\footnote{This name is justified by the fact that, as we will discuss, 
because of  flavour effects the washout of a pre-existing asymmetry
is possible only locking all possible ways out in flavour space, 
i.e. imposing thermal equilibrium in all flavours, 
something that can be indeed regarded as  a 
`strong' thermal equilibrium condition. Note that it also ensures independence
of the initial RH neutrino abundance.} 

Within a minimal type I extension of the SM and assuming a hierarchical RH neutrino
mass spectrum, there is only one mass pattern able to realise successful ST leptogenesis, 
solving the problem of the initial conditions: the tauon $N_2$-dominated scenario \cite{problem}.
\footnote{This conclusion holds for hierarchical RH neutrino spectra, where wash-out
and asymmetry production from each RH neutrino species occurs sequentially. 
It has been claimed that quasi-degenerate RH neutrino spectra might provide
an alternative way to realise ST leptogenesis \cite{problem}.
This claim seems to be supported by recent calculations within specific models \cite{pilateresi,*pilaftsis2}.} 
In this scenario the lightest RH neutrino ($N_1$) mass has to be $M_1 \ll 10^9\,{\rm GeV}$ so that
the $N_1$ wash-out stage occurs  in a three-flavoured regime \cite{bcst,roulet,*losada}. This necessarily implies that   the $N_1$-decays cannot produce a sizeable contribution to the final asymmetry \cite{di,*cmb,predictions}. 
On the other hand the next-to-lightest RH neutrino ($N_2$) mass falls in the range 
$10^{12}\,{\rm GeV} \gg M_2 \gg 10^9 \,{\rm GeV}$  so that the 
asymmetry production from $N_2$-decays occurs in the two-flavoured 
regime \cite{bcst,roulet,*losada} and can be potentially sizeable.  
In this way the pre-existing large tauon asymmetry can be washed-out by the $N_2$-inverse processes 
while at the same time the $N_2$-decays can re-generate an asymmetry in the same tauon 
flavour. This tauon component would then explain the observed asymmetry
if it can escape the $N_1$ wash-out. The latter, on the other hand, has necessarily to suppress
the electron and muon pre-existing asymmetries since they cannot be fully washed-out by 
the $N_2$-inverse  processes. This is because these
can only wash-out the ${\ell}_2$-flavour parallel component but not the orthogonal one,
where ${\ell}_2$ is the charged lepton flavour produced by $N_2$-decays \cite{bcst,nardinir}.  

It is quite interesting and non-trivial that so called $SO(10)$-inspired models 
\cite{buchplum,*nezriorloff,*falcone,branco,afs}  
can indeed realise this seemingly contrived chain of conditions for successful 
ST leptogenesis \cite{strongSO10solution}. This happens for a subset 
 of the whole set of the solutions that realise successful $SO(10)$-inspired  leptogenesis, 
\cite{SO10lep1,SO10lep2}. 
This  subset defines the `ST $SO(10)$-inspired solution' \cite{strongSO10solution}. 

Interestingly, this solution implies a predictive set of constraints on low energy neutrino
parameters that can be almost completely \footnote{We will comment in the
final discussion why `almost'.} tested by (data-taking or planned) low energy neutrino experiments (including
cosmological observations).
 One of the most striking features is that the value of the absolute neutrino mass scale
is quite constrained within a narrow window, such that 
the lightest neutrino mass $m_1 \simeq 20\,{\rm meV}$ (while $\sum m_i \simeq 100\,{\rm meV}$). 
The Majorana phases are also constrained about particular values. 
In this way cancellations in the effective neutrino mass are very mild and 
$m_{ee} \sim 0.8\,m_1 \simeq 15\,{\rm meV}$.  In addition parameters 
tested by neutrino oscillation experiments are also quite definitely predicted:
neutrino masses have to be NO; the atmospheric mixing angle has to lie in the first octant
and the Dirac phase in the fourth quadrant ($\d \sim -\pi/4$) so that
${\rm sign}(J_{CP}) = - {\rm sign}(\eta_B) < 0$.  It is also interesting that the solution
has correctly predicted a non-vanishing value of the reactor mixing angle 
($\theta_{13}\gtrsim 2^{\circ}\mbox{--}3^{\circ})$, as now robustly established by different experiments \cite{T2Ktheta13,*dayabay,*reno}. \footnote{Preliminary results were presented in \cite{talks}.}

This set of predictive constraints has so far been derived numerically, 
from scatter plots with points randomly generated \cite{strongSO10solution}. 
A partial analytical insight was found for 
$SO(10)$-inspired solutions (no strong thermal condition)
 in the hierarchical limit \cite{SO10lep1} , 
for $m_1 \ll 10\,{\rm meV}$, and, therefore, it does not apply to the
strong thermal $SO(10)$-inspired solution that 
corresponds to semi-hierarchical (NO) neutrino masses. 
It has been recently shown analytically \cite{strongthermal} that 
in general strong thermal leptogenesis requires, barring fine tuned models, $m_1 \gtrsim 10\,{\rm meV}$ for NO. 
This is in complete agreement with the analogous lower bound  found numerically in the case  of the strong thermal $SO(10)$-inspired leptogenesis solution.

Encouraged by the recent experimental results, in addition to 
the above mentioned discovery of a non-vanishing reactor 
mixing angle also hints for negative values of $J_{CP}$ \cite{foglilisi2013,nufit14}
and, to a weaker extent, for a possible deviation
of the atmospheric mixing angle from the maximal value \cite{foglilisi2013,nufit14}
(see however \cite{valle}), 
in this paper we perform a systematic analytical derivation of the constraints coming 
from the strong thermal $SO(10)$-inspired solution confirming the numerical results. 
The derivation is made in the approximation $V_L \simeq I$,
where $V_L$ is the matrix describing the misalignment between the neutrino Yukawa basis and the charged
lepton flavour basis but we also discuss the implication of small 
deviations at the level of the CKM quark mixing matrix.  

The paper is organised as follows. In Section 2 we discuss the general 
set of conditions that lead to the solution. In Section 3 we 
discuss the current experimental results on neutrino parameters. In Section 4,  
imposing the $SO(10)$-inspired conditions,  we 
derive compact analytic expressions for the RH neutrino mass spectrum and for the $C\!P$ asymmetries improving and 
extending previous results \cite{branco,afs,SO10lep1}. In particular we are able to determine explicitly all the three  phases associated to the eigenvalues of the Majorana mass matrix, an important ingredient 
to calculate correctly the $C\!P$ asymmetries and, consequently, the final asymmetry. 
We also give an analytical expression for the orthogonal matrix
that can be useful for model building considerations. 
In Section 5  we find an expression for the final asymmetry within $SO(10)$-inspired 
models in terms of the low energy neutrino
parameters showing that it perfectly reproduces the numerical results for $V_L = I$: 
this represents one of the main results of our work.  
We then impose the successful leptogenesis condition and we find, for NO,
a lower bound and an upper bound on $m_1$, an upper bound on the atmospheric mixing angle 
and  different constraints on the low energy phases.  In particular we show
how $SO(10)$-inspired leptogenesis constraints are not symmetric 
under a change of the sign of $J_{CP}$, given by $\sin\d$.
In Section 6 we finally impose the ST condition and, still for NO, we show how
the lightest neutrino mass is constrained within a narrow range 
and how a lower bound on the reactor mixing angle holds.  We also show how this condition 
necessarily selects values of $\d$ in the fourth quadrant. 
In Section 7 we discuss IO neutrino
masses showing how in this case the ST condition cannot be satisfied and, therefore, 
this strictly implies NO neutrino masses.  
In Section 8 we make some final remarks on the different approximations and assumptions 
behind  the results and on the testability of  the solution in next years. 
In Section 9 we summarise our results. 

\section{Set of conditions: the general picture}

The ST $SO(10)$-inspired leptogenesis solution is obtained imposing 
the following set of conditions on the see-saw parameter space: 
\begin{itemize}
\item[(i)] $SO(10)$-inspired conditions on the neutrino Dirac mass matrix; 
\item[(ii)] successful leptogenesis;
\item[(iii)] strong thermal leptogenesis.
\end{itemize}
Let us briefly discuss these conditions in general, showing how (i)  and  (ii)+(iii)  both independently 
select the $N_2$-dominated scenario and how (iii) specifies that the asymmetry has 
to be necessarily tauon dominated \cite{problem}.

\subsection{$SO(10)$-inspired conditions}

In the minimal see-saw mechanism the SM Lagrangian is augmented introducing RH neutrinos with 
Yukawa couplings $h$ and a Majorana mass term $M$. In the (flavour) basis, where both
charged leptons and Majorana mass matrices are diagonal, the leptonic mass terms
 after spontaneous symmetry breaking, can be written as ($\a = e, \m, \t$ and $i=1,2,3$)
\be
- {\cal L}_{M} = \,  \overline{\a_L} \, D_{m_{\ell}}\,\a_R + 
                              \overline{\nu_{\a L}}\,m_{D\a i} \, N_{i R} +
                               {1\over 2} \, \overline{N^{c}_{i R}} \, D_{M} \, N_{i R}  + \mbox{\rm h.c.}\, ,
\ee
where $ D_{m_{\ell}} \equiv {\rm diag}(m_e, m_{\mu}, m_{\tau})$ and 
$D_M \equiv {\rm diag}(M_1, M_2, M_3)$, with $M_1 \leq  M_2 \leq M_3$. 
The neutrino Dirac mass matrix $m_D$ in the flavour basis can then be written
in the bi-unitary parameterisation as 
\be\label{biunitary}
m_D = V^{\dagger}_L \, D_{m_D} \,  U_R  \,  ,
\ee
where $D_{m_D} \equiv {\rm diag}(m_{D1},m_{D2},m_{D3})$ is the neutrino Dirac mass matrix 
in the Yukawa basis and $V_L$ and $U_R$ are the unitary matrices  acting respectively on the 
LH and RH neutrino fields in the transformation from the flavour basis to the Yukawa basis. 

If we parametrise the three eigenvalues in the Dirac mass matrix 
in terms of the up quark masses, writing
\be\label{upquarkmasses}
m_{D1}=\a_1\,m_u \, , \; m_{D2}=\a_2\, m_c \, ,  \;  m_{D3}=\a_3\,m_t \, ,
\ee
we define $SO(10)$-inspired models those respecting the following conditions
\footnote{These conditions can be also satisfied beyond traditional $SO(10)$-models,
 see examples discussed in \cite{volkaslaw}, 
 the 5D-$SO(10)$ model \cite{5DSO10},  the `tetra-model'  
 \cite{tetramodel} or the `A to Z' model \cite{A2Z}. Vice-versa not all $SO(10)$-models necessarily respect them. For example
 they could give rise to a type II see-saw \cite{typeIIrabigoran,*typeIIrabiqaisar}  contribution for the neutrino masses  (e.g. \cite{mohapatra2,*abadalavignac})
 and to alternative leptogenesis scenarios than those considered here.
 It should also be said that traditional (4D, no flavour symmetries) $SO(10)$ models that have been explored as viable realistic models able to fit both quarks and leptons parameters also usually respect these conditions (see discussion in \cite{SO10lep1}). For example if we consider a recent result of a best fit within a non-supersymmetric $SO(10)$ model using 126 and 10-dim Higgs representations including RG corrections \cite{rodejohann}  we obtain:
 $(\a_1,\a_2,\a_3,\theta_{12}^L,\theta_{13}^L,\theta_{23}^L)\simeq(48,8,1,1^{\circ},3^{\circ},4^{\circ})$. 
 The large value of $\alpha_1$  lifts $M_1$. Away from the crossing level solution, where $M_1\simeq M_2$, one still has $M_1\lesssim 10^{9}\,{\rm GeV}$. The best fit hits precisely a crossing level solution at $m_1\simeq 2\,{\rm meV}$ (signalled by the low value of  $m_{ee} \ll 1\,{\rm meV}$). 
 This probably happens since the crossing level allows
the largest possible value for $\theta_{23}$. This is found to be
$\theta_{23}\sim 36^{\circ}$, a value $3\s$ below
the best fit experimental (explaining the
quite high value  $\chi^2_{\rm min} \sim 10$). In addition 
the Dirac phase is found $\d \simeq 0$ and neutrino masses are NO. 
The latest neutrino data increase the tension,  since values $\theta_{23}\sim 36^{\circ}$
are even more strongly disfavoured (cf. eq.~(\ref{expranges})) and values
$\d\sim 0$ are now also disfavoured at $\sim 2\s$ (cf. eq.~(\ref{delta})).
A more promising case seems to be a supersymmetric $SO(10)$ model also including a $120_H$. 
This time the best fit is found for IO and we find that also in this case the model respects
  the $SO(10)$-inspired conditions  (one has $(\a_1,\a_2,\a_3,\theta_{12}^L,\theta_{13}^L,\theta_{23}^L)\simeq(0.3,6,0.5,4^{\circ},0.2^{\circ},13^{\circ}$)).
  Because of the small value of $\a_1$ one finds $M_1 \sim 1\,{\rm TeV}$. Our calculations can be easily extended to treat also this supersymmetric case \cite{susy}.
}
\begin{itemize}  
\item[i)] $I \leq V_L \lesssim V_{CKM}$ \,  ,
\item[ii)] $\a_i = {\cal O}(0.1\mbox{--}10)$ \,  .
\end{itemize}
With the condition i) we mean that the values of the three mixing angles in $V_L$, that we indicate with $\theta_{12}^L, \theta_{13}^L, \theta_{23}^L$ in the usual PDG parametrisation,  are not too larger than the corresponding mixing angles in the CKM matrix and in particular  $\theta_{12}^L \lesssim \theta^{CKM}_{12} \equiv \theta_C \simeq 13^{\circ}$.  
In the see-saw limit, for $M \gg m_D$, the spectrum of neutrino mass eigenstates 
splits into a very heavy set, $N_i \simeq N_{iR} + N_{iR}^c$, with masses almost coinciding  with the Majorana masses $M_i$, and into a light set  $\nu_i \simeq \nu_{iL} + \nu_{i L}^c$,  with a  symmetric mass matrix $m_{\nu}$ given by the see-saw formula
\be\label{seesaw}
m_{\nu} = - m_D \, {1\over D_M} \, m_D^T  \,  .
\ee
This is diagonalised by a unitary matrix $U$,
\be\label{leptonic}
U^{\dagger} \,  m_{\nu} \, U^{\star}  =  - D_m  \,  ,
\ee
where $D_m \equiv {\rm diag}(m_1,m_2,m_3)$ with $m_1 \leq m_2 \leq m_3$,
corresponding to the PMNS leptonic mixing matrix, in a way that we can write
\be\label{diagonalseesaw}
 D_m =  U^{\dagger} \, m_D \, {1\over D_M} \, m_D^T  \, U^{\star}     \,   .
\ee
When the current experimental information from neutrino oscillation experiments 
on the leptonic mixing matrix and on the neutrino masses is taken into account, 
the RH neutrino mass spectrum, barring regions around crossing level solutions \cite{afs}, turns out to be highly hierarchical
with approximately $M_1 : M_2 : M_3 = \a_1^2\,m_u^{\, 2} : 
\a_2^2\,m_c^{\, 2} : \a_3^2\,m_t^{\, 2}$, implying $M_1 \ll 10^9\,{\rm GeV}$,
$10^{9}~\,{\rm GeV}~\lesssim~ M_2 ~\lesssim 10^{12}\,{\rm GeV}$ and  
$M_3 \gg 10^{12}\,{\rm GeV}$.
\footnote{For recent models realising crossing level solutions, where either two or all three RH neutrino masses
are  quasi-degenerate and $C\!P$ asymmetries are enhanced, 
see \cite{mohapatra,*buccella,*altameloni}.}
In this way the lightest  RH neutrino is too light to contribute significantly 
to the final asymmetry when successful leptogenesis is imposed. 
The heaviest RH neutrino also gives vanishing or in any case negligible contribution, since either it is not thermalised at all or, even if it is thermalised, its total $C\!P$ asymmetry is strongly suppressed. In this situation
the only RH neutrino species that can give a sizeable asymmetry able to explain the observed one is $N_2$ and   
in this way, the $SO(10)$-inspired conditions naturally realise the 
$N_2$-dominated scenario \cite{geometry}. This is crucially possible thanks to flavour effects 
\cite{bcst,roulet,*losada} that 
enlarge the regions where the lightest RH neutrino wash-out is negligible and the $N_2$-asymmetry can survive\cite{vives,bounds}.

In addition to a strong hierarchy of the mass spectrum,
in the approximation $V_L \simeq I$, the $N_2$-flavoured $C\!P$ 
asymmetries are also strongly hierarchical with 
$\ve_{2\t} \gg \ve_{2\mu} \gg \ve_{2e}$ and this results into a
tauon dominated final asymmetry \cite{SO10lep1}. 
Relaxing the approximation $V_L \simeq I$ this conclusion partially changes 
since a muon-dominated type solution becomes also possible. 
This, however, exists only  for  large values of the lightest neutrino mass 
$m_1 \gtrsim 0.01 \, {\rm eV}$ so that it is fair to say that 
$SO(10)$-inspired conditions typically imply a tauon $N_2$-dominated scenario. 
In the next Section we will discuss this result in detail.  

\subsection{Successful ST leptogenesis}

Let us now shortly discuss the two conditions imposed by leptogenesis,
successful and  ST leptogenesis conditions. It is convenient to discuss them separately, 
in a way to highlight more clearly the different steps
leading to the tauon $N_2$-dominated scenario as the only way to realise
successful ST leptogenesis.  For illustrative purposes it is convenient to start from
the ST condition. 

\subsubsection{ST leptogenesis condition}

Since we are assuming a hierarchical RH neutrino mass spectrum, with $M_3 \gtrsim 3\,M_2$ and 
$M_2 \gtrsim 3\,M_1$, the  decays and wash-out processes associated to each RH neutrino species occur sequentially,  with no overlap \cite{beyond}.  In this situation a pre-existing 
asymmetry, while temperature drops down in the expanding very early Universe, 
undergoes a multiple stage wash-out, involving sequentially each RH neutrino species, 
starting from the $N_3$-washout stage (if $N_3$ is thermalised) and ending with the $N_1$-washout stage. 

However, since in the one-flavoured regime (for $M_i \gtrsim 10^{12}\,{\rm GeV}$) 
each RH neutrino species is able to wash-out only a component of the asymmetry in a 
flavour space direction ${\ell}_i \propto (|m_{D e i}|^2, |m_{D \mu i}|^2, |m_{D \t i})|^2)$ \cite{bcst} 
and since, in general the three flavour directions 
$({\ell}_1, {\ell}_2, {\ell}_3)$ do not form an orthonormal basis
\footnote{On the other hand if they would form an orthonormal basis, there would be no interference
among the RH neutrinos and successful leptogenesis would be simply impossible since all $C\!P$ asymmetries
would vanish \cite{densitymatrix}.},  for simple geometric reasons the only possibility to fully wash-out a generic pre-existing asymmetry
is that (at least) the $N_1$ wash-out stage occurs in the three flavoured regime 
\cite{nardinir}. In this case
the most straightforward way to realise ST leptogenesis would be to impose strong $N_1$ wash-out in each flavour. This condition can be expressed as $K_{1\a} \gg 1$, having defined,  for any $\a=e,\m,\t $, 
the flavour decay parameters
\be\label{Kialpha}
K_{i\a}\equiv {\G_{i\a}+\overline{\G}_{i\a}\over H(T=M_i)}= 
{|m_{D\a i}|^2 \over M_i \, m_{\star}} \,  ,
\ee
where  $\Gamma_{i\a}=\Gamma (N_i \ra \phi^\dagger \, l_\alpha)$ 
and $\bar{\Gamma}_{i \a}=\Gamma (N_i \ra \phi \, \bar{l}_\alpha)$ are the
zero temperature limit of the flavoured decay rates into $\a$ leptons
and anti-leptons in the three-flavoured regime and $m_{\star}\simeq 1.1 \times 10^{-3}\,{\rm eV}$
is the equilibrium neutrino mass.
This simple set of conditions is, however, too restrictive to 
allow also  successful leptogenesis,
since the lightest RH neutrino washout would act strongly not only on all 
components of the pre-existing asymmetry but also on the
leptogenesis contribution from $N_2$-decays. 
\footnote{On the other hand this set up might be useful to have the wash-out of a  pre-existing $B-L$ asymmetry in some baryogenesis model, e.g. electroweak baryogenesis, occurring at some energy scale lower  than the $N_1$ wash-out.}

\subsubsection{Successful leptogenesis condition}

The only way to realise successful ST leptogenesis is to have the 
wash-out of the pre-existing asymmetry occurring in two separate steps. 
In a first step, imposing $K_{2\t} \gg 1$,  the tauon component of the pre-existing asymmetry 
is $N_2$ washed-out in the two-flavoured regime, implying 
$10^9\,{\rm GeV}\ll M_2 \ll 10^{12}\,{\rm GeV}$.   In the second step 
the $N_1$-washout stage has still to occur in the three flavoured regime
 ($M_1 \ll 10^9\,{\rm GeV}$) in a way that, imposing $K_{1e}, K_{1\m} \gg 1$,  it 
 can eventually suppress also the electron and muon components of the pre-existing asymmetries.
This time, however, the tauon asymmetry generated by the $N_2$-decays at the end of the $N_2$ wash-out stage, the genuine contribution from leptogenesis, can be sufficiently large to explain the observed asymmetry
if at the same time the $N_1$ wash-out in the tauon flavour is weak, i.e. if $K_{1\t} \lesssim 1$.
In this way the successful ST leptogenesis necessarily leads to a tauon $N_2$-dominated scenario, 
where the final asymmetry is dominated by the tauon flavour component produced by the 
out-of-equilibrium $N_2$-decays.   

It is quite interesting, and highly non-trivial, that $SO(10)$-inspired models naturally 
realise this kind of leptogenesis scenario and can, therefore, also  potentially 
realise successful ST leptogenesis. 
However, as we have seen, successful ST leptogenesis also requires the additional conditions
\be\label{STcond}
K_{1e}, K_{1\m}, K_{2\t} \gg 1  \, ,
\ee
and it is then to be verified whether these  can be met within $SO(10)$-inspired models. 
 
Summarising, we have a situation where 
both $SO(10)$-inspired and successful strong thermal leptogenesis might  be simultaneously realised for an 
interesting (intersection) region in the space of see-saw parameters. If this region exists, then it is clearly very interesting to understand what are the corresponding constrains on the low energy neutrino parameters.
In the following discussion,  our main objective will be to show analytically not only that such region exists, 
confirming the numerical results \cite{strongSO10solution}, 
but also that  it indeed implies definite constraints on the low energy neutrino 
parameters if the pre-existing $B-L$ asymmetry  is sufficiently large.  
We will derive these constraints in the approximation $V_L \simeq I$.
 Finally we will compare the analytical results
with the numerical results obtained in \cite{strongSO10solution} and also discuss how
the constraints get slightly relaxed going beyond the approximation $V_L \simeq I$.

%%%%%%%%%%%%%%%%%%%%%%%%%%%%%%%%%%%%%%%%%%%%%%%%%%%%%%%%%%%%%%%%%

\section{Low energy neutrino data}

As we will see the final asymmetry from $SO(10)$-inspired leptogenesis 
depends in such a  non trivial way on low energy neutrino parameters that
the successful leptogenesis conditions strongly links them, in  
a way that any experimental information on one parameter usually produces constraints
also on the other parameters. The main goal is to be able to place constraints 
that can be tested by future experiments. 
In this Section we briefly review the current experimental information on low energy neutrino parameters 
that we will employ for the derivation of the constraints. 

Neutrino oscillation experiments measure two mass squared differences, 
$\D m^2_{\rm tam}$ and $\D m^2_{\rm sol}$.  
Neutrino masses can then be either NO,  with 
$m^2_2 - m^2 _1 =  \Delta m^2_{\rm sol}$ and $m^2_3 - m^2_2 = \D m^2_{\rm atm}$,  or IO, with 
$m_3^2 - m_2^2 = \Delta m^2_{\rm sol}$ and $m^2_2 - m^2_1 = \D m^2_{\rm atm}$.
For example, in a recent global analysis \cite{nufit14} it is found
$m_{\rm atm}\equiv \sqrt{m^{\,2}_3 - m_1^{\,2}} \simeq 0.0495\,{\rm eV}$
and  $m_{\rm sol}\equiv \sqrt{\Delta m^2_{\rm sol}} \simeq 0.0087\,{\rm eV}$.

Finally, the cosmological observations place an upper bound on the sum of the neutrino masses and recently the Planck collaboration found $\sum_i m_i \lesssim 0.23 \cite{planck2013}
\,{\rm eV}$ that, combined with the measurements of 
$m_{\rm sol}$ and $m_{\rm atm}$,  translates into the upper bound
\be\label{m1upperbound}
m_1 \lesssim 0.07\,{\rm eV}  \,  .
\ee
For NO the leptonic mixing matrix can be parameterised in the usual standard way
\footnote{In the PDG parameterization the matrix of Majorana phases 
is defined as ${\rm diag}\left(1, e^{i\,{\a_{21}\over 2}}, e^{i\,{\a_{31}\over 2}}\right)$
and, therefore, one simply has $\a_{31} = 2(\s - \r)$ and $\a_{21} = -2\,\r$.}
\begin{equation}\label{Umatrix}
U^{\rm (NO)}=\left( \begin{array}{ccc}
c_{12}\,c_{13} & s_{12}\,c_{13} & s_{13}\,e^{-{\rm i}\,\d} \\
-s_{12}\,c_{23}-c_{12}\,s_{23}\,s_{13}\,e^{{\rm i}\,\d} &
c_{12}\,c_{23}-s_{12}\,s_{23}\,s_{13}\,e^{{\rm i}\,\d} & s_{23}\,c_{13} \\
s_{12}\,s_{23}-c_{12}\,c_{23}\,s_{13}\,e^{{\rm i}\,\d}
& -c_{12}\,s_{23}-s_{12}\,c_{23}\,s_{13}\,e^{{\rm i}\,\d}  &
c_{23}\,c_{13}
\end{array}\right)
\, {\rm diag}\left(e^{i\,\rho}, 1, e^{i\,\sigma}
\right)\, ,
\end{equation}
($s_{ij}\equiv \sin\theta_{ij}, c_{ij} \equiv \cos \theta_{ij}$) while for IO, within our convention for labelling  light neutrino masses and adopting the usual
definition for the thee mixing angles $\theta_{ij}$, 
%that proves to be more convenient within our context than the usual PDG one, 
the columns of the leptonic mixing matrix have to be permuted in a way that
\be
U^{\rm (IO)} =  \left( \begin{array}{ccc}
s_{13}\,e^{-{\rm i}\,\d} & c_{12}\,c_{13} & s_{12}\,c_{13}  \\
s_{23}\,c_{13} & -s_{12}\,c_{23}-c_{12}\,s_{23}\,s_{13}\,e^{{\rm i}\,\d} &
c_{12}\,c_{23}-s_{12}\,s_{23}\,s_{13}\,e^{{\rm i}\,\d} \\
c_{23}\,c_{13} & s_{12}\,s_{23}-c_{12}\,c_{23}\,s_{13}\,e^{{\rm i}\,\d}
& -c_{12}\,s_{23}-s_{12}\,c_{23}\,s_{13}\,e^{{\rm i}\,\d}
\end{array}\right)
\, {\rm diag}\left(e^{i\,\sigma}, e^{i\,\rho}, 1  \right) \,  .
\ee

 The mixing angles, respectively the reactor, the solar and the atmospheric ones, 
are now measured with the following best fit values and $1\s$ ($3\s$) ranges 
\cite{foglilisi2013} for NO and IO respectively,
\bea\label{expranges}
\theta_{13} & = &  8.8^{\circ}\pm 0.4^{\circ} \, \;\;  (7.6^{\circ}\mbox{--}9.9^{\circ}) 
 \;\; \mbox{\rm and} \;\;
 \theta_{13}  =    8.9^{\circ}\pm 0.4^{\circ} \, \;\; (7.7^{\circ}\mbox{--}9.9^{\circ}) \,  ,
 \\ \nonumber
\theta_{12} & = &  33.7^{\circ}\pm 1.1^{\circ} \,  \;\;  (30.6^{\circ}\mbox{--}36.8^{\circ}) 
\;\; \mbox{\rm and} \;\;
\theta_{12}  =   33.7^{\circ}\pm 1.1^{\circ} \,  \;\;  (30.6^{\circ}\mbox{--}36.8^{\circ}) \,  , 
\\ \nonumber
\theta_{23} & = &  {41.4^{\circ}}^{+1.9^{\circ}}_{-1.4^{\circ}} \,  \;\;  
(37.7^{\circ}\mbox{--}52.3^{\circ}) 
\;\; \mbox{\rm and} \;\;
\theta_{23}  =   
                 {42.4^{\circ}}^{+8.0^{\circ}}_{-1.8^{\circ}}  
                  \;\;  (38.1^{\circ}\mbox{--}52.3^{\circ})  \,  .
 \eea 
It is interesting that current experimental data also start to put constraints on the
Dirac phase and the following best fit values and $1\s$ errors are found  for NO and IO respectively,
\be\label{delta}
\d/\pi = -0.61^{+0.38}_{-0.27} \,\,  
\;\;\mbox{\rm and} \;\;
\d/\pi = -0.69^{+0.29}_{-0.33} \,  ,
\ee
while all values $[-\pi,+\pi]$ are still allowed at $3\,\s$.

 \section{From $SO(10)$-inspired conditions to RH neutrino masses and $C\!P$ flavoured asymmetries}

In this section we show how the $SO(10)$-inspired conditions imply, in the hierarchical case, 
a $N_2$-dominated  RH neutrino mass spectrum \cite{branco}. 
Only for particular conditions on the low energy neutrino parameters, there exist crossing level solutions, 
in vicinity of which two or even all three RH neutrino masses are quasi-degenerate \cite{afs}.  
We derive compact analytic expressions both for the RH neutrino masses and for their $C\!P$ asymmetries and compare them with the numerical results for some selected examples.  

\subsection{RH neutrino masses}

Inserting the bi-unitary parameterisation eq.~(\ref{biunitary}) into the  diagonalised seesaw formula eq.~(\ref{diagonalseesaw}) one obtains an expression for the  
(symmetric) inverse Majorana mass matrix in the Yukawa basis, 
$M^{-1} = U_R \, D^{-1}_M \, U^{T}_R$, in terms  of the unitary matrix $V_L$, the 
low energy neutrino mass matrix $m_{\nu}=-U\,D_m\,U^T$ and the three neutrino Yukawa eigenvalues $m_{D i}$,
\be\label{Mm1}
M^{-1} = D^{-1}_{m_D}  \, V_L \, U\, D_m \, U^T \, V_L^T \, D^{-1}_{m_D} \,  .
\ee
This can be also easily inverted obtaining for the Majorana mass matrix 
in the Yukawa basis, $M = U^{\star}_R \, D_M \, U_R^{\dagger}$, 
\be\label{M}
M = D_{m_D} \, V^{\star}_L \, U^{\star} \, D_m^{-1} \, U^{\dagger} \, V_L^{\dagger} \, D_{m_D} \,  .
\ee
From these expressions,  either from $M^{-1}$ or from $M$, one can derive the RH neutrino mass spectrum 
and the RH neutrino mixing matrix $U_R$, as a function of the 9 low energy neutrino parameters in $m_{\nu}$ 
(6 mixing parameters in $U$ and 3 light neutrino masses $m_i$),  the 6 parameters in the unitary matrix $V_L$ and the 3 Dirac neutrino masses $m_{D i}$. 

This can be done diagonalising the hermitian matrix $M^{\dagger}\,M = U_R\,D_M^2\,U_R^{\dagger}$ 
(or equivalently $M^{-1}\,(M^{-1})^{\dagger}=U_R\,D_M^{-2}\,U_R^{\dagger}$). 
For a given $U_R$, any matrix $\widetilde{U}_R = U_R\,D_{\phi}^{-1}$, where 
\be
D_\phi \equiv (e^{-i\,{\Phi_1\over 2}},e^{-i\,{\Phi_2\over 2}},e^{-i\,{\Phi_3\over 2}})  
\ee 
is a generic diagonal unitary matrix, also diagonalises $M$ and $M^{-1}$.  However, going back to the (Takagi) diagonalisation 
$M=U^{\star}_R \, D_M \, U_R^{\dagger}$ and given a $\widetilde{U}_R$, 
one can unambiguously fix \cite{SO10lep2}
\be\label{phases}
D_{\phi} = \sqrt{D_M\,\widetilde{U}_R^{\dagger}\,M^{-1}\,\widetilde{U}_R^{\star}}  \,  .
\ee 
If one is not in the vicinity of crossing level solutions, where at least two RH neutrino masses become 
degenerate, the RH neutrino mass spectrum is strongly hierarchical and analytical expressions can be easily found  \cite{branco,afs}.
Here we adopt a slightly different procedure that yields simplified and more compact expressions in terms both of the light neutrino mass and of the inverse light neutrino mass matrix entries. If we start from  the eq.~(\ref{M}) for $M$,  in the approximation $V_L \simeq I$, we can write
\be\label{Mdiag}
U^{\star}_R\,D_M\,U^{\dagger}_R \simeq D_{m_D}  \, U^{\star} \, D_m^{-1} \, U^{\dagger} \,  D_{m_D}  \,  .
\ee
Considering that from the definition of $U$ (cf. eq.~(\ref{leptonic})) one easily finds
\be\label{inversemnu}
m_{\nu}^{-1} = - U^{\star}\,D_m^{-1} \, U^{\dagger} \, ,
\ee
the eq.~(\ref{Mdiag}) can be also written more compactly as 
\be\label{takagidiag}
M = U^{\star}_R \, D_M \, U^{\dagger}_R  \simeq  
- D_{m_D}  \, m_{\nu}^{-1} \,  D_{m_D} \,  .
\ee
This equation shows that $M_{i3}/M_{33} = M_{3i}/M_{33}\propto m_{Di}/m_{D3}$ and, therefore, in first approximation the LH side is in a block diagonal form and, neglecting terms ${\cal O}(m_{D1}/m_{D3},m_{D2}/m_{D3})$ 
one finds 
\be\label{M3}
M_3 \simeq m^2_{D3}\,|(m_{\nu}^{-1})_{\t\t}| = m^2_{D3}\,
\left| {(U^{\star}_{\t 1})^2\over m_1}+ {(U^{\star}_{\t 2})^2\over m_2} + {(U^{\star}_{\t 3})^2\over m_3} \right| 
\propto \a_3 ^2 \, m_{t}^2 \,  .
\ee
At the same time the phase $\Phi_3$ is also specified and one simply has
\be\label{Phi3}
\Phi_3 = {\rm Arg}[-(m_{\nu}^{-1})_{\t\t}] \, .
\ee
The same procedure can be adopted for $M^{-1}$, rewriting the eq.~(\ref{Mm1}) 
in the approximation $V_L \simeq I$  and imposing the Takagi diagonalisation
\be\label{invMtakagi}
M^{-1} = U_R\, D_M^{-1} \, U_R^T \simeq D^{-1}_{m_D}  \, U\, D_m \, U^T \, D^{-1}_{m_D} = 
-  D^{-1}_{m_D}  \, m_{\nu} \, D^{-1}_{m_D} \, .
\ee
This time the RH side is approximately in a block-diagonal form with
$M^{-1}_{i1}/M^{-1}_{11} = M^{-1}_{1i}/M^{-1}_{11} \propto m_{D1}/m_{Di}$, so that 
the largest $M^{-1}$ eigenvalue, $1/M_1$, can be written as $1/M_1 \simeq |m_{\nu ee}|/m_{D 11}^2$ and, therefore,
\be\label{M1}
M_1 \simeq {m_{D 1}^2 \over |m_{\nu ee}|} = {m_{D1}^2 \over 
|m_1\,U^2_{e1} + m_2 \, U^2_{e2} + m_3\, U^2_{e3}|} \propto \a_1^2 \, m_u^2 \, .
\ee 
Also in this case the procedure allows to specify the phase $\Phi_1$,
\be
\Phi_1 = {\rm Arg}[-m_{\nu ee}^{\star}] \, . 
\ee
Finally, from the approximate expressions eq.~(\ref{M3}) for $M_3$ and eq.~(\ref{M1}) for $M_1$,  one can also easily find an approximate expression for $M_2$. 
From the see-saw formula eq.~(\ref{diagonalseesaw}) one has
\be\label{sumphases}
m_1 \, m_2 \, m_3 = {m^2_{D1}\,m^2_{D2}\,m^2_{D3}\over M_1 \, M_2 \, M_3} \, 
e^{i\,(2\,\widetilde{\Phi}_R-2\,\Phi_U-\sum_i\,\Phi_i)}  ,
\ee
where $\widetilde{\Phi}_R \equiv {\rm Arg}[{\rm det}(\widetilde{U}_R)]$ 
and $\Phi_U \equiv {\rm Arg}[{\rm det}(U)]=\r+\s$, implying $\sum_i \Phi_i = 
2\,(\widetilde{\Phi}_R - \Phi_U)$.
 In this way we can write
\be\label{M2}
M_2 \simeq {m^2_{D2}\over m_1\,m_2\, m_3}\,{|m_{\nu ee}| \over |(m_{\nu}^{-1})_{\t\t}|}
= m^2_{D2} \, { |m_1\,U^2_{e1} + m_2 \, U^2_{e2} + m_3\, U^2_{e3}| \over 
|m_2\,m_3\,U^{\star \, 2}_{\t 1} + m_1\,m_3\, U^{\star \, 2}_{\t 2} + m_1 \, m_2 \,  U^{\star \, 2}_{\t 3}|} 
\propto \a_2^{\, 2} \, m_{c}^2 \,  ,
\ee
and for the phase $\Phi_2 = 2\,(\widetilde{\Phi}_R - \Phi_U) - \Phi_3 - \Phi_1$ one finds
\be\label{Phi2}
\Phi_2 = {\rm Arg}\left[{m_{\nu ee}\over (m_{\nu}^{-1})_{\t\t}}\right] +2\,\widetilde{\Phi}_R-2\,(\rho+\s) \,  .
\ee
It is easy to see from the above general expressions, that in the hierarchical limit, $m_1 \ll m_{\rm sol}$
(remember that we are assuming NO), the RH neutrino masses tend to the following simple expressions \cite{branco,afs}
\bea\label{Milowm1}
M_1  & \simeq &   {m^2_{D1} \over |m_{\rm sol} \,s^2_{12}\,c^2_{13}+
m_{\rm atm}\, s^2_{13}\, e^{i\,(2\,\s-\d)}|}
 \approx   \, {m^2_{D1} \over m_{\rm sol} \,s^2_{12}} \, ,  \\ \nonumber
M_2 & \simeq & {m^2_{D2}\,|m_{\rm sol}\,s^2_{12}\,c^2_{13}+ m_{\rm atm}\,s^2_{13}\,e^{i\,(2\,\s-\d)}| \over m_{\rm sol}\,m_{\rm atm} \, |s_{12}\,s_{23} - c_{12}\,c_{23}\,s_{13}\,e^{-i\,\d}|^2} 
  \approx  \, {m^2_{D2} \over m_{\rm atm} \, s^2_{23}} \, ,  \\  \nonumber
M_3 & \simeq  & {m^2_{D3} \, |s_{12}\,s_{23} - c_{12}\,c_{23}\,s_{13}\,e^{-i\,\d}|^2 \over m_1}    \approx   {m^2_{D3} \over m_1}\, s^2_{12} \, s^2_{23} \,  ,
\eea
where the last ones are obtained within the (rough) approximation $s_{13}\simeq 0$.
In Fig.~1 we compare the found approximated analytic expressions for the RH neutrino masses (cf. eqs.~(\ref{M3}), (\ref{M1}) and (\ref{M2})) with the numerical solutions for the simple four sets of parameters yielding level crossings  for special values of $m_1$  as discussed in \cite{afs} (note that for simplicity 
$\theta_{13}=0$ and $\theta_{23}=\p/4$).  
For the up quark masses at the leptogenesis scale, 
\footnote{
In the case of $SO(10)$-inspired models this is approximately given by
$T_{L} \sim (3$--$10)\times 10^{10}\,{\rm GeV}$ as we will
show later.}
we adopted  the values $(m_u, m_c, m_t) = (1\,{\rm MeV}, \, 400\,{\rm MeV}, 100\,{\rm GeV})$ \cite{quarkmasses}.
It can be noticed how the analytic solutions (dashed black lines) track perfectly the numerical ones 
(solid coloured lines) except in the close vicinity
of those values of $m_1$ where the RH neutrino masses become  quasi-degenerate and
the validity of the adopted approximations breaks down.
\begin{figure}
\vspace*{-19mm}
\begin{center}
\psfig{file=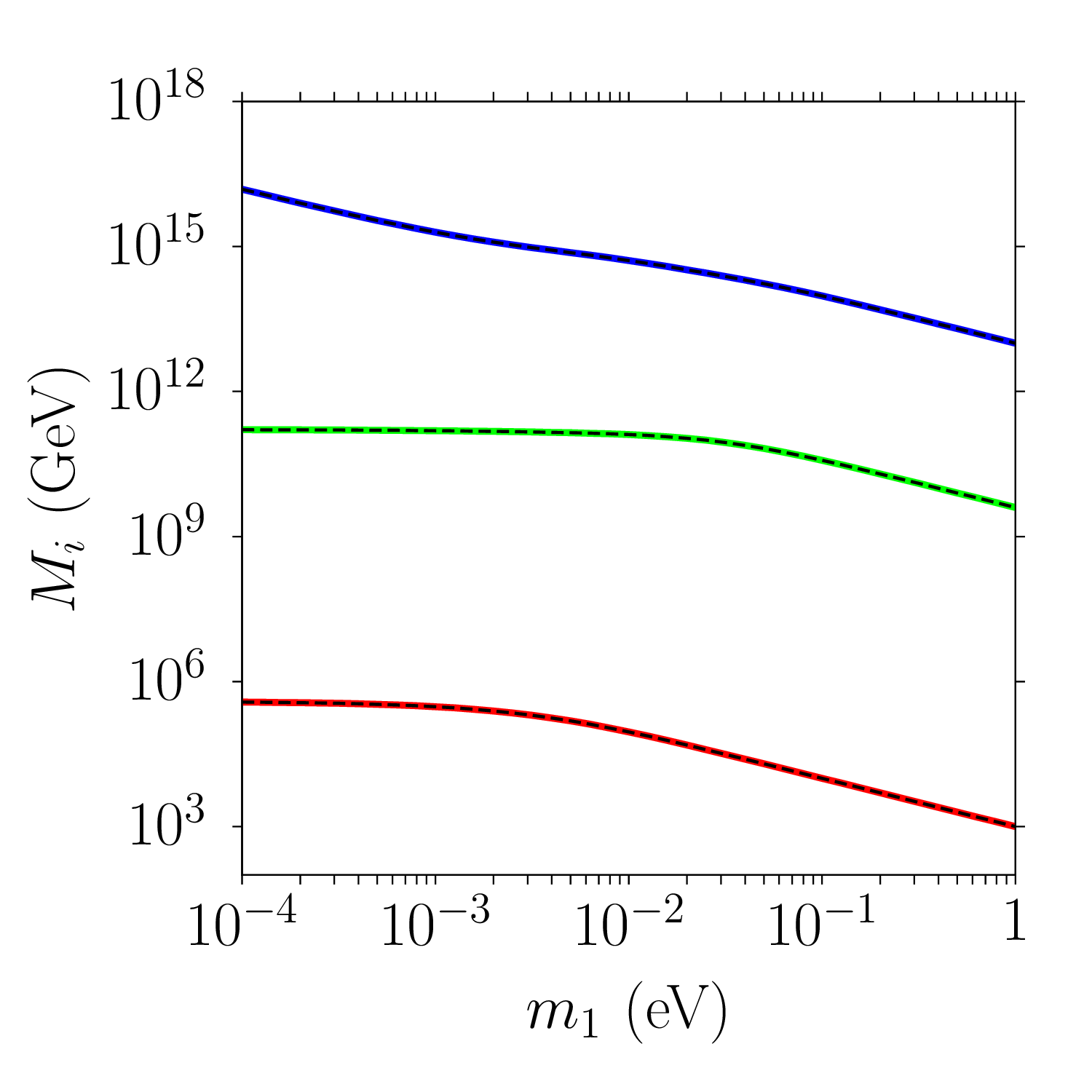,height=54mm,width=60mm}
\hspace{8mm}
\psfig{file=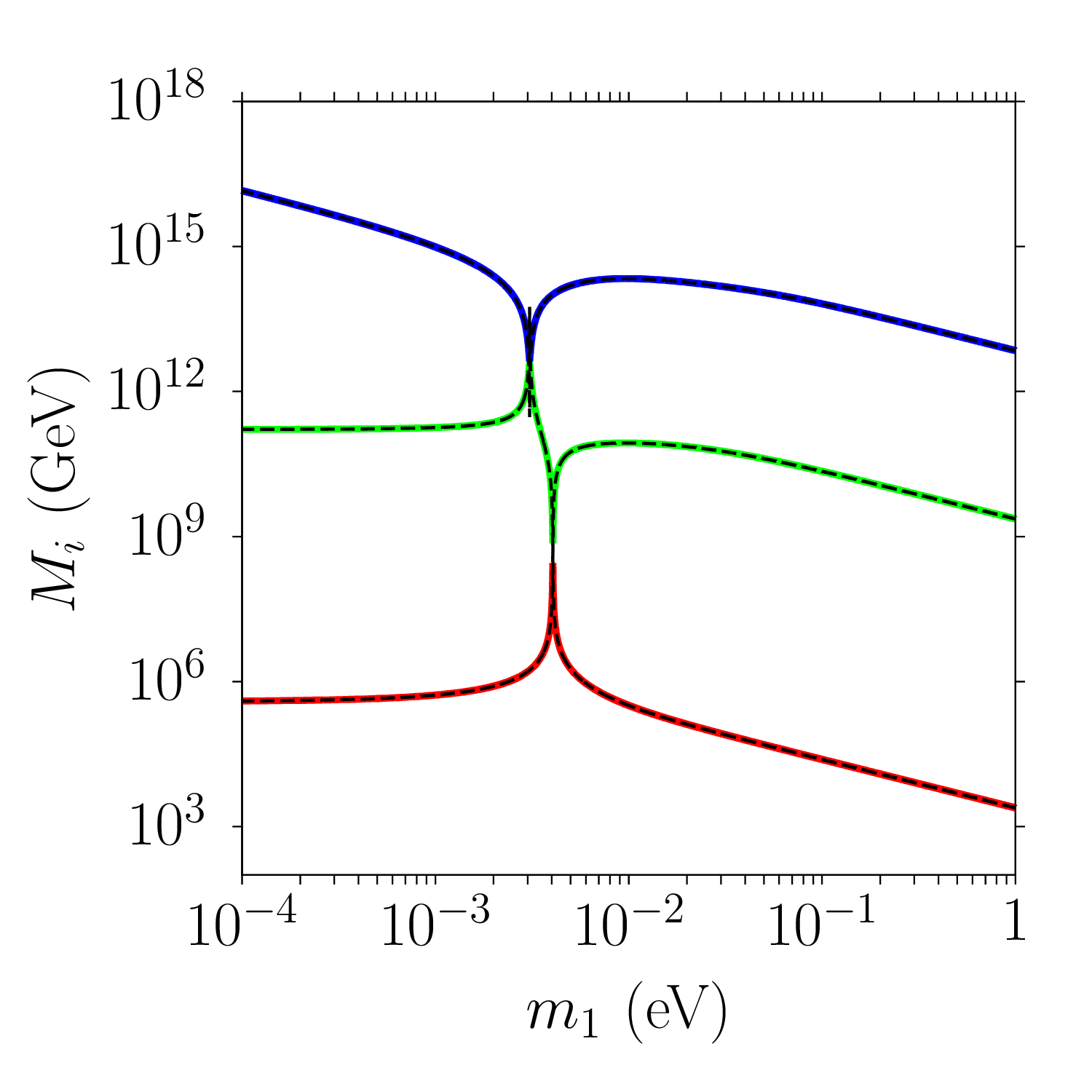,height=54mm,width=60mm}
\\
\vspace*{-3mm}
\psfig{file=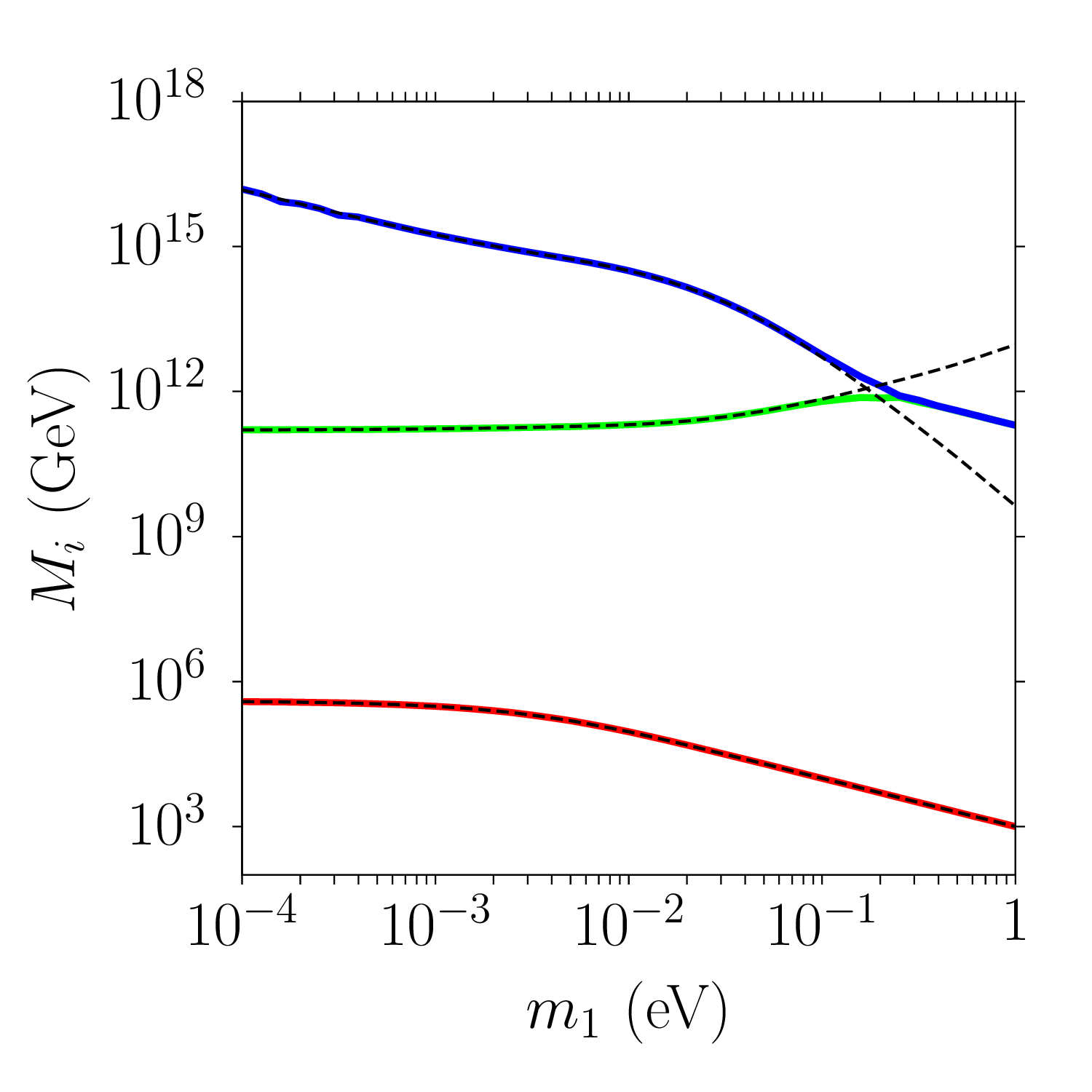,height=54mm,width=60mm} 
\hspace{8mm}
\psfig{file=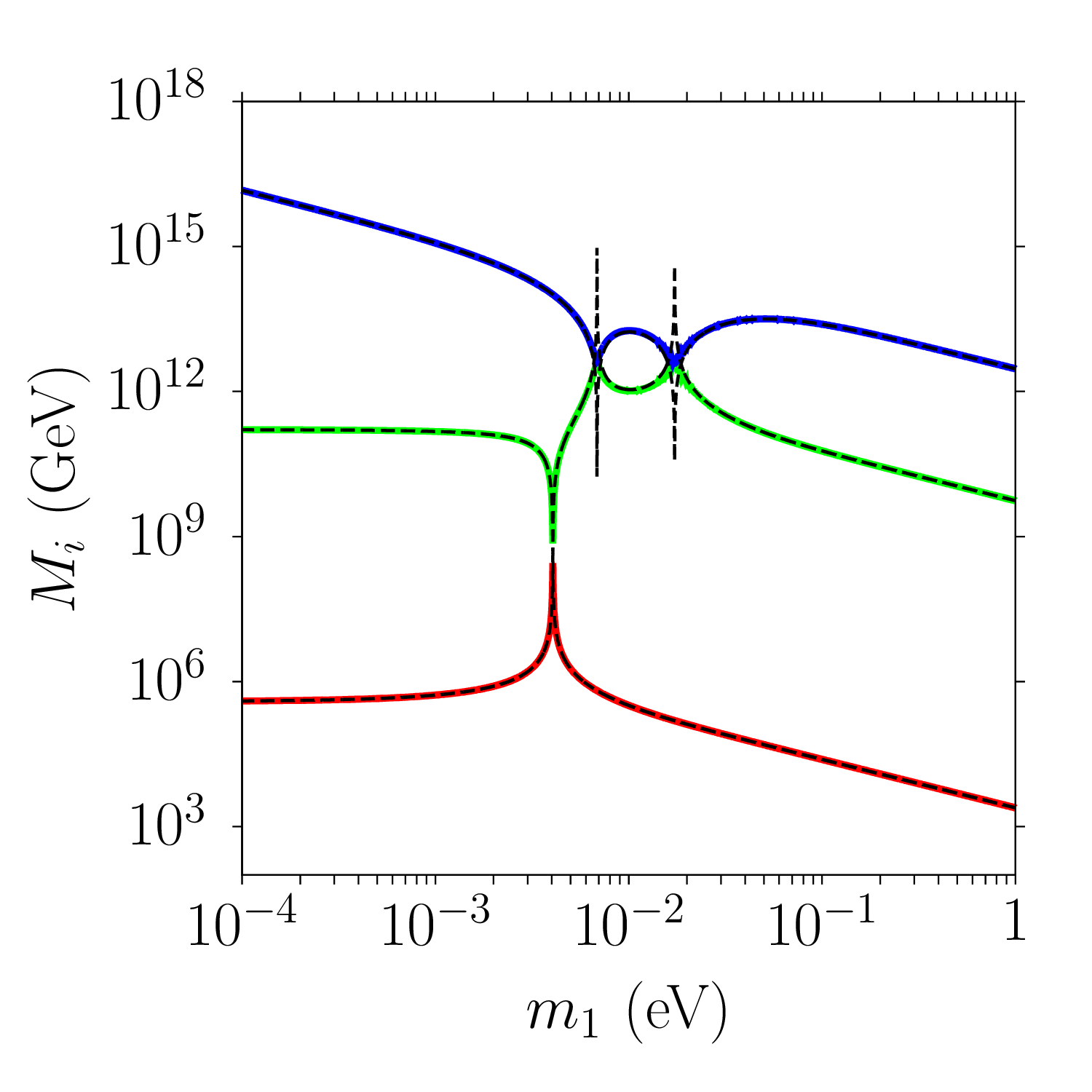,height=54mm,width=60mm}
\end{center}
\vspace{-10mm}
\caption{Comparison between the numerical solutions for the RH neutrino masses
(solid lines) and the analytical solutions eqs. (\ref{M3}), (\ref{M2}) and 
(\ref{M1}) (dashed lines). The solutions are obtained for $\theta_{13}=0$, 
$\theta_{23}=45^{\circ}$, $\theta_{12}=33^{\circ}$, 
$\a_1 = \a_2 = \a_3 = 1$, $V_L = I$ and for $(\rho,\s)=(0,0), (\pi/2,0),(0,\pi/2),(\pi/2,\pi/2)$ 
from top left to bottom right respectively.}
\end{figure}
 In the panels of Fig.~2 we show the same comparison but this time with 
 three solutions realising successful (non resonant) $SO(10)$-inspired leptogenesis, around some (indicated) 
 values of $m_1$ for NO. The first two cases realise two different types of tauon $N_2$-dominated leptogenesis solutions so called type A and type B \cite{SO10lep1,SO10lep2} that we will 
 fully describe analytically in Section 5.
The third case is a strong thermal $SO(10)$-inspired solution \cite{,strongSO10solution} 
(realised for $m_1 \simeq 10\,{\rm meV}$
and $N^{\rm p,i}_{B-L}=0.001$). As we will discuss in Section 6, this can only emerge within 
type A solutions.  As one can see this time there are  no level crossings  
 and the analytic solutions perfectly track the numerical ones for any value of $m_1$. 
\begin{figure}
\begin{center}
\psfig{file=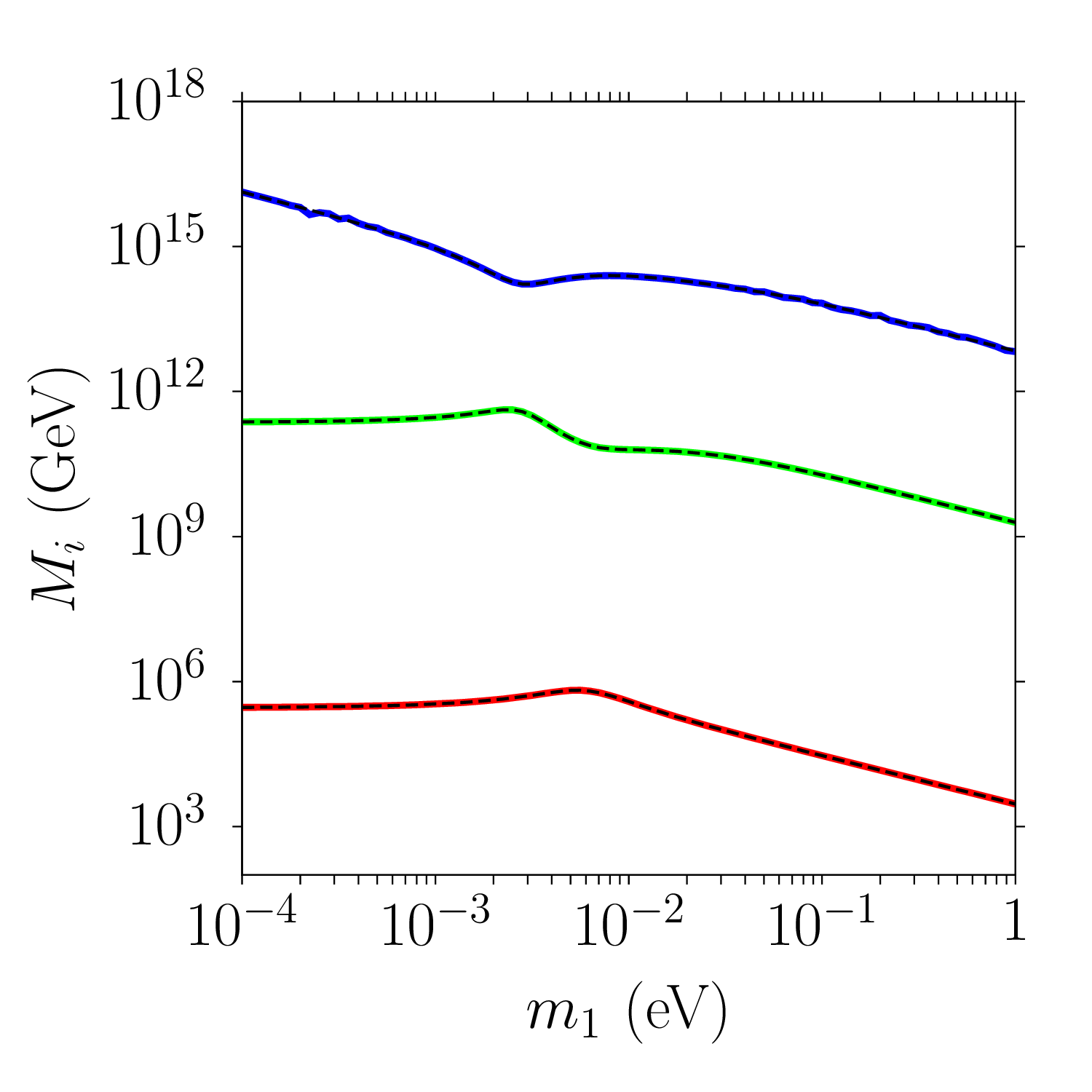,height=42mm,width=49mm}
\hspace{4mm}
\psfig{file=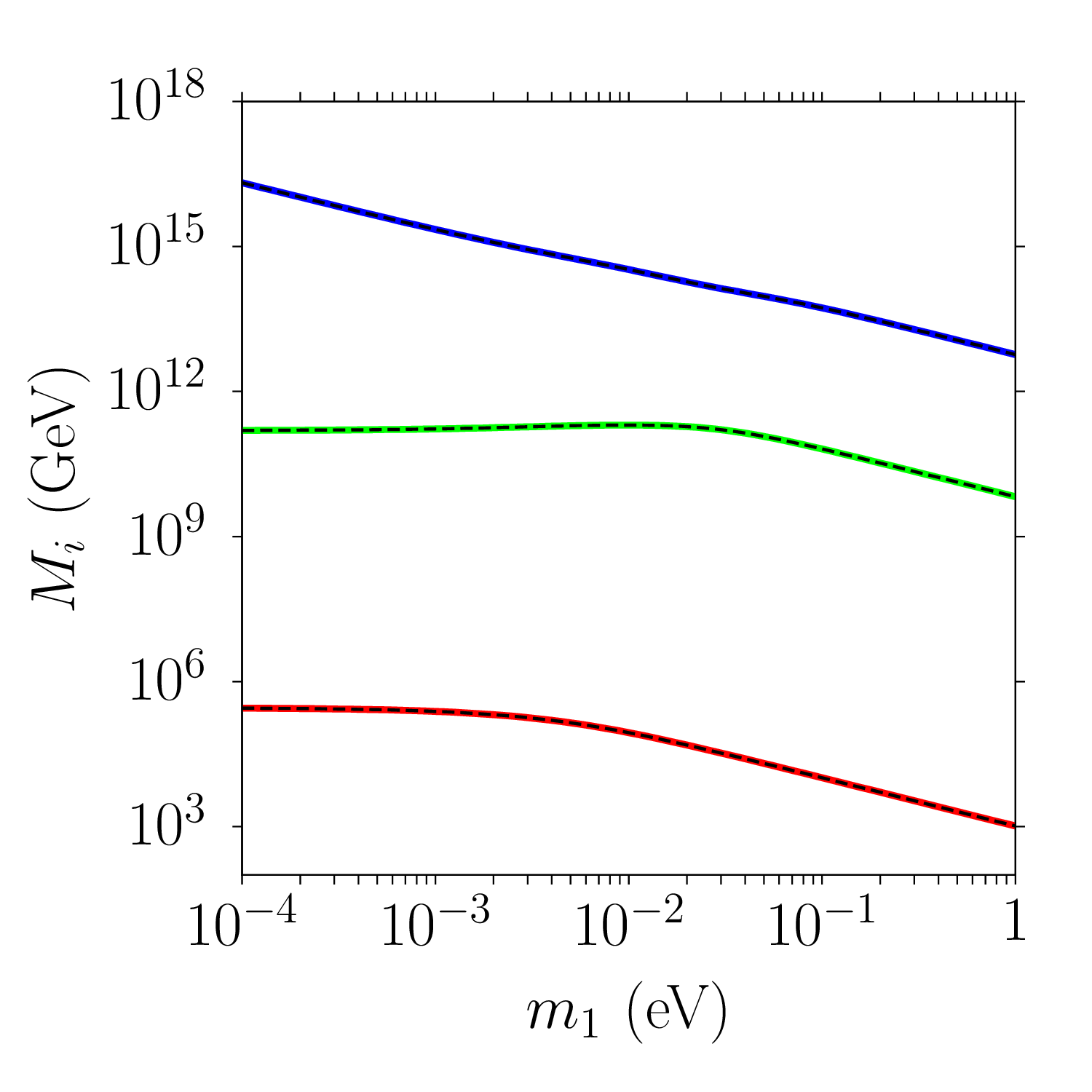,height=42mm,width=49mm} 
\hspace{4mm}
\psfig{file=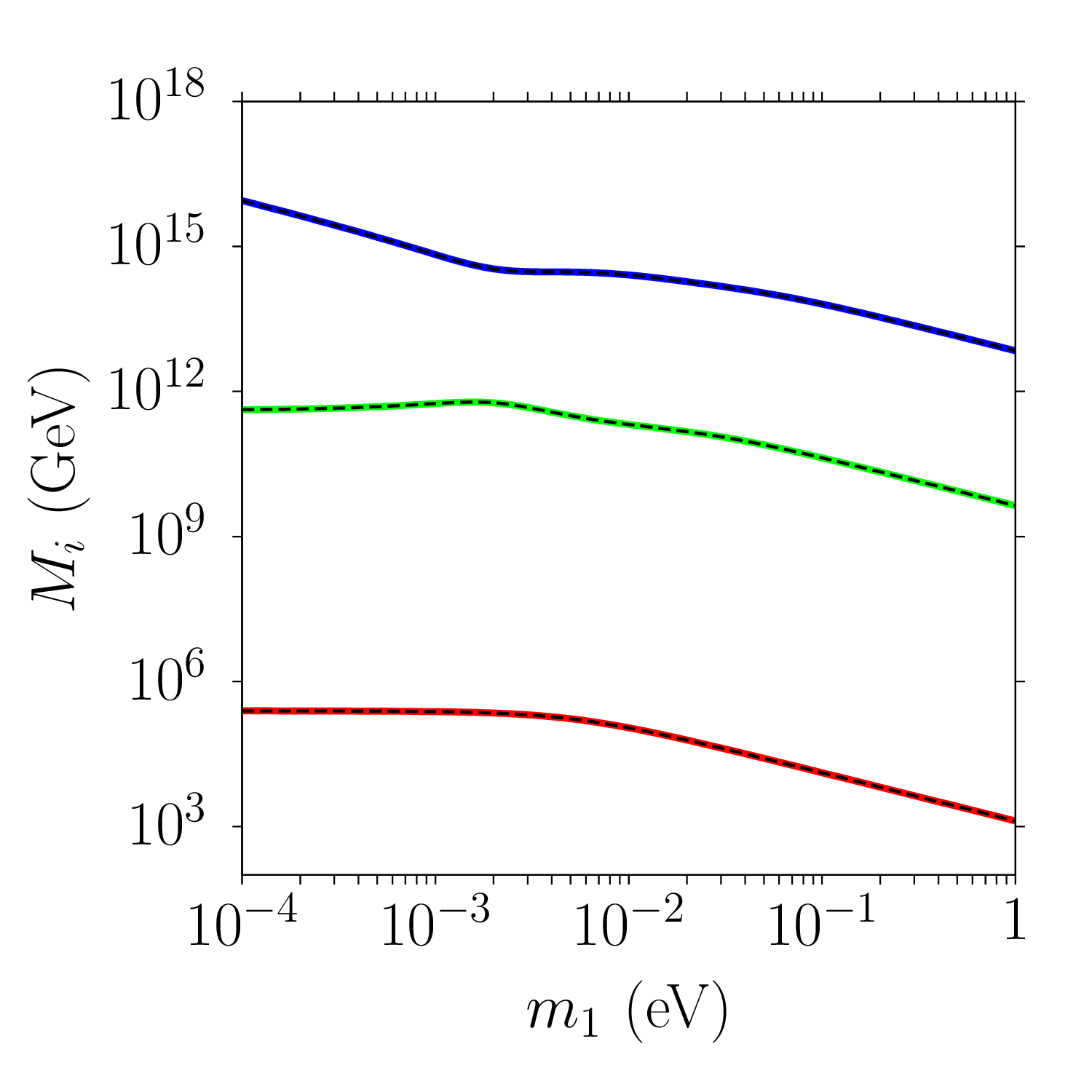,height=42mm,width=49mm} 
\end{center}
\vspace{-10mm}
\caption{
Comparison of the analytical expressions for the RH neutrino masses 
(cf. eqs.(\ref{M3}), (\ref{M2}), (\ref{M1}), dashed lines) with the numerical solutions (solid lines) 
versus $m_1$ for the three following sets of parameters: $V_L = I$,
$(\a_1,\a_2,\a_3) = (1,5,1)$, 
$\theta_{13}=(7.55^{\circ},8.14^{\circ},9.2^{\circ})$,
$\theta_{12}=(35.2^{\circ},34.75^{\circ},35.0^{\circ})$,
$\theta_{23}=(46.2^{\circ},42.1^{\circ},40.0^{\circ})$,
$\delta/\pi=(0.275,0.067,-0.24)$,
$\rho/\pi=(0.54,1.080,0.24)$,
$\s/\pi=(1.14,0.94,0.80)$. These three solutions 
are examples of  $\tau_A$, $\tau_B$
and strong thermal solutions respectively and realise successful leptogenesis 
for $m_1 \simeq (2.5,300,10)\,{\rm meV}$. All three cases are for NO.}
\end{figure}
 These results show explicitly how we can safely adopt the 
 analytic solutions in our following discussion, though it should be 
 made clear that the comparison is made for $V_L = I$ and, therefore,
 at this stage we are not testing the validity of the approximation $V_L \simeq I$
that will instead be discussed in 8.1. 

It is also possible to find an analytical approximate expression for 
the RH neutrino mixing matrix $U_R$. From the discussion above it should be
clear that $U_R$ is of the form $U_R= I + \xi$, where $\xi_{ii}=0$
and the $\xi_{i\neq j}$ leading terms are suppressed   $\propto m_{Di}/{m_{Dj}}$ with $j > i$ in a way that $U_R$ is well approximated by 
\be\label{URapp}
U_R \simeq \left( \begin{array}{ccc}
1 & -{m_{D1}\over m_{D2}} \,  {m^\star_{\nu e \mu }\over m^\star_{\nu ee}}  & 
{m_{D1}\over m_{D3}}\,
{ (m_{\n}^{-1})^{\star}_{e\t}\over (m_{\n}^{-1})^{\star}_{\t\t} }   \\
{m_{D1}\over m_{D2}} \,  {m_{\nu e \mu }\over m_{\nu ee}} & 1 & 
{m_{D2}\over m_{D3}}\, 
{(m_{\n}^{-1})_{\m\t}^{\star} \over (m_{\n}^{-1})_{\t\t}^{\star}}  \\
 {m_{D1}\over m_{D3}}\,{m_{\nu e\t }\over m_{\nu e e}}  & 
- {m_{D2}\over m_{D3}}\, 
 {(m_\nu^{-1})_{\m\t}\over (m_\nu^{-1})_{\t\t}} 
  & 1 
\end{array}\right) \, D_{\Phi} ,
\ee
equivalent to the expression in \cite{afs} but where we identified 
neutrino mass matrices combinations with entries of the inverse neutrino mass matrix,
something that will prove very useful when we will extract 
the constraints on the low energy neutrino parameters. 
Details on the derivation can be found in the Appendix. It should be noticed that  the phases
$\Phi_i$ are now specialised in a way that 
$\widetilde{\Phi}_R \simeq 0$, so that the eq.~(\ref{Phi2}) for $\Phi_2$ becomes
\be\label{Phi2bis}
\Phi_2 =  {\rm Arg}\left[{m_{\nu ee}\over (m_{\nu}^{-1})_{\t\t}}\right] -2\,(\rho+\s) \, .
\ee
 It can be also useful to calculate the orthogonal matrix $\O$
within $SO(10)$-inspired models. Starting from the orthogonal parameterisation 
for the neutrino Dirac mass matrix in the charged lepton basis \cite{ci}, 
$m_D=U\,\sqrt{D_m}\,\O\,\sqrt{D_M}$ where $\O\,\O^T =I$, 
and comparing with the bi-unitary 
parameterisation eq.~(\ref{biunitary}), one finds straightforwardly 
an expression for the orthogonal parameterisation \cite{SO10lep1}
$\O =  D_m^{-{1\over 2}}\,U^{\dagger}\,V^{\dagger}_L\,
D_{m_D}\,U_R \, D_M^{-{1\over 2}}$, 
that in the approximation $V_L \simeq I$ simplifies into
$\O \simeq  D_m^{-{1\over 2}}\,U^{\dagger}\,
D_{m_D}\,U_R \, D_M^{-{1\over 2}} $, 
that in term of the entries can be written as 
\be\label{Oij}
\O_{ij} \simeq {1\over \sqrt{m_i \, M_j}} \, 
\sum_k m_{D k} \, U^{\star}_{ki}\,U_{R k j} \, .
\ee
From the eq.~(\ref{URapp}) one can then find this approximate  expression for $\O$ (see Appendix),
\be\label{Omegaapp}
\O \simeq 
\left( \begin{array}{ccc}
- {\sqrt{m_1\,|m_{\nu ee}|}\over m_{\nu ee}}\,U_{e 1}  & 
\sqrt{m_2\,m_3\,|(m_{\nu}^{-1})_{\t\t}| \over |m_{\nu ee}|}\,
\left(U^{\star}_{\m 1} - U^{\star}_{\t1}\,{{(m_{\nu}^{-1})_{\m\t}}\over (m_{\nu}^{-1})_{\t\t}}\right) & 
{U^{\star}_{31}\over \sqrt{m_1\,|(m_{\nu}^{-1})_{\t\t}|}} \\
- {\sqrt{m_2\,|m_{\nu ee}|}\over m_{\nu ee}}\, U_{e 2} & 
\sqrt{m_1\,m_3\,|(m_{\nu}^{-1})_{\t\t}| \over |m_{\nu ee}|}\,
\left(U^{\star}_{\m 2} - U^{\star}_{\t2}\,{{(m_{\nu}^{-1})_{\m\t}}\over (m_{\nu}^{-1})_{\t\t}}\right)  
& {U^{\star}_{32}\over \sqrt{m_2\,|(m_{\nu}^{-1})_{\t\t}|}}  \\
- {\sqrt{m_3\,|m_{\nu ee}|}\over m_{\nu ee}}\,U_{e 3} & 
\sqrt{m_1\,m_2\,|(m_{\nu}^{-1})_{\t\t}| \over |m_{\nu ee}|}\,
\left(U^{\star}_{\m 3} - U^{\star}_{\t 3}\,{{(m_{\nu}^{-1})_{\m\t}}\over (m_{\nu}^{-1})_{\t\t}}\right)  
& {U^{\star}_{33}\over \sqrt{m_3\,|(m_{\nu}^{-1})_{\t\t}|}}  
\end{array}\right)  \,  D_{\Phi},
\ee 
that in the limit $m_1 \ra 0$ correctly reduces to 
the two RH neutrino limit form \cite{Turzinsky}
\be
\O \xrightarrow{m_1 \ra 0} 
\left( \begin{array}{ccc}
0 & 0 & 1 \\ 
1 + {\cal O}(\theta_{13}) & {\cal O}(\theta_{13}) & 0 \\
 {\cal O}(\theta_{13}) & 1 + {\cal O}(\theta_{13}) 
  & 0 
\end{array}\right)  \,  .
\ee
As we said in the introduction and clearly shown in the examples of Fig.~2, barring regions around
crossing level  solutions, the $SO(10)$-inspired RH neutrino mass spectrum
naturally realises the $N_2$-dominated scenario. This is because $N_1$ is too light to produce a sizeable asymmetry. At the same time, since
$M_3 \gg 10^{12}\,{\rm GeV}$  and the $N_3$ total asymmetry is strongly
suppressed as $\ve_{3} \propto (M_2/M_3)^2$, 
the  $N_3$-decays contribution to the final asymmetry is also negligible.
In this way the only possibility to reproduce the final asymmetry relies on the 
$N_2$-production that occurs in the two-flavoured regime since 
$M_2 \sim 10^{10\mbox{--}11}\,{\rm GeV}$. 

In the $N_2$-dominated scenario the contribution to the asymmetry from 
leptogenesis can  be calculated as  the sum of the 
three (charged lepton) flavoured asymmetries $\D_\a \equiv B/3 - L_\a$.
Normalising the abundance $N_X$ of a generic quantity $X$ in a way that
in the ultra-relativistic thermal equilibrium the abundance of RH neutrinos 
$N^{\rm eq}_{N_i}(T \gg M_1) = 1$,  
the final asymmetry produced by the decays of the ($N_2$) RH neutrinos
can then be written, in terms of the $C\!P$ asymmetries 
$\ve_{2\a}$ and  the efficiency factors $\k(K_{2\a})$ at the production, 
as \cite{vives,bounds,fuller,densitymatrix}
\bea\label{twofl} \nonumber
N_{B-L}^{\rm lep, f} & \simeq &
\left[{K_{2e}\over K_{2\tau_2^{\bot}}}\,\ve_{2 \tau_2^{\bot}}\kappa(K_{2 \tau_2^{\bot}}) 
+ \left(\ve_{2e} - {K_{2e}\over K_{2\tau_2^{\bot}}}\, \ve_{2 \tau_2^{\bot}} \right)\,\kappa(K_{2 \tau_2^{\bot}}/2)\right]\,
\, e^{-{3\pi\over 8}\,K_{1 e}}+ \\ \nonumber
& + &\left[{K_{2\mu}\over K_{2 \tau_2^{\bot}}}\,
\ve_{2 \tau_2^{\bot}}\,\kappa(K_{2 \tau_2^{\bot}}) +
\left(\ve_{2\mu} - {K_{2\mu}\over K_{2\tau_2^{\bot}}}\, \ve_{2 \tau_2^{\bot}} \right)\,
\kappa(K_{2 \tau_2^{\bot}}/2) \right]
\, e^{-{3\pi\over 8}\,K_{1 \mu}}+ \\
& + &\ve_{2 \tau}\,\kappa(K_{2 \tau})\,e^{-{3\pi\over 8}\,K_{1 \tau}} \,  ,
\eea
where $K_{2\tau_2^{\bot}} \equiv K_{2e} + K_{2\mu}$ and 
$\ve_{2\tau_2^{\bot}} \equiv \ve_{2e} + \ve_{2\mu}$.
%As we will show soon, the strong thermal condition implies $K_{1e}, K_{1\m} \gg 1$ and, %therefore, 
%in this case the contribution to the asymmetry from leptogenesis 
%simply reduces to 
The baryon-to-photon number ratio can then be simply calculated 
as $\eta_B \simeq 0.01 \, N_{B-L}^{\rm lep,f}$. This is true assuming that any
contribution from external sources can be neglected, a point that we will  
address in detail when we will discuss the strong thermal leptogenesis condition. 

\subsection{Flavoured $C\!P$ asymmetries}

It is now interesting to calculate the $N_2$ (flavoured) $C\!P$ asymmetries 
within $SO(10)$-inspired models and in particular to see how these are linked to the 
low energy neutrino parameters. Defining them as 
\be
\ve_{2\a}\equiv -{\G_{2\alpha}-\overline{\G}_{2\alpha}
\over \G_{2}+\overline{\G}_{2}} \,  ,
\ee
these can be calculated from \cite{crv}
\be\label{eps2a}
\ve_{2\a} \simeq
\overline{\ve}(M_2) \, \left\{ {\cal I}_{23}^{\a}\,\x(M^2_3/M^2_2)+
\,{\cal J}_{23}^{\a} \, \frac{2}{3(1-M^2_2/M^2_3)}\right\}\, ,
\ee
where we introduced
\be
\overline{\ve}(M_2) \equiv {3\over 16\,\pi}\,{M_2\,m_{\rm atm} \over v^2} \, , \hspace{3mm} \xi(x)=\frac{2}{3}x\left[(1+x)\ln\left(\frac{1+x}{x}\right)-\frac{2-x}{1-x}\right] \,  ,
\ee
\be
{\cal I}_{23}^{\a} \equiv   {{\rm Im}\left[m_{D\a 2}^{\star}
m_{D\a 3}(m_D^{\dag}\, m_D)_{2 3}\right]\over M_2\,M_3\,\mtt\,m_{\rm atm} }\,   
\hspace{5mm}
\mbox{\rm and}
\hspace{5mm}
{\cal J}_{23}^{\a} \equiv  
{{\rm Im}\left[m_{D\a 2}^{\star}\, m_{D\a 3}(m_D^{\dag}\, m_D)_{3 2}\right] 
\over M_2\,M_3\,\mtt\,m_{\rm atm} } \,{M_2\over M_3}   ,
\ee
with $\mtt \equiv (m_D^{\dag}\, m_D)_{2 2}/M_2$.
%The quantities ${\cal I}_{23}^{\a}$ and ${\cal J}_{23}^{\a}$ can  be expressed in the %orthogonal 
%parameterisation as \cite{bounds,diraclep}
%\be
%{\cal I}_{23}^{\a} =   {\rm Im} \Big[ \sum\limits_{k,h,l}
%{m_{k}\,\sqrt{m_{h}\,m_{l}} \over \mtt \, m_{\rm atm}}\,
%\,\O^*_{k2}\,\O_{k3}\,\O^*_{h2}\,\O_{l 3}\,U_{\alpha h}^* \, U_{\alpha l} \Big] \, ,
%\ee
%\be
%{\cal J}_{23}^{\a} =
%{\rm Im} \Big[ \sum\limits_{k,h,l} {m_{k}\,\sqrt{m_{h}m_{l}} \over \mtt \, m_{\rm %atm}}\,
%\, \O^*_{k3}\,\O_{k2}\,\O^*_{h2}\,\O_{l 3} \, U_{\alpha h}^* \, U_{\alpha l}\Big] \, .  
%\ee
%We can also conveniently define $\ve_{2\tau^{\bot}} \equiv \ve_{2e}+\ve_{2\mu}$ and
%$K_{2\tau^{\bot}}\equiv K_{2e}+K_{2\m}$, where $\tau^{\bot}$ indicates a $\tau$ %orthogonal
%flavour component  that is a coherent superposition of electron and muon components,
%in this specific  case those ones of the leptons ${\ell}_2$ produced in the $N_2$ %decays.
Since in our case  $M_3 \gg M_2$,  we can use the approximation $\xi(M^2_3/M_2^2)~\simeq~1$ and neglect the second term $\propto {\cal J}_{23}^{\a}$. Moreover, making use of the bi-unitary parameterisation (cf. eq.~(\ref{biunitary})) and the approximation $V_L \simeq I$,
one arrives to the following approximated expression for the flavoured
$C\!P$ asymmetries in $SO(10)$-inspired models,
\be\label{CPanalytic}
\ve_{2\a} \simeq \overline{\ve}(M_2) \, {m^2_{D\a}\over m^2_{D3}\,|U_{R 32}|^2 + m^2_{D2}}\,
{|(m_{\nu}^{-1})_{\t\t}|^{-1} \over \,m_{\rm atm}}\,
{\rm Im}[U^{\star}_{R \a 2}\,U_{R \a 3}\,U^{\star}_{R 3 2}\,U_{R 3 3}]  
  \,  .
\ee
Using the approximated expression eq.~(\ref{URapp}) for $U_R$  and the 
relations (\ref{upquarkmasses}), one finds the following hierarchical pattern for the $\ve_{2\a}$'s:
\be\label{CPhierarchy}
\ve_{2\t}:\ve_{2\m}:\ve_{2e} = \a_3^{\,2}\,m^2_t : \a_2^{\, 2}\,m^2_c :
\a_1^{\,2}\,m^2_u \, {\a_3 m_t \over a_2 \,m_c} \, 
{\a_1^{\,2}\,m_u^2 \over \a_2^{\,2}\,m^2_c} \, .
\ee 
As one can see, while $\ve_{2\m}$ is suppressed by about
four orders of magnitude ($\sim m^2_c /m^2_t$) compared to $\ve_{2\t}$, 
the electronic $C\!P$ asymmetry is suppressed even by about seven orders of magnitude compared to 
$\ve_{2\m}$. For this reason the electron contribution to the final asymmetry 
is always completely negligible.  
\footnote{It is curious to notice that since the contribution to $\ve_{2e}$ from the interference with
$N_3$ is so suppressed, actually it becomes comparable to the term coming from the interference with $N_1$
that we are neglecting in the eq.~(\ref{eps2a}).}
This is well shown in Fig.~3 where, for the same four
sets of parameters of Fig.~2, the flavoured (and total) $C\!P$ asymmetries
are plotted versus $m_1$, comparing the numerical result (solid lines) with the analytic expressions eq.~(\ref{CPanalytic}) (dashed lines). 
\begin{figure}
\begin{center}
\psfig{file=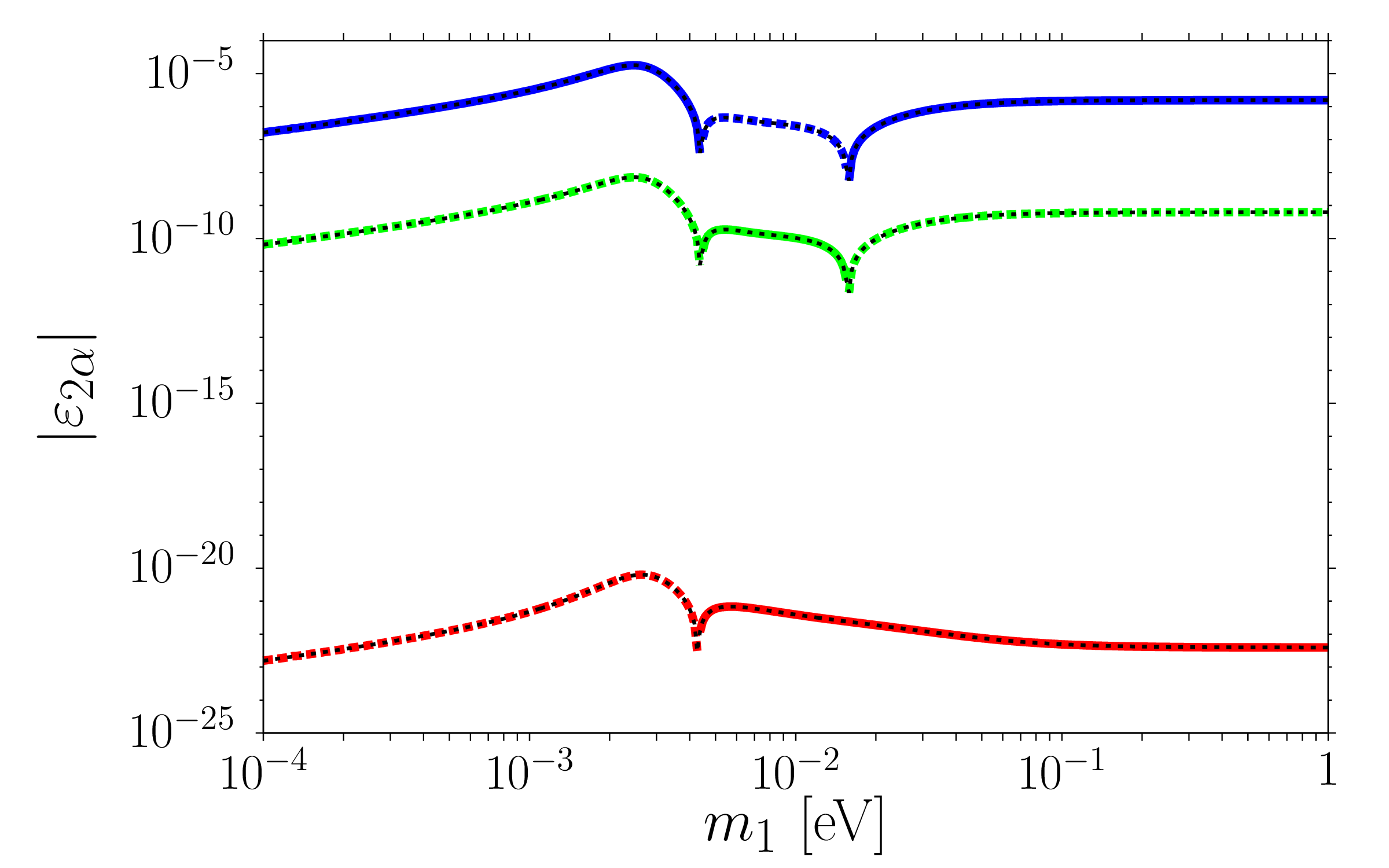,height=39mm,width=46mm}
\hspace{5mm}
\psfig{file=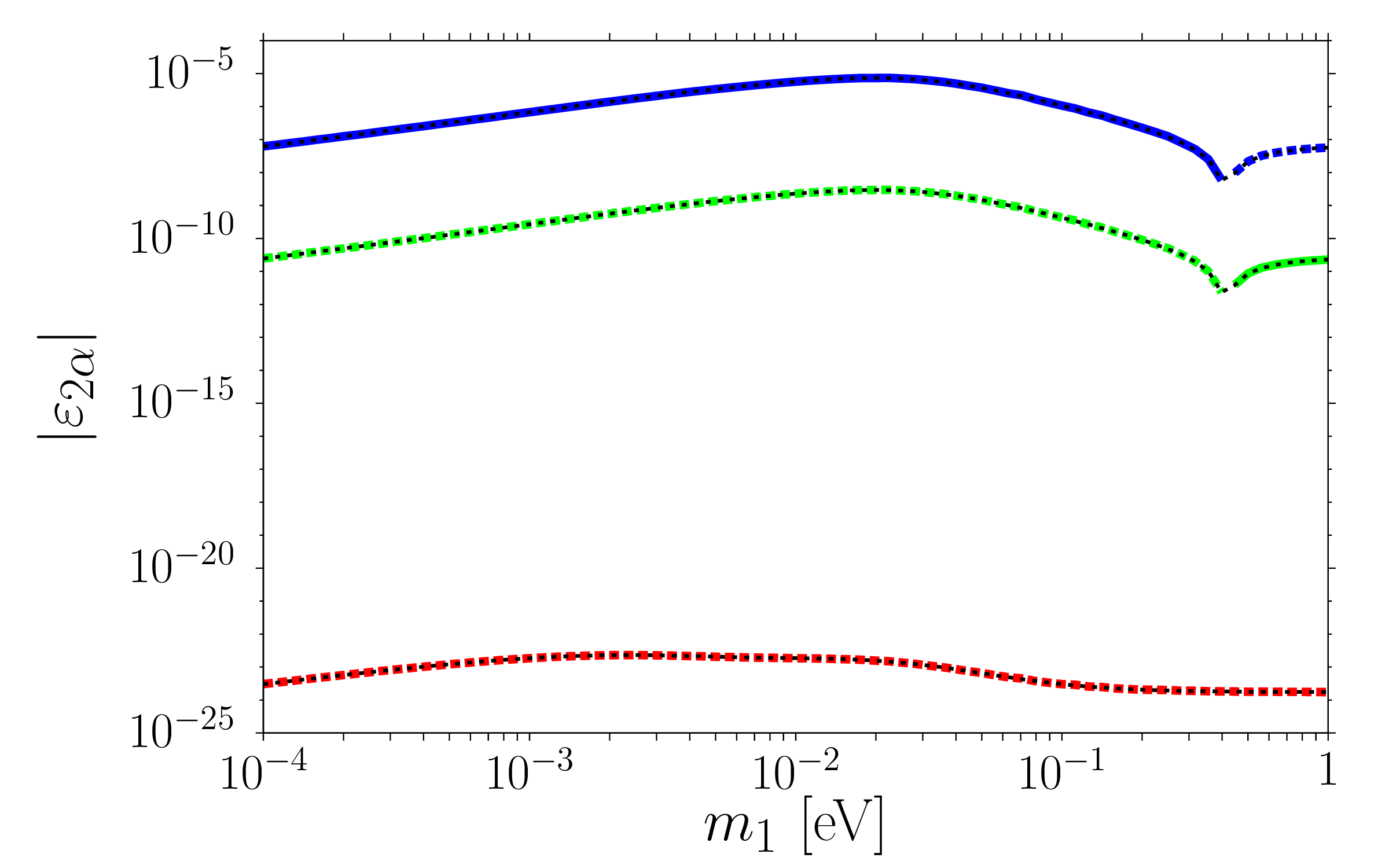,height=39mm,width=46mm} 
\hspace{5mm}
\psfig{file=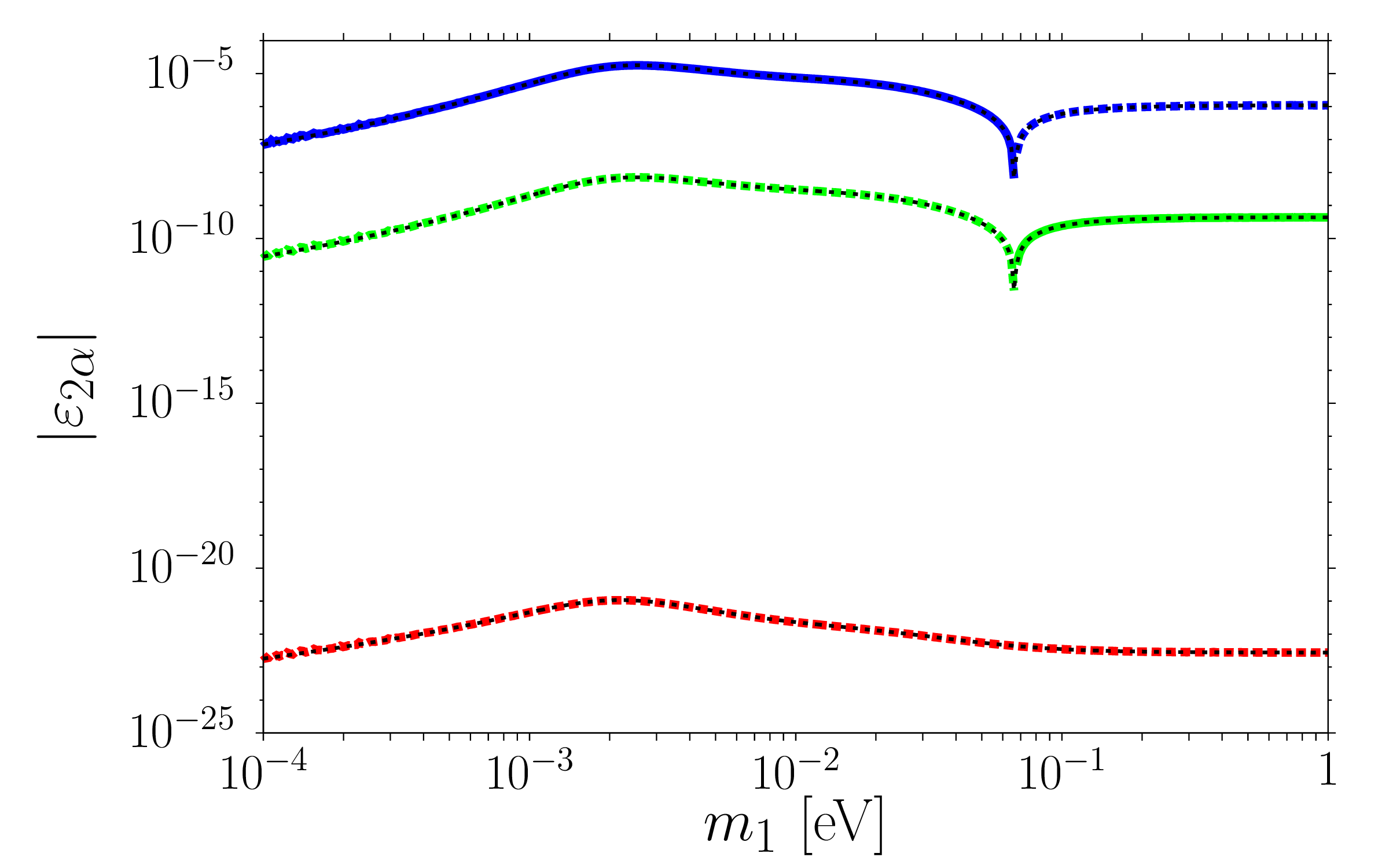,height=39mm,width=46mm} 
\end{center}
\vspace{-5mm}
\caption{Plots of the $C\!P$ flavoured asymmetries corresponding respectively to the same three sets of parameters of Fig.~2. The solid coloured lines are the numerical curves
(blue, green and red lines correspond respectively to tauon, muon and electron flavours).
The dashed lines are the analytical expressions eqs.~(\ref{CPanalytic}).}
\end{figure}
One can see how the analytic expressions again reproduce very well the numerical results. 
In particular one can recognise the hierarchical pattern eq.~(\ref{CPhierarchy}).

\section{Successful leptogenesis condition}

In this Section we finally impose the successful leptogenesis condition finding some first 
interesting constraints on the low energy parameters.  In this respect we extend 
the results found in the hierarchical (LH) neutrino masses limit \cite{SO10lep1,SO10lep2}
to  arbitrary values of $m_1$.  The final asymmetry should be calculated using the 
eq.~(\ref{twofl}).  However, as we have seen, in the approximation $V_L \simeq I$ the tauon $C\!P$
asymmetry is by far the dominant one and the inclusion of the wash-out at the production 
cannot change the $\t$-dominance as a contribution to the final $B-L$ asymmetry.
However, it should be stressed that this result holds using the $V_L \simeq I$ approximation, while
relaxing this approximation, a muon-dominated solution also appears 
for $m_1 \gtrsim 10\,{\rm meV}$ \cite{SO10lep2}. In any case
the ST condition will select the tauon-dominated solution anyway 
and for this reason we can neglect the muon-dominated solution in our discussion. 
%\footnote{We will be back on this point later on, when we will discuss how the results change when the %approximation $V_L \simeq I$ is relaxed.} 
In this way the expression for the final asymmetry eq.~(\ref{twofl}) greatly simplifies into
\be\label{NBmLlepftau}
N_{B-L}^{\rm lep,f} \simeq \ve_{2 \tau}\,\kappa(K_{2 \tau})\,e^{-{3\pi\over 8}\,K_{1 \tau}} \,  .
\ee
Using the explicit expressions eqs.~(\ref{URapp}) and (\ref{Omegaapp}) for the $U_R$ and $\O$ matrices respectively, we are now able to express the final $B-L$ asymmetry in $SO(10)$-inspired models
in terms of the  $\a_i$ and the low energy neutrino parameters. 

\subsection{Final asymmetry in terms of the low energy neutrino parameters}

%\subsubsection{The source term: $N_2$ tauon $C\!P$ asymmetry}

Let us start from the derivation of an expression of $\ve_{2\tau}$. First of all we can
specialise the eq.~(\ref{CPanalytic}) to the case $\a=\t$, obtaining 
\be
\ve_{2\tau} \simeq {3\over 16\,\pi}\,{M_2 \over v^2} 
{m^2_{D 3}\over m^2_{D3}\,|U_{R 32}|^2 + m^2_{D2}}\,
{1 \over |(m_{\nu}^{-1})_{\t\t}|}\,
{\rm Im}[(U^{\star}_{R 3 2}\,U_{R 3 3})^2]    \,  .
\ee
Using then the expressions found for $U_R$ and $M_2$, we arrive to
\be\label{ve2tausinalpha}
\ve_{2\tau} \simeq {3\over 16\,\pi}\, {\a_2^2\,m_c^2 \over v^2}\, {|m_{\nu ee}|\,
(|m^{-1}_{\nu \t \t}|^2 + |m^{-1}_{\nu \m \t}|^2)^{-1} \over m_1\,m_2\,m_3}
\, {|(m_{\nu}^{-1})_{\m \t}|^2 \over |(m_{\nu}^{-1})_{\t \t}|^2} 
\, \sin \a_L \,  ,
\ee
where we have introduced the {\em effective $SO(10)$-inspired 
leptogenesis phase $\a_L$} (in the approximation $V_L = I$)
\be
\a_L \equiv  2\,{\rm Arg}[(m^{-1}_{\nu})_{\t\t}]  - 2\,{\rm Arg}[(m^{-1}_{\nu})_{\m\t}]
+ \Phi_2 - \Phi_3 \,  ,
\ee 
%Using the expressions (\ref{leptonic}) and (\ref{inversemnu})
%one can easily find for the $m_{\nu}$ and $m_{\nu}^{-1}$ entries 
%\be
%(m_{\nu})_{\a \b}  = - \sum_{k} U_{\a k}\,U_{\b k}\, m_k \, \;\;\;\; \mbox{\rm and} \;\;\;\;\;
%(m^{-1}_{\nu})_{\a \b}  = - \sum_{k} U^{\star}_{\a k}\,U^{\star}_{\b k}\,{1\over m_k} \,  .  
%\ee
and where from the eqs.~(\ref{Phi3}) and (\ref{Phi2bis}) one has
\be
\Phi_2 - \Phi_3 = {\rm Arg}\left[m_{\nu ee}\right]
-2\, {\rm Arg}\left[(m_{\nu}^{-1})_{\t\t}\right]  + \pi -2\,(\rho+\s) \,  ,
\ee
so that we can write for the effective leptogenesis phase
\be
\a_L =  {\rm Arg}\left[m_{\nu ee}\right]  - 2\,{\rm Arg}[(m^{-1}_{\nu})_{\m\t}] + \pi -2\,(\rho+\s)  \,  .
\ee 

%\subsubsection{Wash-out at the production}

Let us now calculate the efficiency factor at the production $\k(K_{2\t})$. 
First of all, from the 
expression eq.~(\ref{Kialpha}), one can easily find, for $V_L \simeq I$,  
a general expression for the $K_{i\a}$'s, 
\be\label{Kiaapp}
K_{i\a} \simeq {m^2_{D\a}\over m_{\star}\,M_i} \, |U_{R \a i}|^2 \,  .
\ee 
From this one we can then obtain a specific expression for
\be\label{K2tau}
K_{2\t} \simeq {m^2_{D3} \over m_{\star} \, M_2} \, |U_{R 3 2}|^2  
\simeq {m_1\,m_2\,m_3 \over m_{\star}}\, 
{|(m_{\nu}^{-1})_{\m \t}|^2 \over |m_{\nu ee}|\, |(m_{\nu}^{-1})_{\t \t}|}  \,  ,
\ee
where in the second approximation we made use of the eqs.~(\ref{URapp})
and (\ref{M2}). 

%\subsubsection{Lightest RH neutrino wash-out}

From the general expression eq.~(\ref{Kiaapp}) we can also write an expression for $K_{1\t}$
describing the exponential suppression of the lightest RH neutrino wash-out 
(cf. eq.~(\ref{NBmLlepftau}))
\be\label{K1tau}
K_{1\t}   \simeq  {m^2_{D3} \over m_{\star} \, M_1} \, |U_{R 3 1}|^2  
\simeq {|m_{\nu e\t}|^2 \over m_{\star}\,|m_{\nu ee}|}  
= {|m_1\,U_{e1}\,U_{\t 1}+m_2\,U_{e2}\,U_{\t2} +m_3\,U_{e3}\,U_{\t3}|^2 
\over m_{\star}\,|m_1 \, U_{e1}^2 + m_2\,U_{e2}^2 + m_3\,U_{e3}^2|}\,   .
\ee
From this one we can then obtain an explicit expression in terms of the mixing angles and low energy phases
that will prove useful,
\be\label{K1tauexplicit}
K_{1\t} \simeq {|c_{13}\,c_{12}\,s_{12}\,s_{23}\,(m_1\,e^{2\,i\,\rho}-m_2) 
+ s_{13}\,c_{13}\,c_{23}\,(m_3\,e^{i\,(2\s-\d)} - m_2\,s^2_{12}\,e^{i\,\d} - m_1\,c^2_{12}\,e^{i\,(2\,\rho+\d)})|^2\over m_{\star}\,|m_1\,c^2_{12}\,c^2_{13}\,e^{2\,\,i\,\rho}+ m_2\,s^2_{12}\,c^2_{13}+ m_3\,s^2_{13}\,e^{2\,i\,(\s-\d)}|} \, .
\ee

%\subsection{Constraints on low energy neutrino parameters}

We can then finally put together all the results finding, from the eq.~(\ref{NBmLlepftau}),
an expression in terms of the low energy neutrino parameters that can be written as
\bea\label{NBmLflow}
N_{B-L}^{\rm lep,f} & \simeq & 
{3\over 16\,\pi}\, {\a_2^2\,m_c^2 \over v^2}\, {|m_{\nu ee}|\,
(|m^{-1}_{\nu \t \t}|^2 + |m^{-1}_{\nu \m \t}|^2)^{-1} \over m_1\,m_2\,m_3}\,
{|m^{-1}_{\n\t\t}|^2\over |m^{-1}_{\n\m\t}|^2}\,\sin\a_L    \\  \nonumber
& \times & \kappa\left({m_1\,m_2\,m_3 \over m_{\star}}\, 
{|(m_{\nu}^{-1})_{\m \t}|^2 \over |m_{\nu ee}|\, |(m_{\nu}^{-1})_{\t \t}|} \right)  \\  \nonumber
& \times & 
e^{-{3\pi\over 8}\,{|m_{\nu e\t}|^2 \over m_{\star}\,|m_{\nu ee}|}  }  \,  .
\eea 
It is interesting to notice that:
\begin{itemize}
\item The asymmetry does not depend on $\a_1$ and on $\a_3$ \cite{SO10lep1}.
This is an important point since the only left non-observable parameter is $\a_2$ on which however
one can place a lower bound and, within $SO(10)$-inspired models cannot be in any case too large. 
\item The effective neutrinoless double beta decay mass $m_{ee} \equiv |m_{\nu ee}|$ plays  a direct role and it can be noticed that successful leptogenesis implies the existence
of a lower bound. 
\item The successful leptogenesis condition links together all low energy neutrino parameters 
constraining them to lie on a  hypersurface described by the only left theoretical parameter $\a_2$.  
\end{itemize}
In Fig.~4 we have plotted $\eta_B$ vs. $m_1$ for the same three sets of parameters of Figs. 2 and 3 comparing
the numerical results (blue solid lines) with the analytical results (black dashed lines) obtained from the 
eq.~(\ref{NBmLflow}). As one can see the analytical results perfectly match the numerical ones. 
\begin{figure}
\begin{center}
\psfig{file=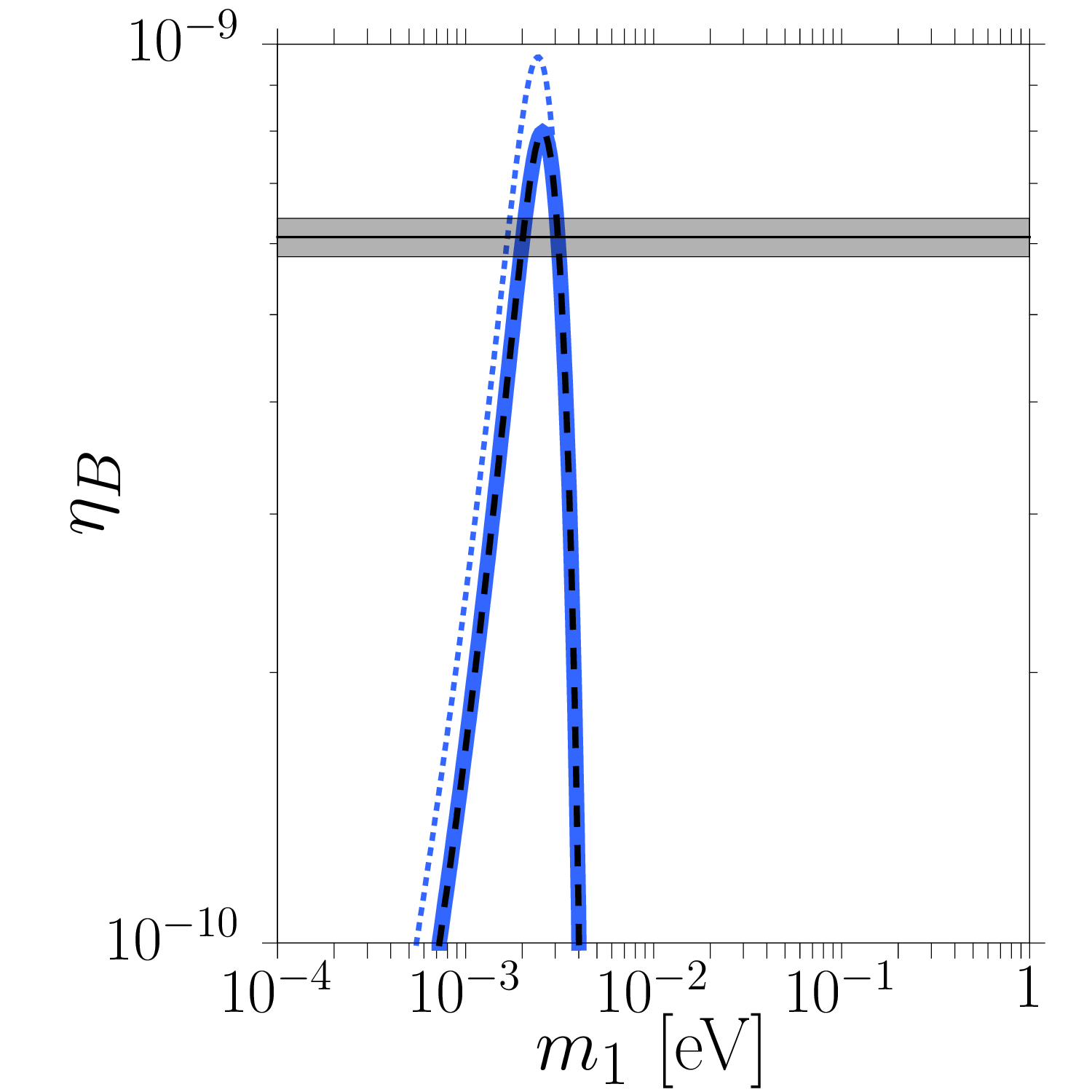,height=39mm,width=46mm}
\hspace{5mm}
\psfig{file=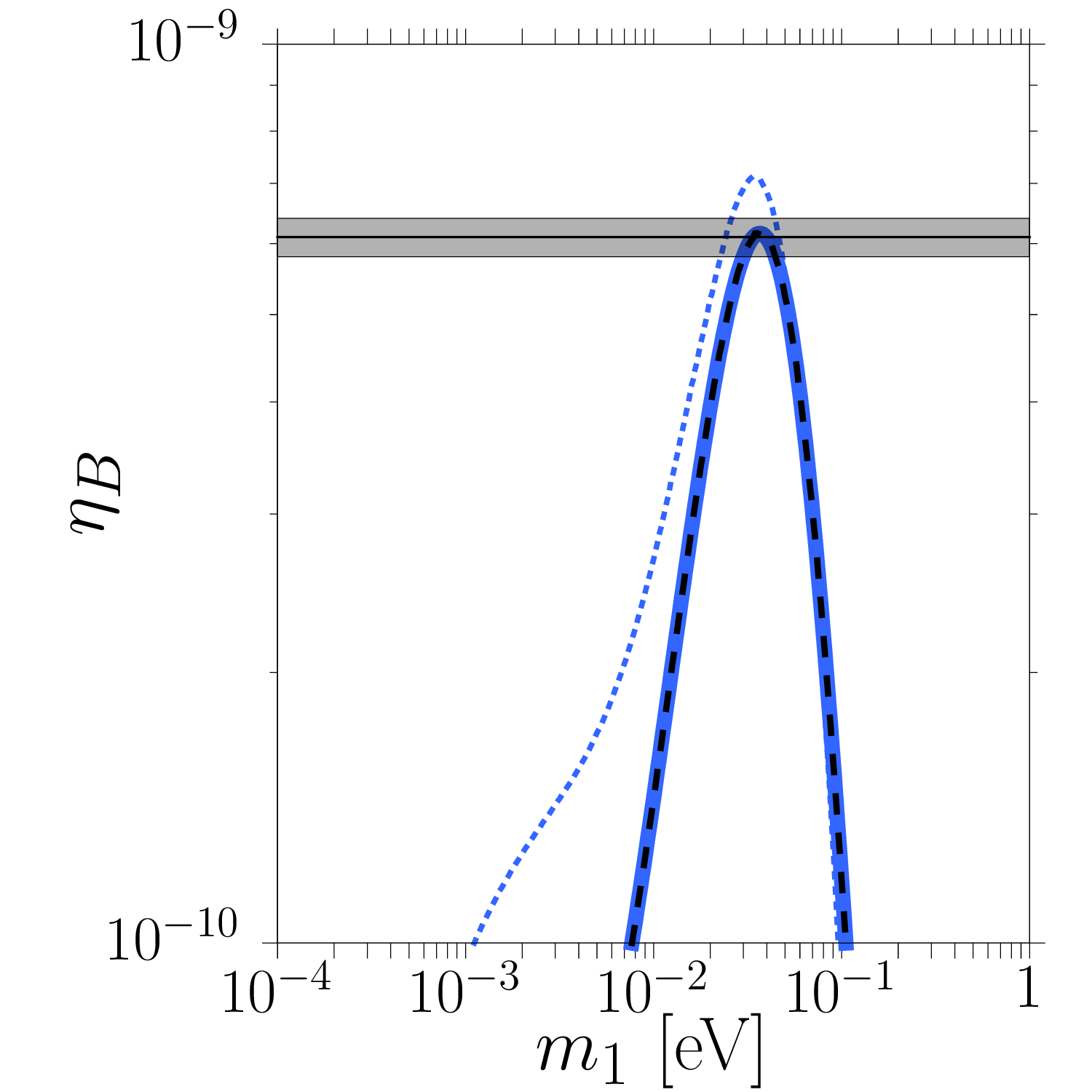,height=39mm,width=46mm} 
\hspace{5mm}
\psfig{file=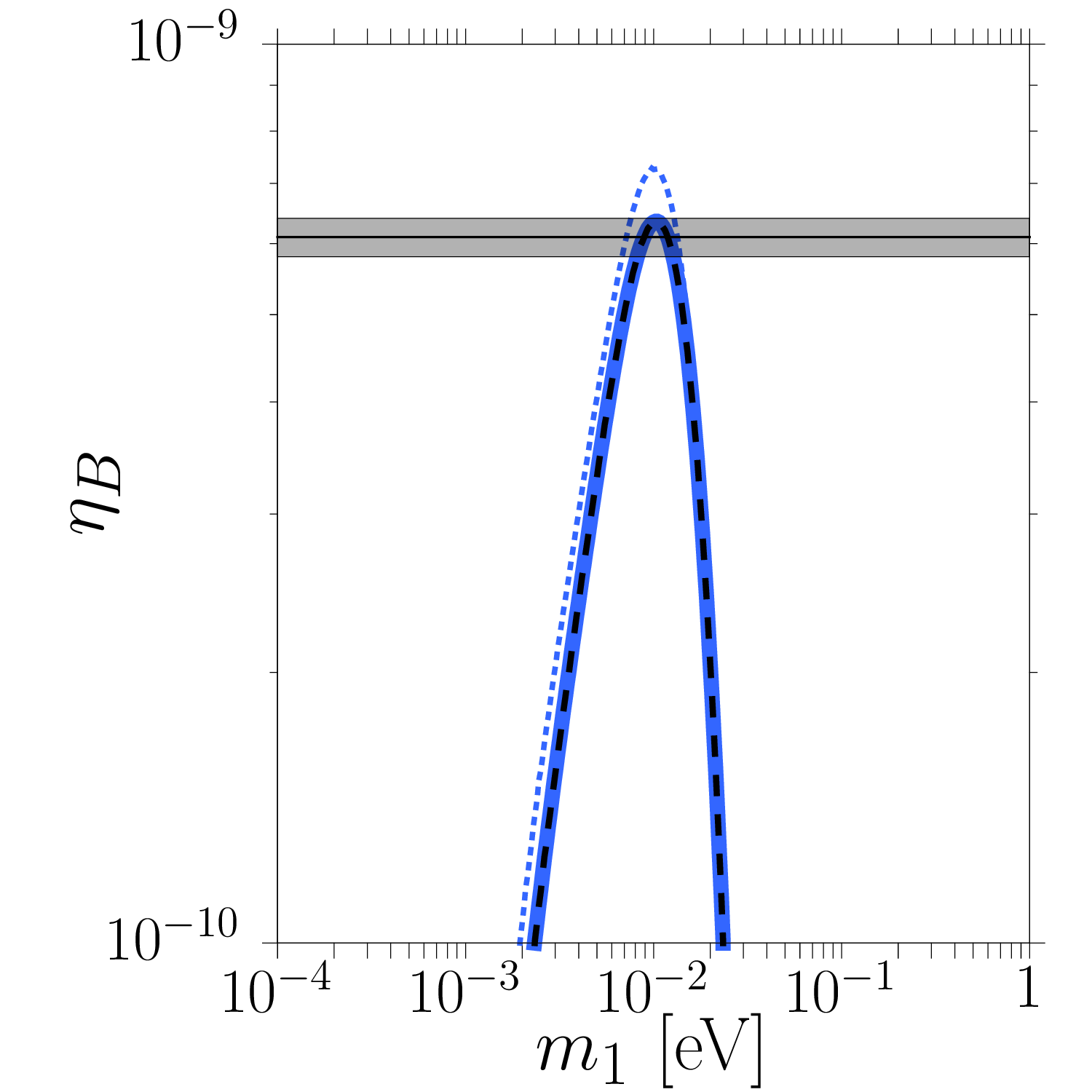,height=39mm,width=46mm} 
\end{center}
\vspace{-5mm}
\caption{Plots of the final $\eta_B$ for the same three sets of parameters of Figs. 2 and 3. 
The numerical results (blue solide lines) are compared with with the analytical results (black dashed lines) obtained using the  eq.~(\ref{NBmLflow}). The dotted lines are obtained switching on $V_L \neq I$. Using the same 
parameterization for $V_L$ as for $U$, in the three cases from left to right one has 
$\theta_{12}^L =(0.79^{\circ},4.1^{\circ},0.1^{\circ})$, $\theta_{13}^L=(0,0.05^{\circ},0.07^{\circ})$, 
$\theta_{23}^L=(2.3^{\circ},2.3^{\circ},2.3^{\circ})$, $\delta_L/\pi =(0.2,0.63,1.22)$, 
$\rho_L/\pi=(1.65,0.85,0.79)$ and $\sigma_L/\pi=(1.05,1.1,0.94)$.}
\end{figure}

We also made a more general comparison between the constraints derived from the 
analytical expression  eq.~(\ref{NBmLflow}) and the numerical constraints (for $V_L = I$).
In Fig.~5 we show, with orange points, the results of a scatter plot for $V_L =I$ imposing
successful $SO(10)$-inspired leptogenesis for $\a_2=5$. 
\begin{figure}
\begin{center}
\psfig{file=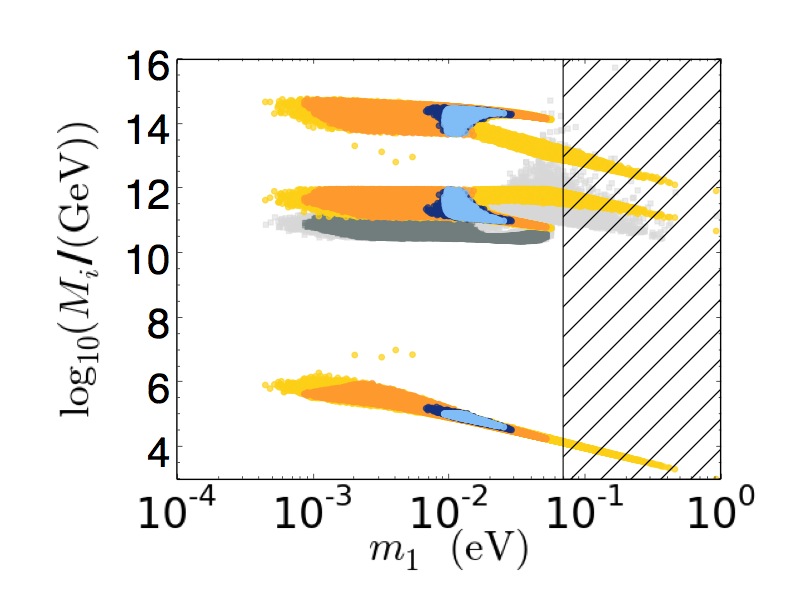,height=50mm,width=56mm}
\hspace{-7mm}
\psfig{file=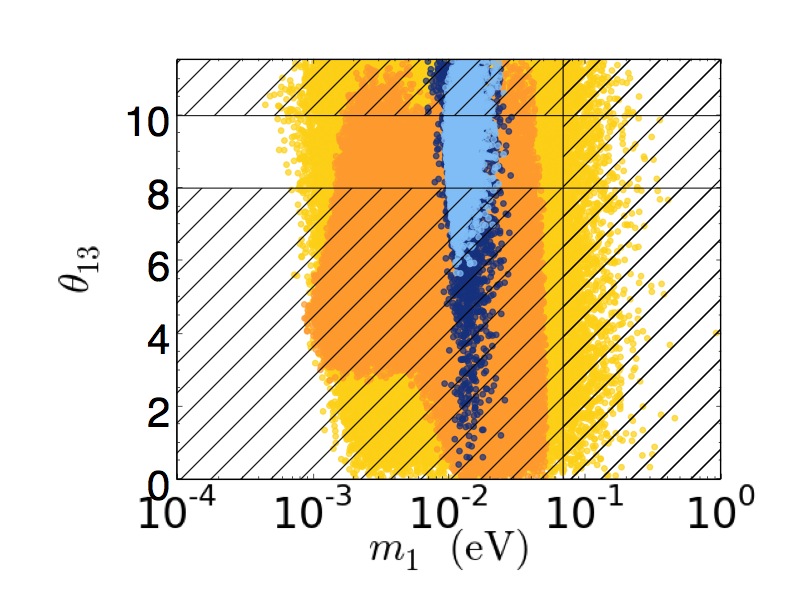,height=50mm,width=56mm}
\hspace{-7mm}
\psfig{file=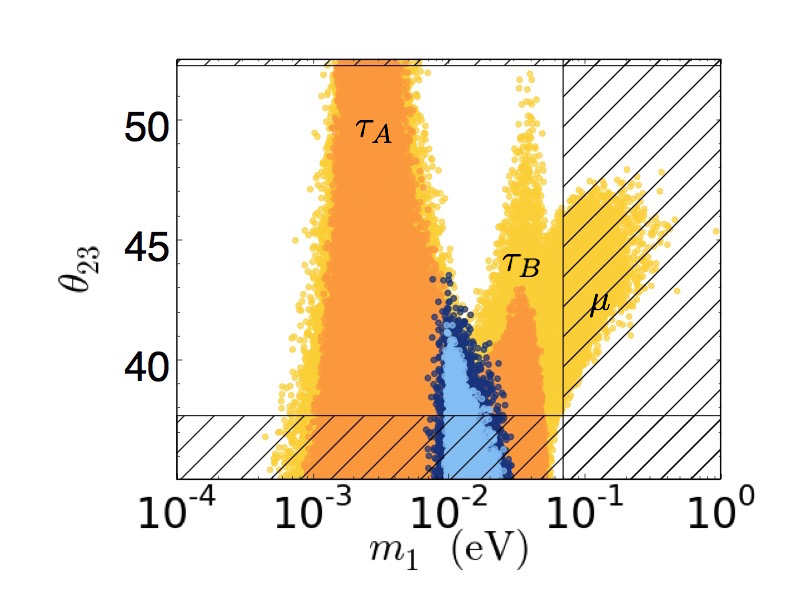,height=50mm,width=56mm} \\
\psfig{file=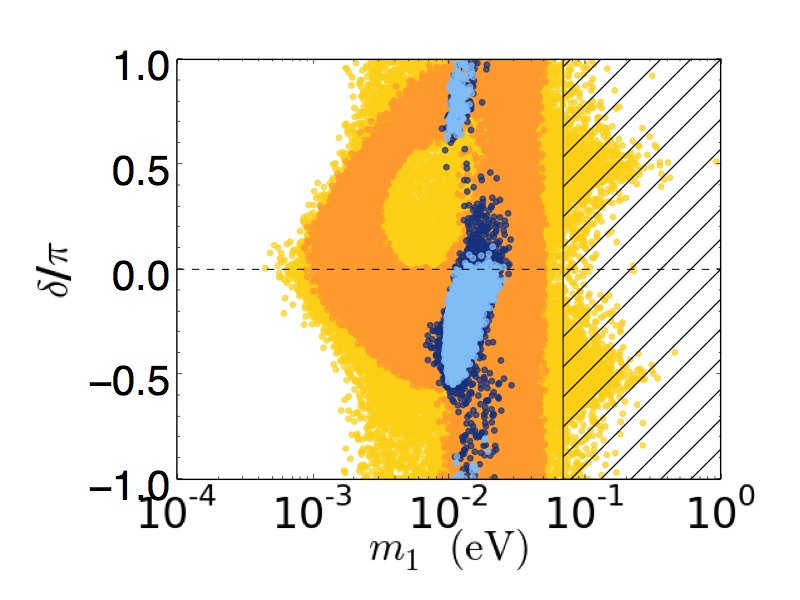,height=50mm,width=56mm}
\hspace{-7mm}
\psfig{file=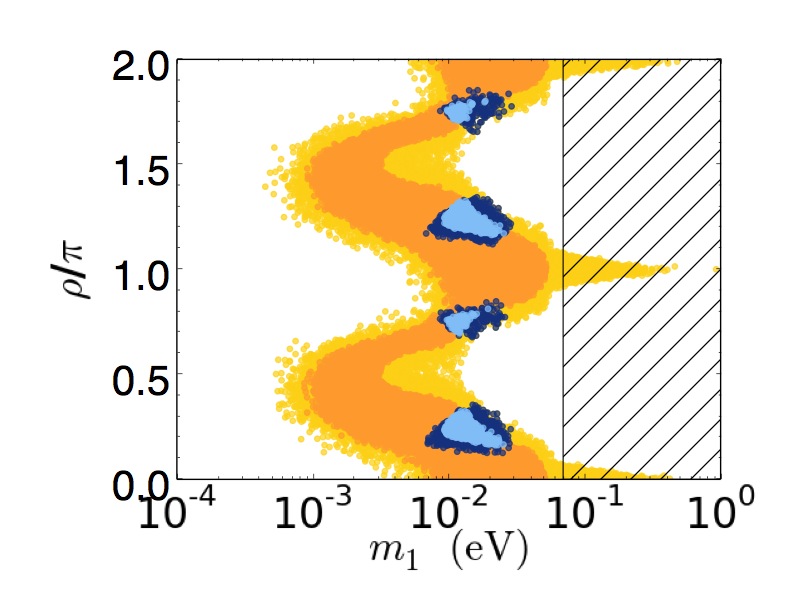,height=50mm,width=56mm}
\hspace{-7mm}
\psfig{file=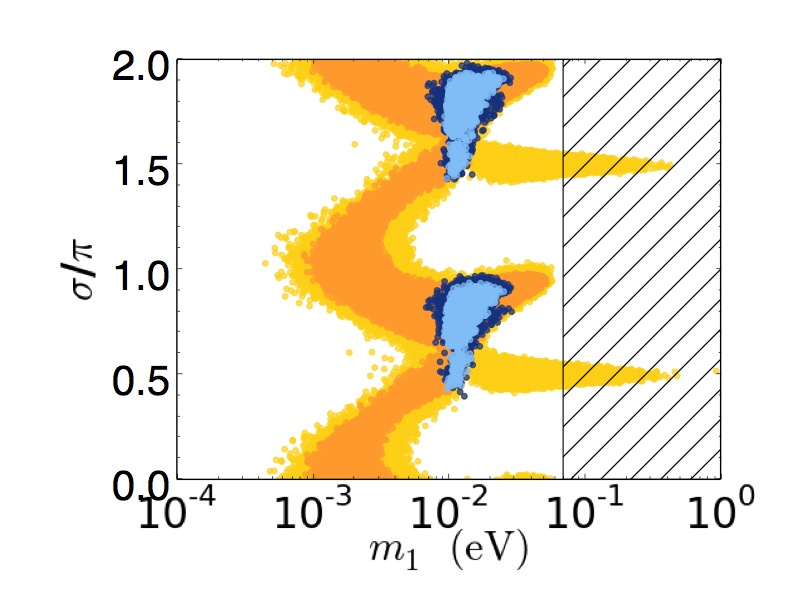,height=50mm,width=56mm} \\
\psfig{file=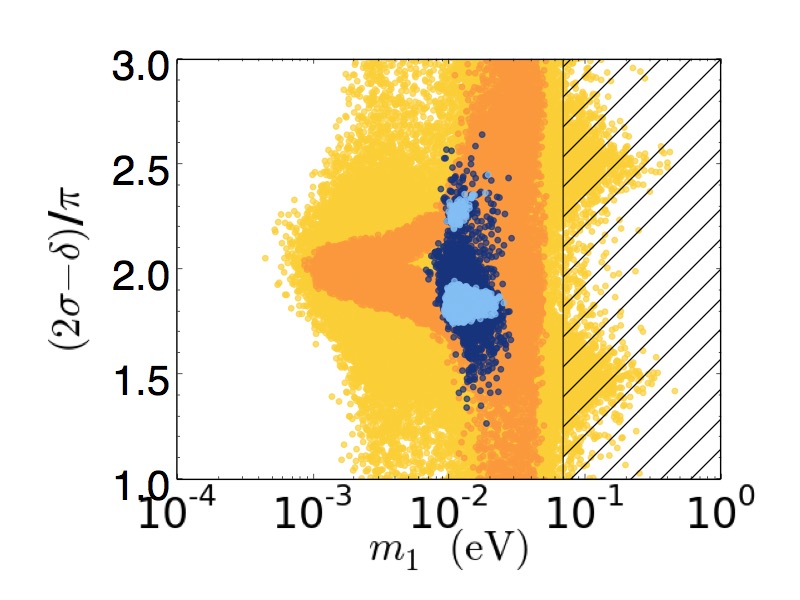,height=50mm,width=56mm}
\hspace{-7mm}
\psfig{file=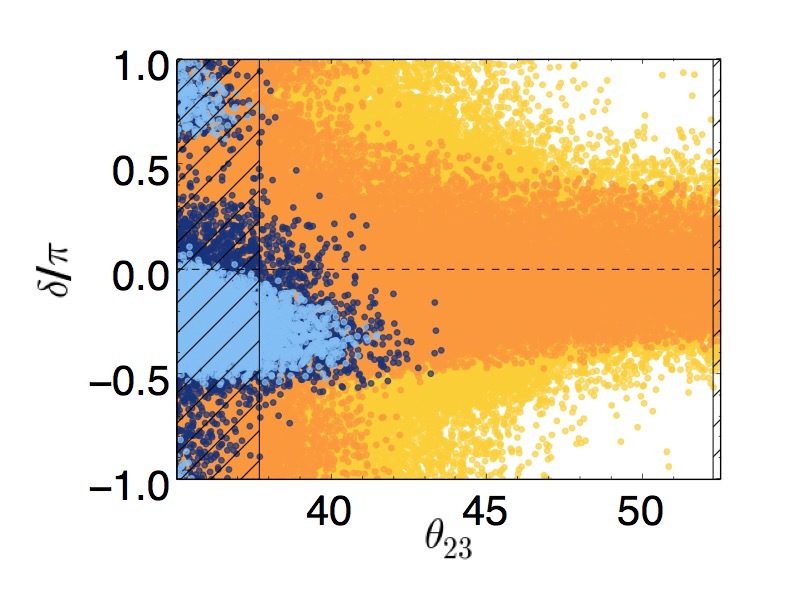,height=50mm,width=56mm}
\hspace{-7mm}
\psfig{file=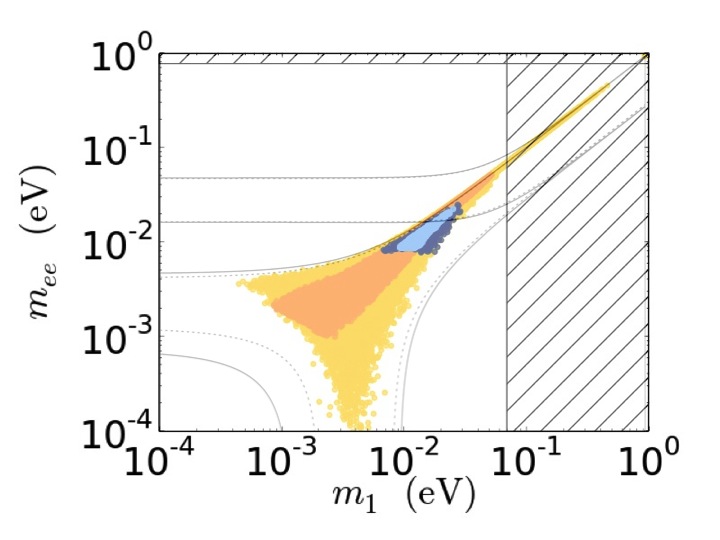,height=50mm,width=56mm}
\end{center}
\vspace{-11mm}
\caption{Scatter plots in the low energy neutrino parameter space 
projected on different selected planes for NO and $\a_2 = 5$.
The orange points respect the successful leptogenesis condition
$\eta_B^{\rm lep} > \eta_B^{\rm CMB} > 5.9 \times 10^{-10}$
for $V_L =I$ where $\eta_B^{\rm lep}$ is calculated from the eq.~(\ref{NBmLlepftau})
using a numerical determination of  RH neutrino masses, mixing matrix and phases.  
The mixing angles vary within the ranges eqs.~(\ref{ranges}).  The blue points are those
respecting the additional ST condition within the approximation $V_L = I$ (light blue)
or for $I \leq V_L \leq V_{CKM}$ (dark blue).
The dashed regions indicate either the values of $m_1$ excluded by 
the CMB upper bound (cf. eq.~(\ref{m1upperbound})), or the values of $m_{ee}$
excluded by $0\nu\b\b$ experiments, or
the values of $\theta_{13}$ excluded by current 
determination at $3\s$ (cf. eq.~(\ref{expranges})).  
In the bottom right panel the dashed (solid) black lines indicate the general (no leptogenesis) allowed bands, both for NO and IO, in the plane $m_{ee}$ vs. $m_1$
for $\theta_{13}$ in the range in eq.~(\ref{ranges}).}
\label{constrNO}
\end{figure}
The asymmetry is calculated   
from the eq.~(\ref{NBmLlepftau}) where RH neutrino masses and mixing matrix $U_R$ are calculated
numerically.  The mixing angles are uniformly random generated within the same ranges
adopted in \cite{SO10lep1},
\be\label{ranges}
0 \leq \theta_{13} \leq 11.5^{\circ} \, , \;\;
35^{\circ}\leq \theta_{23} \leq 52^{\circ} \, , \;\;
31.3^{\circ} \leq \theta_{12} \leq 36.3^{\circ} \,  ,
\ee 
with the only
exception of $\theta_{23}$ that is allowed to be slightly lower, as adopted in \cite{strongSO10solution}. 
The results confirm those obtained in \cite{SO10lep1,SO10lep2}, simply here 
a much higher (about thousand times) amount of points has been obtained 
and the constraints are much sharper. 

We have then produced corresponding scatter plots using directly 
the analytical expression for the final asymmetry eq.~(\ref{NBmLflow}). 
The results are shown in Fig.~6 and as one can see they perfectly 
reproduce the numerical results shown in Fig.~5 (orange points). 
\begin{figure}
\begin{center}
\psfig{file=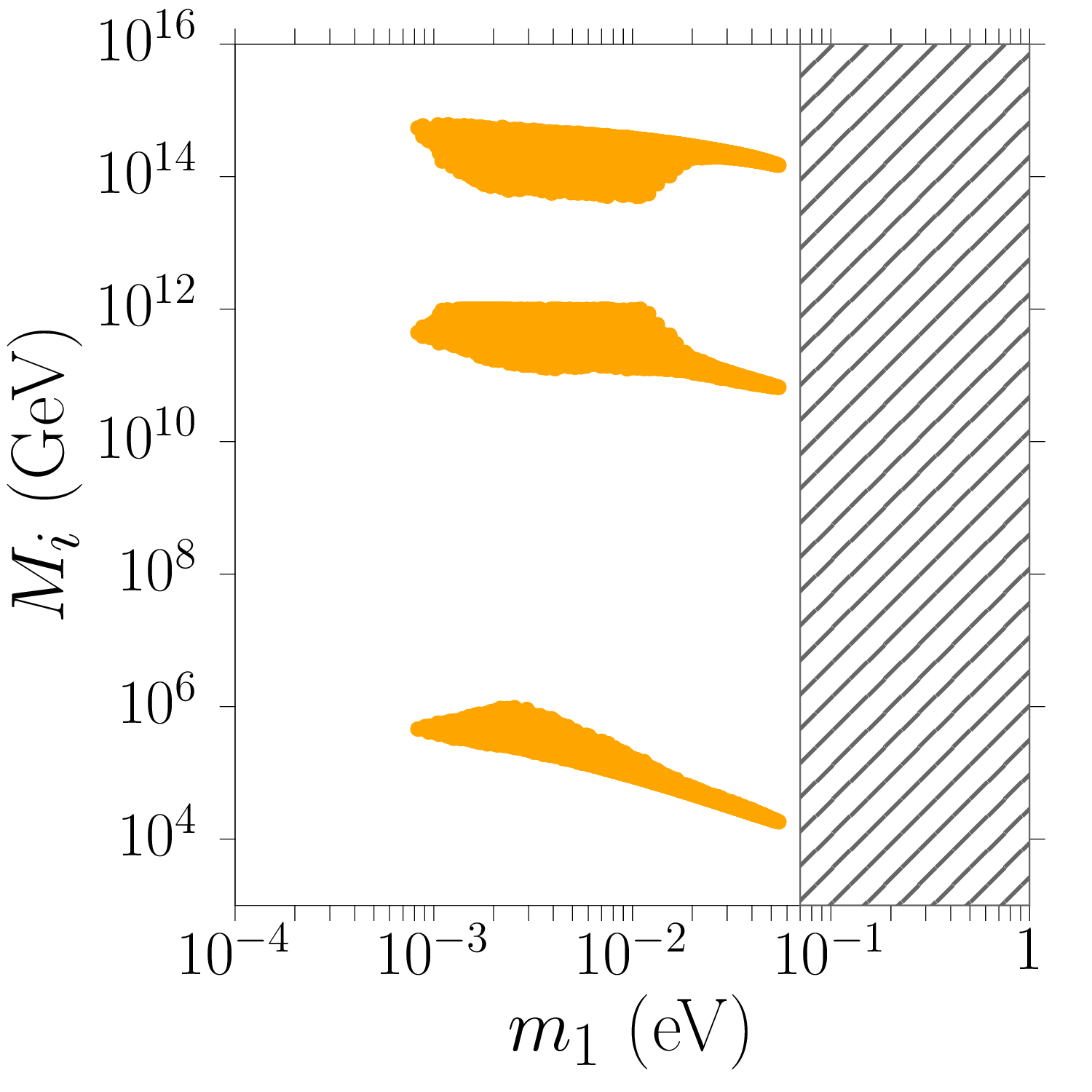,height=50mm,width=56mm}
\hspace{-7mm}
\psfig{file=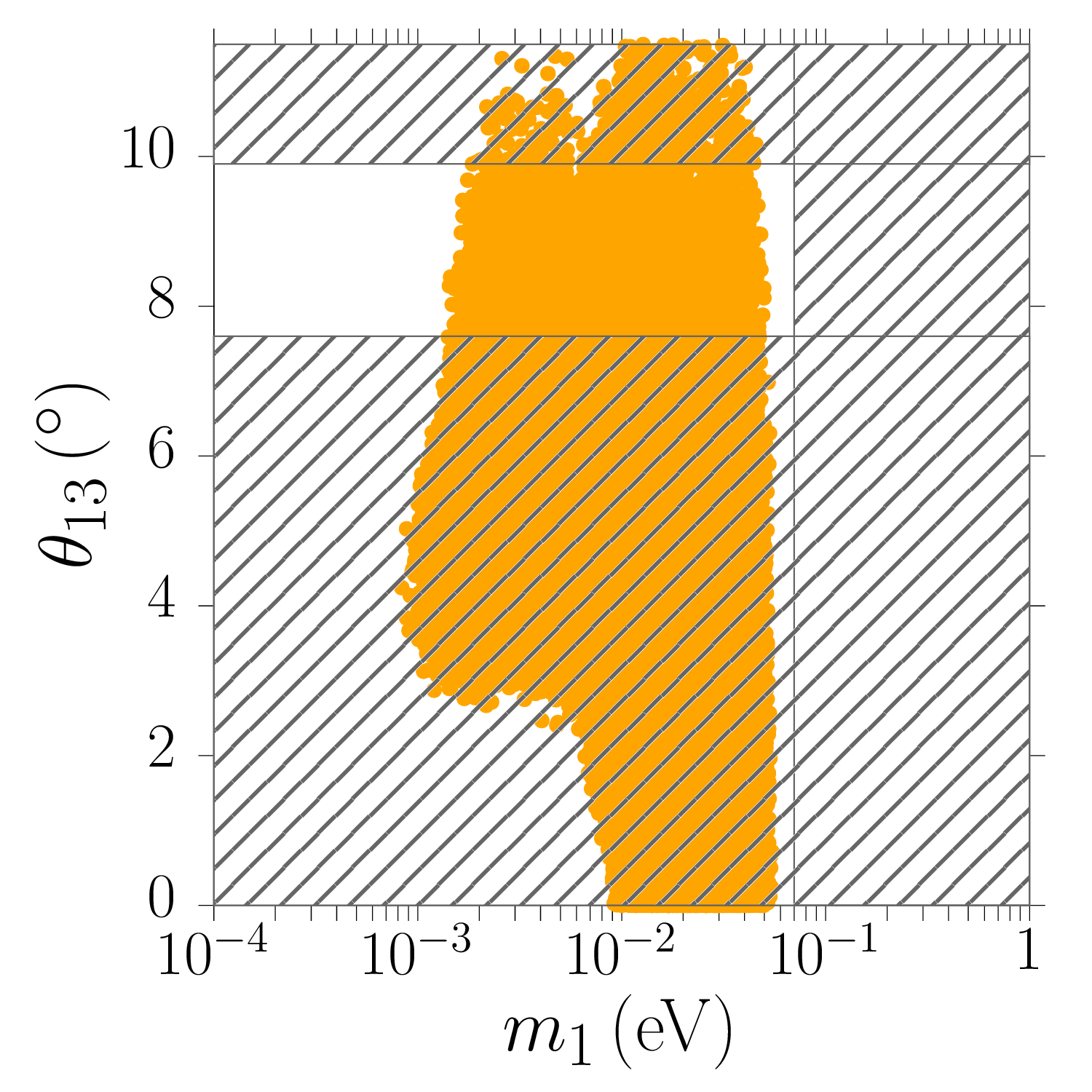,height=50mm,width=56mm}
\hspace{-7mm}
\psfig{file=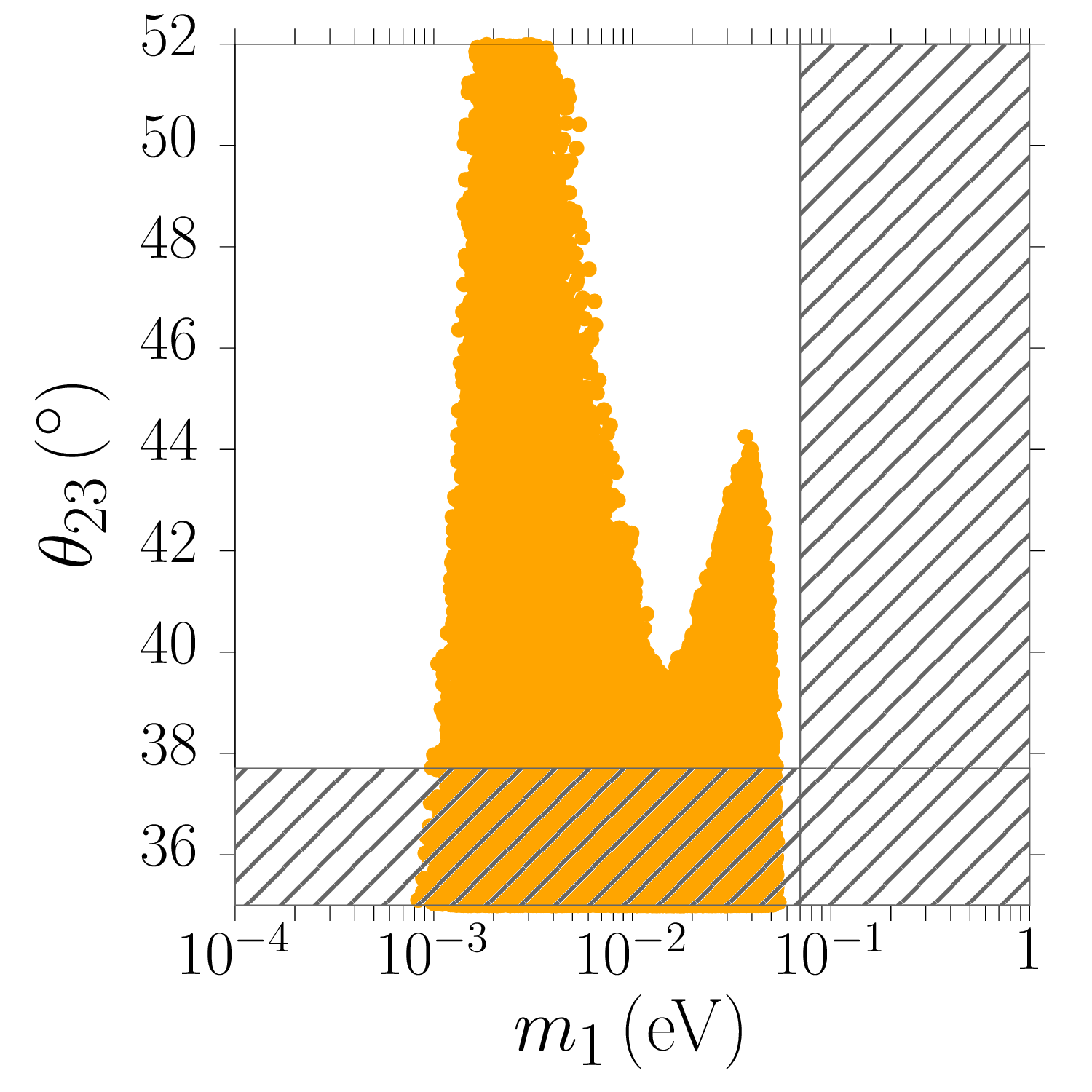,height=50mm,width=56mm} \\
\psfig{file=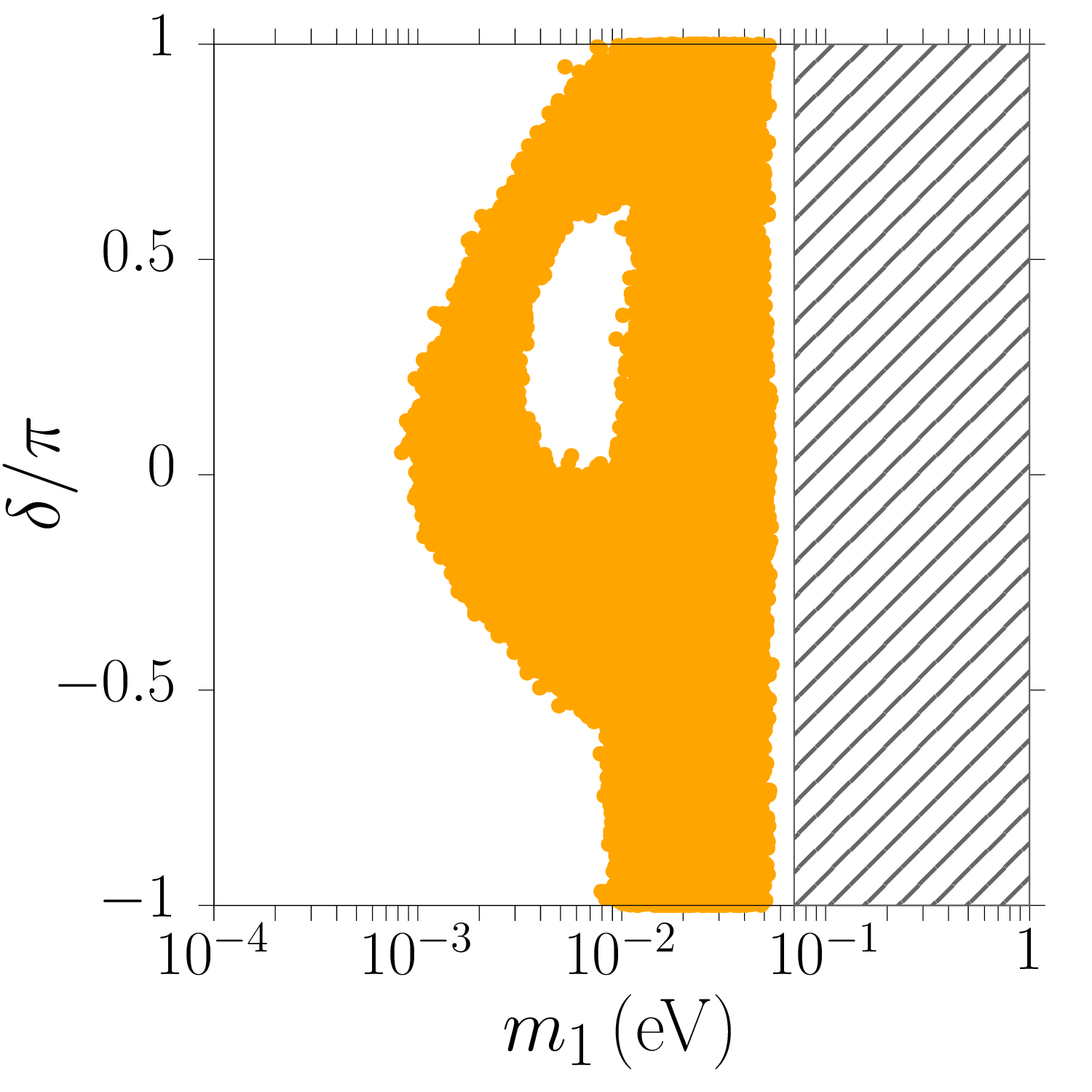,height=50mm,width=56mm}
\hspace{-7mm}
\psfig{file=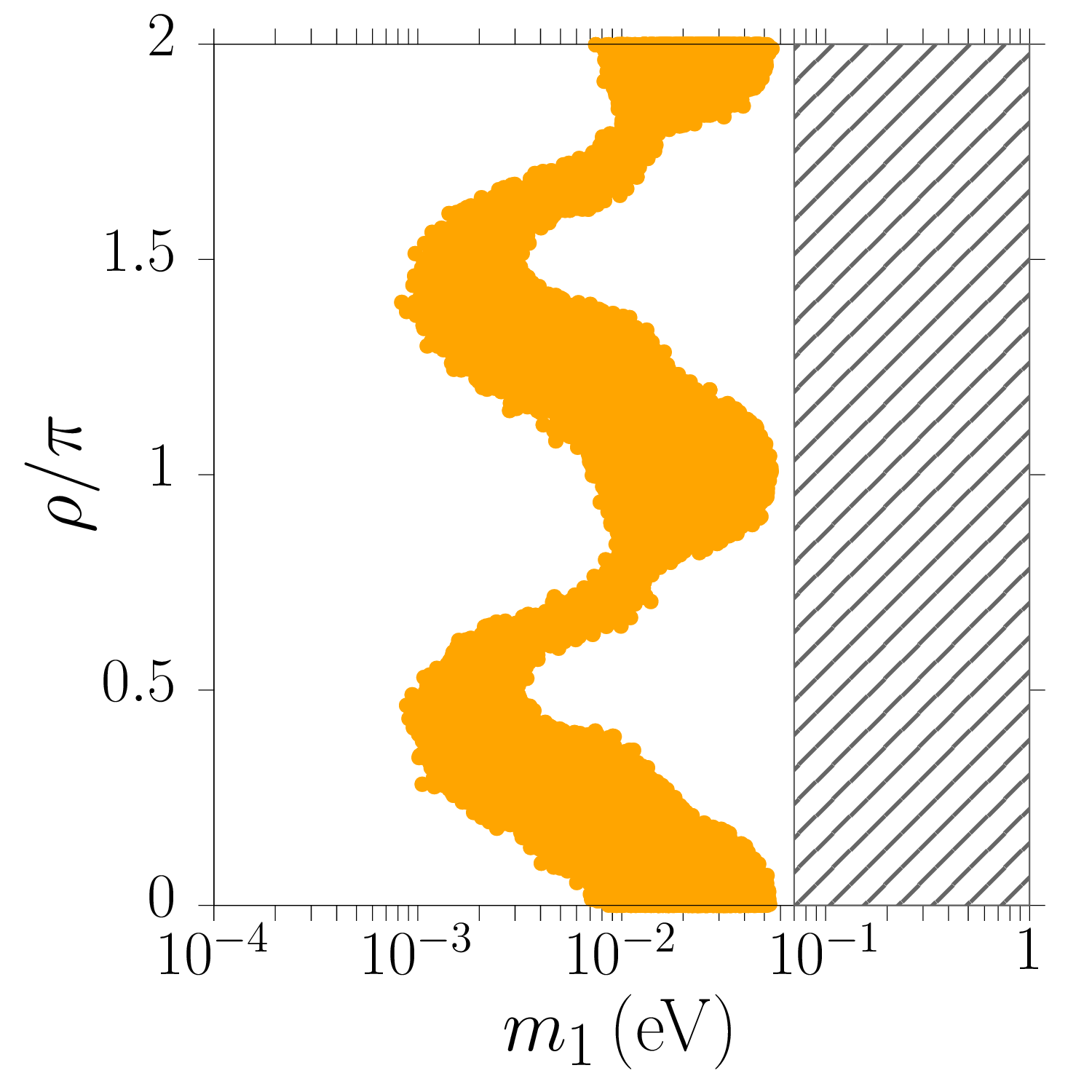,height=50mm,width=56mm}
\hspace{-7mm}
\psfig{file=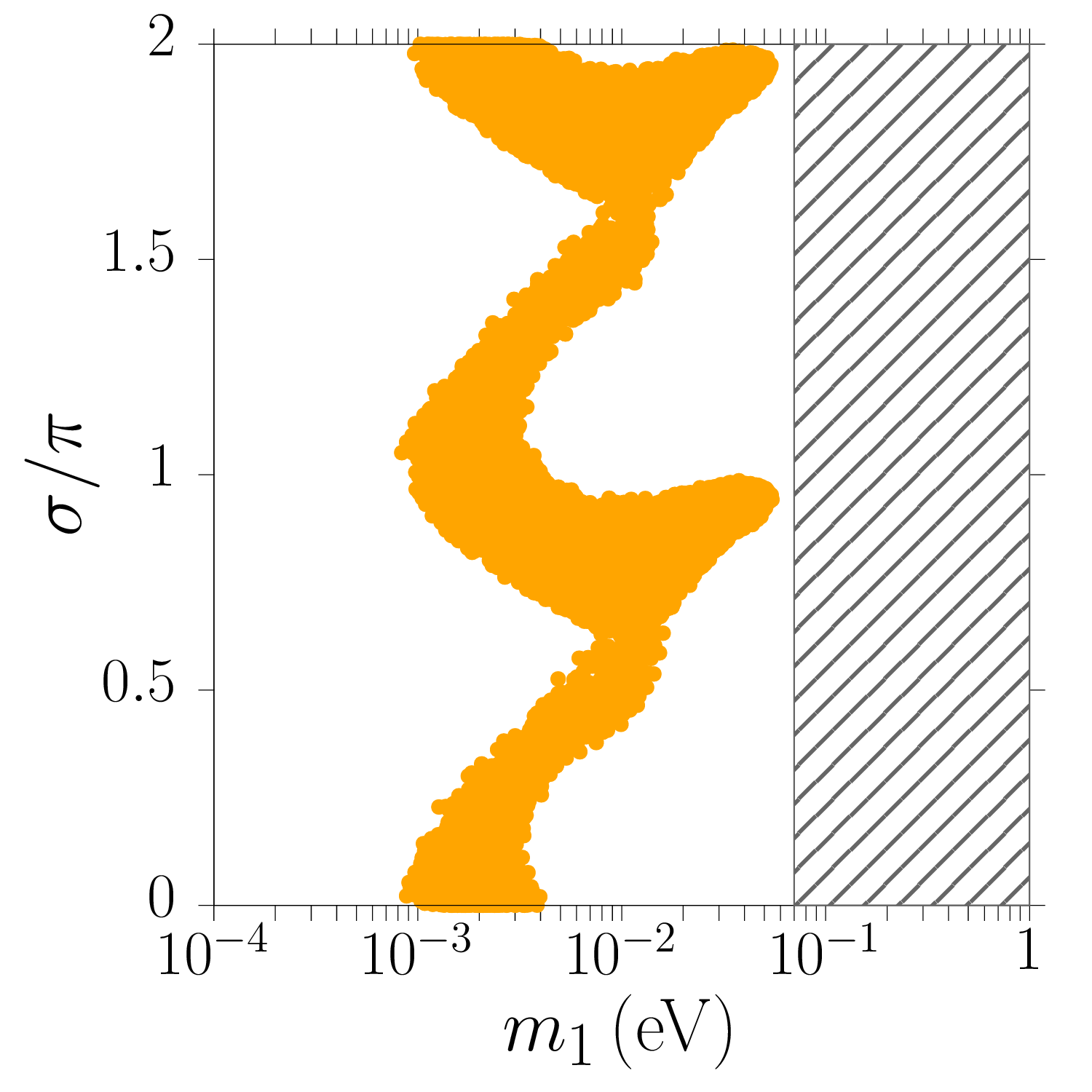,height=50mm,width=56mm} \\
\psfig{file=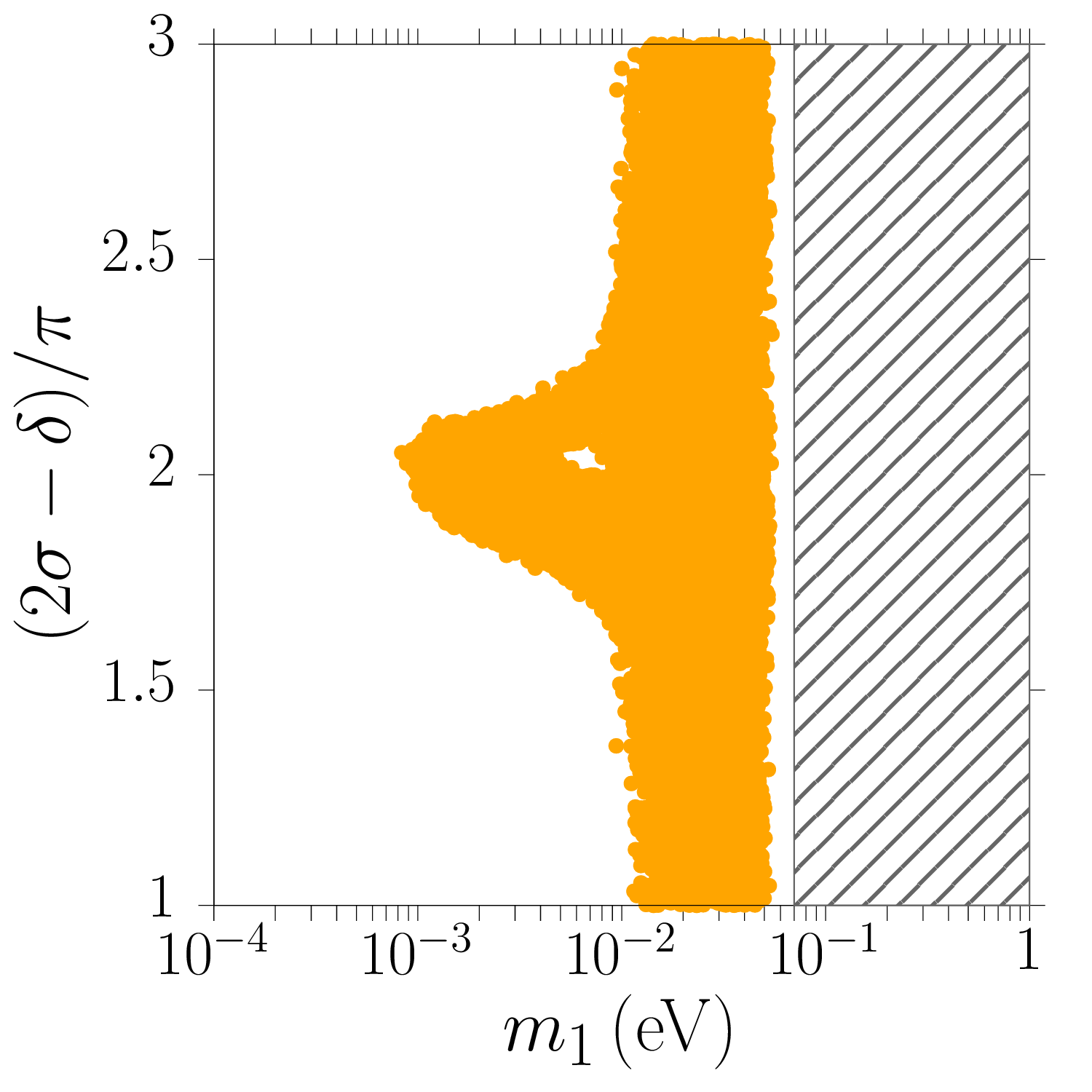,height=50mm,width=56mm}
\hspace{-9mm}
\psfig{file=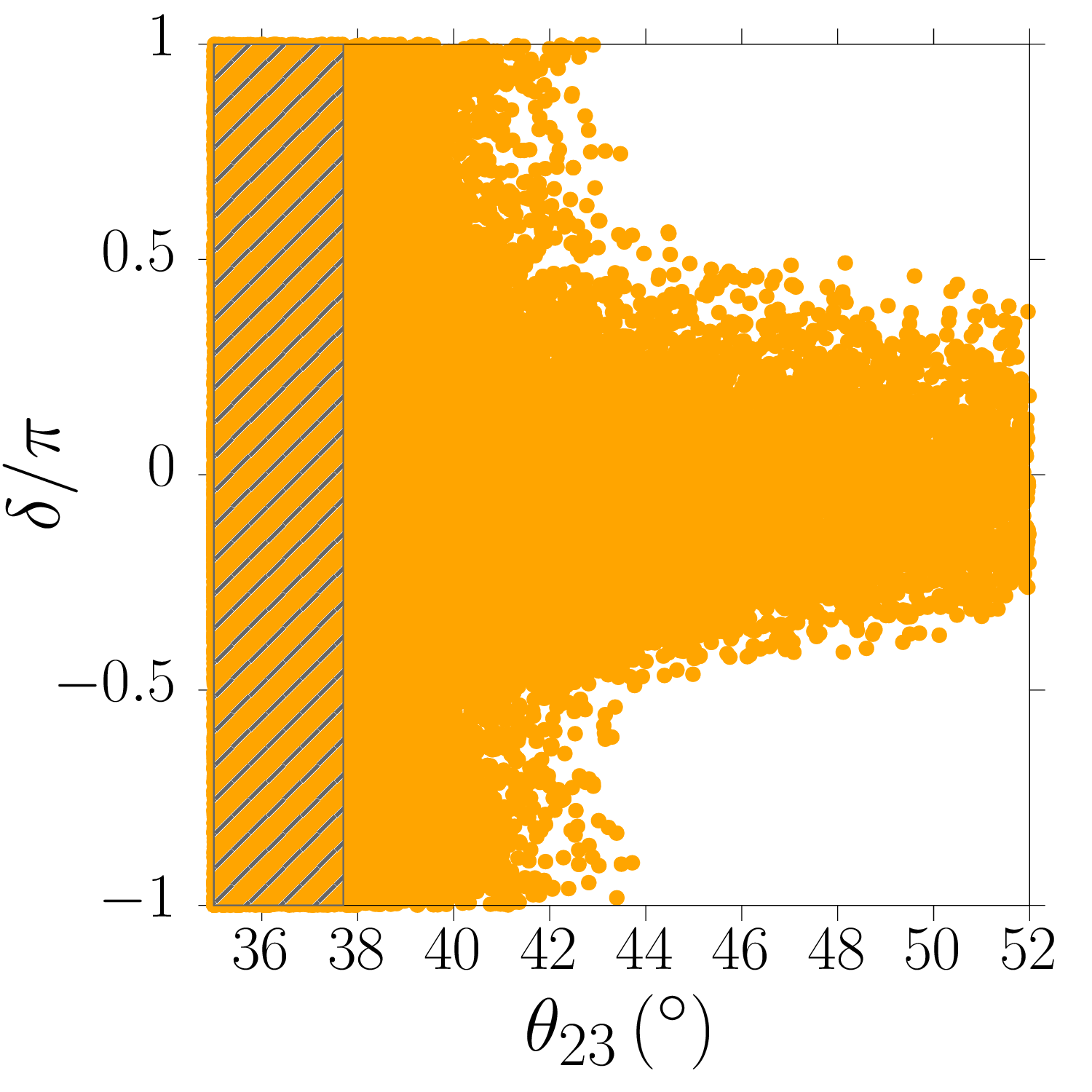,height=50mm,width=56mm}
\hspace{-5mm}
\psfig{file=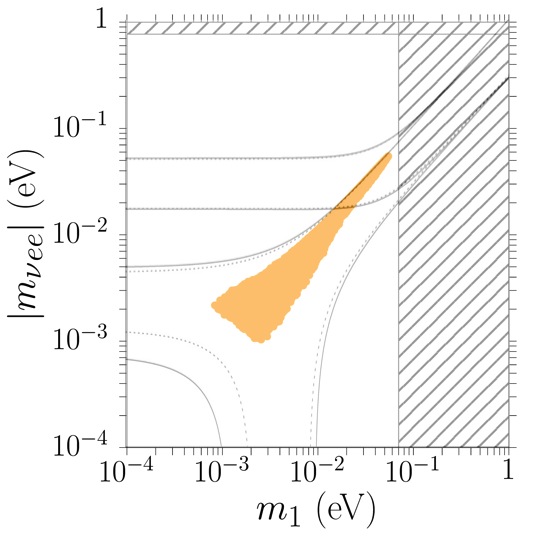,height=50mm,width=56mm}
\end{center}
\vspace{-11mm}
\caption{Scatter plots in the low energy neutrino parameter space 
projected on different selected planes for NO and $\a_2 = 5$ 
respecting the successful leptogenesis condition
$\eta_B^{\rm lep} > \eta_B^{\rm CMB} > 5.9 \times 10^{-10}$
and obtained from the analytical expression eq.~(\ref{NBmLflow}) for the final asymmetry.
Same ranges and conventions as in Fig.~5 are adopted. These analytical results should be 
compared with the numerical results of Fig.~5 (orange points).}
\label{constrNO}
\end{figure}
We have also checked that one has a lower bound $\a_2 \gtrsim 3$ 
confirming the numerical result found in \cite{SO10lep1}. 
We can then conclude that the eq.~(\ref{NBmLflow}) provides
a very precise analytical way to calculate  the final asymmetry in $SO(10)$-inspired models
(we are excluding crossing level solutions from our analysis) in the approximation $V_L \simeq I$
and can be regarded as one of the main results of our investigation.
Indeed it can reliably be applied in all models where $SO(10)$-inspired conditions hold 
in order impose the successful leptogenesis condition using directly predictions on low energy neutrino
data (the only additional parameter that has to be calculated is $\a_2$).

Having done this important cross check, we can now safely proceed further deriving analytical constraints on the low energy neutrino parameters imposing successful leptogenesis fully 
trusting our eq.~(\ref{NBmLflow}).  Because of the intricate 
dependence of the eq.~(\ref{NBmLflow}) on the low energy neutrino parameters, we need to 
understand the behaviour at low and high $m_1$ values, respectively for 
$m_1 \ll m_{\rm sol}$ and $m_1 \gg m_{\rm sol} \simeq 10\,{\rm meV}$ 
and then match the results for intermediate
values $m_1 \simeq m_{\rm sol} \simeq 10\,{\rm meV}$. 

\subsection{Lower bound on $m_1$}

Let us now calculate the final asymmetry in the limit $m_1 \ra 0$ showing that
this tends to vanish and, therefore, that successful $SO(10)$-inspired leptogenesis implies a lower bound
on the absolute neutrino mass scale \cite{SO10lep2}. 
It is convenient to start from $K_{1\t}$. In the limit 
$m_1/m_{\rm sol} \ra 0$ the eq.~(\ref{K1tau}) simplifies into
\be\label{K1taulowm1}
K_{1\t} \simeq {1\over m_{\star}}\,{|m_{\rm atm}\,s_{13}\,c_{13}\,c_{23}\,e^{i\,(2\s -\d)} - m_{\rm sol}\,c_{13}\,s_{12}\,c_{12}\,s_{23}|^2
\over |m_{\rm sol}\,s^2_{12}\,c_{13}^2 + m_{\rm atm}\,s^2_{13}\,e^{2i\,(\s -\d)}|}  \, .
\ee
From this result, we can see that the condition $K_{1\t}\lesssim 1$
is verified for $2\s - \d \simeq 2\,\pi\,n$ and 
\be\label{t13lb}
s_{13} \gtrsim {m_{\rm sol}\over m_{\rm atm}}\,s_{12}\,c_{12}\,\tan \theta_{23} \gtrsim 0.06\, ,
\;\; \mbox{\rm implying} \;\; \theta_{13} \gtrsim 3^{\circ} \, ,
\ee
a lower bound
that confirms the results of the scatter plots, first obtained in \cite{SO10lep1},
shown in the top central panel of Figs.~5 and 6 in the plane $m_1-\theta_{13}$.

We can now consider the low $m_1$ limit of $\ve_{2\t}$, obtaining from the eq.~(\ref{ve2tausinalpha}),
\bea
\ve_{2\t}  & \simeq &  {3\over 16\,\pi}\,{\a_2^2 \, m_c^2 \over v^2} \, 
{m_1 \over m_{\rm sol}\,m_{\rm atm}} \, {|m_{\rm sol}\,U_{e2}^2+m_{\rm atm}\,U^2_{e3}| \, |U_{\m 1}|^2\over 
|U_{\t 1}|^2\, (|U_{\t 1}|^4 + |U_{\m 1}|^2\,|U_{\t 1}|^2)} \,
\sin \a_L \\
& \simeq &
{3\over 16\,\pi}\,{\a_2^2 \, m_c^2 \over v^2} \, 
{m_1 \over m_{\rm sol}\,m_{\rm atm}} \, {|m_{\rm sol}\,s^2_{12}\,c^2_{13}
+m_{\rm atm}\,s^2_{13}\,e^{2\,i\,(\s-\d)}|\,c^2_{23}\over s^4_{12} \, s^4_{23}} \, \sin \a_L \, ,
\eea
where the asymptotic limit for the effective leptogenesis phase is given by
$\a_L \simeq  2\,(\r - \s)$.

Notice that we have retained the term $\propto s_{13}$ in $m_{ee}$
since, as we have seen, it has to be non-vanishing. The expression is maximised 
for $\s -\d \simeq n\,\pi$ with $n$ integer and clearly for $\sin\a_L =1$, finding
\be
\ve_{2\t} \lesssim  {75\over 16\,\pi}\,{\a_2^2 \, m_c^2 \over v^2} \,  \left({\a_2\over 5}\right)^2 
\, {m_1 \over m_{\rm atm}} 
{c^2_{23}\over s^2_{12}\,s^4_{23}}\, \left(1+{m_{\rm atm}\,s^2_{13}\over m_{\rm sol}\,s^2_{12}}\right).
\ee

 Finally the  the asymptotic  limit for $K_{2\t}$ for $m_1/m_{\rm sol} \ll 1$ 
 is given by $K_{2\t} \simeq c^2_{23}\,{m_{\rm atm}/ m_{\star}} \simeq 25$ \cite{SO10lep1}.
This shows that in the low $m_1$ limit the wash-out at the production is strong and
in this case one can use a simple approximation for the efficiency factor \cite{predictions},
$\k(K_{2\t}) \simeq {0.5 / K_{2\t}^{1.2}} \simeq 0.01$.  
Combining together the results found for the three terms, one finds that in the low $m_1$ limit 
the baryon-to-photon number ratio is maximised by
\bea
\eta_B^{\rm lep}  <  \eta_{B}^{\rm max} & \simeq &
10^{-4}\,{75\over 16\,\pi}\,{\a_2^2 \, m_c^2 \over v^2} \,  \left({\a_2\over 5}\right)^2 
\, {m_1 \over m_{\rm atm}} 
{c^2_{23}\over s^2_{12}\,s^4_{23}}\, \left(1+{m_{\rm atm}\,s^2_{13}\over m_{\rm sol}\,s^2_{12}}\right)  \\
&  \equiv & m_1 \, \left({\a_2\over 5}\right)^2  \, f(\theta_{12},\theta_{13},\theta_{23}) \,  .
\eea 
Imposing finally the successful leptogenesis condition implying 
$\eta_{B}^{\rm max} \gtrsim \eta_B^{\rm CMB}$, one obtains
the  lower bound 
\be\label{lbm1}
m_1 \gtrsim   6\times 10^{-10}\, \left({5 \over \a_2 }\right)^2 \,[f(\theta_{12},\theta_{13},\theta_{23}) ]^{-1}\, 
\gtrsim 0.08\,{\rm meV}\,\left({5 \over \a_2 }\right)^2   \,  ,
\ee
where the last inequality has been obtained for the values of the mixing angles 
within the ranges eq.~(\ref{ranges}) that minimise $[f(\theta_{12},\theta_{13},\theta_{23}) ]^{-1}$. 
The result is  
in very good agreement with the results of the scatter plots shown in Fig.~5 (orange points) and Fig.~6
and confirms, in more detail,  the value obtained in \cite{SO10lep1}. 

Finally it should be noticed that the three conditions for maximal asymmetry
on the three low energy phases, $2\,\s-\d \simeq m\,\pi$, $\s-\d \simeq n\,\pi$
and $\sin[2\,(\rho-\s)] \simeq 1$, with $n,m$ integers, imply that in the low $m_1$ limit one has
$\s = (2\,n-m)\,\pi$ and $\d=2\,\pi\,n$, results
that are confirmed by the results, at low $m_1$, 
of the scatter plot shown in the two panels in Fig.~5 (and Fig.~6 as well)
for $\s$ and $\d$ vs. $m_1$ (orange points). 
One also finds $\r=\pi/4 + q\,\pi$,with $q$ integer. It can be seen
however that at small $m_1$ the value of $\rho$ is actually $\rho \simeq 0.35 \, \pi + q\,\pi$.
The reason for this shift is understood from the more complete 
expression eq.~(\ref{K1tauexplicit}) for $K_{1\t}$. For $\rho =\pi/2$ the term 
$m_1\,e^{2i\pi\rho} = -m_1$ adds to the term $-m_2$ in a way that $K_{1\t}\lesssim 1$
for slightly lower values of $s_{23}$. However, because of the strong dependence 
$\ve_{2\t} \propto s_{23}^{-4}$, a shift of $\rho$ towards $\pi/2$ maximises the asymmetry even though
 the phase $\a_L$ is not maximal. We will be back soon on the fact that value of $\theta_{23}$  
 cannot be too large. 
 
\subsection{Upper bound on $m_1$}

Together with a lower bound on $m_1$, there is also an upper bound on $m_1$.
We can work in the quasi-degenerate neutrino limit $m_1 \simeq m_2 \simeq m_3$ and then check whether the upper bound does indeed fall in the quasi-degenerate regime. 

Let us first calculate separately the quasi-degenerate limit of $K_{1\t}, K_{2\t}$ and 
$\ve_{2\t}$, the three relevant quantities  determining the final asymmetry.
For $K_{1\t}$ from the eq.~(\ref{K1tauexplicit}) one can immediately see that if
$\rho = n\,\pi$, with $n$ integer,  then 
\be
K_{1\t}\simeq s_{13}\,c^2_{23}\,{m_1\over m_{\star}}\,\left|e^{i(2\s-\d)}-s^2_{12}-c^2_{12}e^{i\d}\right|^2 
\lesssim 0.015 \, {m_1\over m_{\star}}\,\left|e^{i(2\s-\d)}-s^2_{12}-c^2_{12}\, e^{i\d}\right|^2  \,  .
\ee
This expression shows that for $m_1 \lesssim 0.1\,{\rm eV}$ 
one has $K_{1\t}\lesssim 4$, where the maximum is saturated for $\s = 2\pi\,m$ and $\d =\pi/2 + k\,\pi$,
so in any case it cannot be too large and it can be always made vanishing.

Let us now calculate the asymptotic limit of $\ve_{2\t}$ for $\rho\simeq n\,\pi$. 
We can neglect all sub-dominant terms $\propto s^2_{13}$ containing $\d$ in a way
that the dependence on $\d$ cancels out.
First of all notice that $m_{ee}  \equiv |m_{\nu ee}| \ra m_1$ an asymptotic limit that  
is clearly visible in the panels of Fig.~5 and 6.   At the same time one has, 
\bea
\left|(m_{\nu}^{-1})_{\t\t}\right|^2 & \ra & {1\over m_1^2}  \,   
\left| s^2_{23} + c^2_{23}\,e^{-2\,i\,\s}\right|^2  \,  ,\\
\left|(m_{\nu}^{-1})_{\m\t}\right|^2 & \ra & {s^2_{23}\,c^2_{23}\over m_1^2} \,\left|e^{-2\,i\,\s} -1\right|^2 .
\eea
In this way putting all terms together one finds from the eq.~(\ref{ve2tausinalpha})  
the following asymptotic limit for $\ve_{2\t}$ 
\be
\ve_{2\t} \ra {3\over 16\,\pi}\, {\a_2^2\,m_c^2 \over v^2}\, 
{s^2_{23}\,c^2_{23}\,|e^{-2i\s}-1|^2 / |s^2_{23}+c^2_{23}\,e^{-2i\s}|^2 \over 
(|s^2_{23}+c^2_{23}\,e^{-2i\s}|^2 + s^2_{23}\,c^2_{23}\,|e^{-2i\s} -1|^2)}
\, \sin \a_L \,  ,
\ee
where the asymptotic limit of $\a_L$ is  given by $\a_L \ra  -4\,\s $.

Finally we can calculate the asymptotic limit of $K_{2\t}$ from the eq.~(\ref{K2tau}),
finding
\be
K_{2\tau} \ra {m_1\over m_{\star}}\,{s^2_{23}\,c^2_{23}\,|e^{-2i\s}-1|^2 \over 
|s^2_{23}+c^2_{23}\,e^{-2\,i\,\s}|} \, .
\ee
Putting all together in the eq.~(\ref{NBmLlepftau}) for $N_{B-L}^{\rm lep,f}$, using the approximation 
$\k(K_{2\t}) \simeq (1+2\,K_{2\t}^{1.2})^{-1}$ and considering that $\eta_B \simeq 0.01\,N_{B-L}^{\rm p,f}$
one finds as asymptotic limit for $\eta_B$,
\be
\eta_B \ra  \simeq {0.03\over 16 \,\pi}\, {\a_2^2\,m_c^2 \over v^2}\, 
{s^2_{23}\,c^2_{23}\,|e^{-2i\s}-1|^2 / |s^2_{23}+c^2_{23}\,e^{-2i\s}|^2 \over 
(|s^2_{23}+c^2_{23}\,e^{-2i\s}|^2 + s^2_{23}\,c^2_{23}\,|e^{-2i\s} -1|^2)}
\, {\sin \a_L\over 1 + 2\,K_{2\t}^{1.2}}   \,   e^{-{3\pi\over 8}\,K_{1\t}} \,  .
\ee
where, remember, we assumed $\rho = n\,\pi$. 

Here we can notice that the asymptotic limit depends mainly on the value of $\s$,
since in any case $\rho\simeq n\,\pi$ in order to have $K_{1\t}\lesssim 1$
and here the residual dependence on $\d$ can be neglected.  Guessing that the value of $\s$ 
that minimise $K_{2\t}$ is such that $2\s \ll 1$ and using simply $\sin\a_L \lesssim 1$, 
one has that $\eta_B$,  is maximised by
 \be
 \eta_B \lesssim {0.03\over 16 \,\pi}\, {\a_2^2\,m_c^2 \over v^2}\, 
\, {x \over 1 + 2\, \left({m_1\over m_{\star}}\right)^{1.2}\, x^{1.2} } \lesssim 
{0.01\over 192 \,\pi}\, {\a_2^2\,m_c^2 \over v^2}\, {m_{\star}\over m_{1}}\,  ,
 \ee
  where we defined $x \equiv s^2_{23}\,c^2_{23}\,|e^{-2i\s} -1|^2$ and maximized on it finding
  $x = 2.5^{1.2}\,(m_{\star}/m_1)$, that indeed implies $\s \ll 1$ as guessed.
 Imposing successful leptogenesis one then straightforwardly finds
 the upper bound 
 \be\label{ubm1}
 m_1 \lesssim  m_{\star}\,\left[{2.5^{1.2} \times 10^8\over 6\times 32\,\pi }\,{\a_2^2\,m_c^2 \over v^2}\right]\lesssim 52\,{\rm meV}  \,  ,
 \ee
very well reproducing the result from the scatter plots (Fig.~5, Fig.~6 and Fig.~7) even though it does not
fall in a full quasi-degenerate regime. This upper bound is shown in Fig.~7 (dot-dashed line).
 In conclusion the upper bound on $m_1$ is mainly due to an interplay between 
 minimising simultaneously both $K_{1\t}$ and $K_{2\t}$ while maximising the 
 $C\!P$ asymmetry. 
 \begin{figure}
\begin{center}
\psfig{file=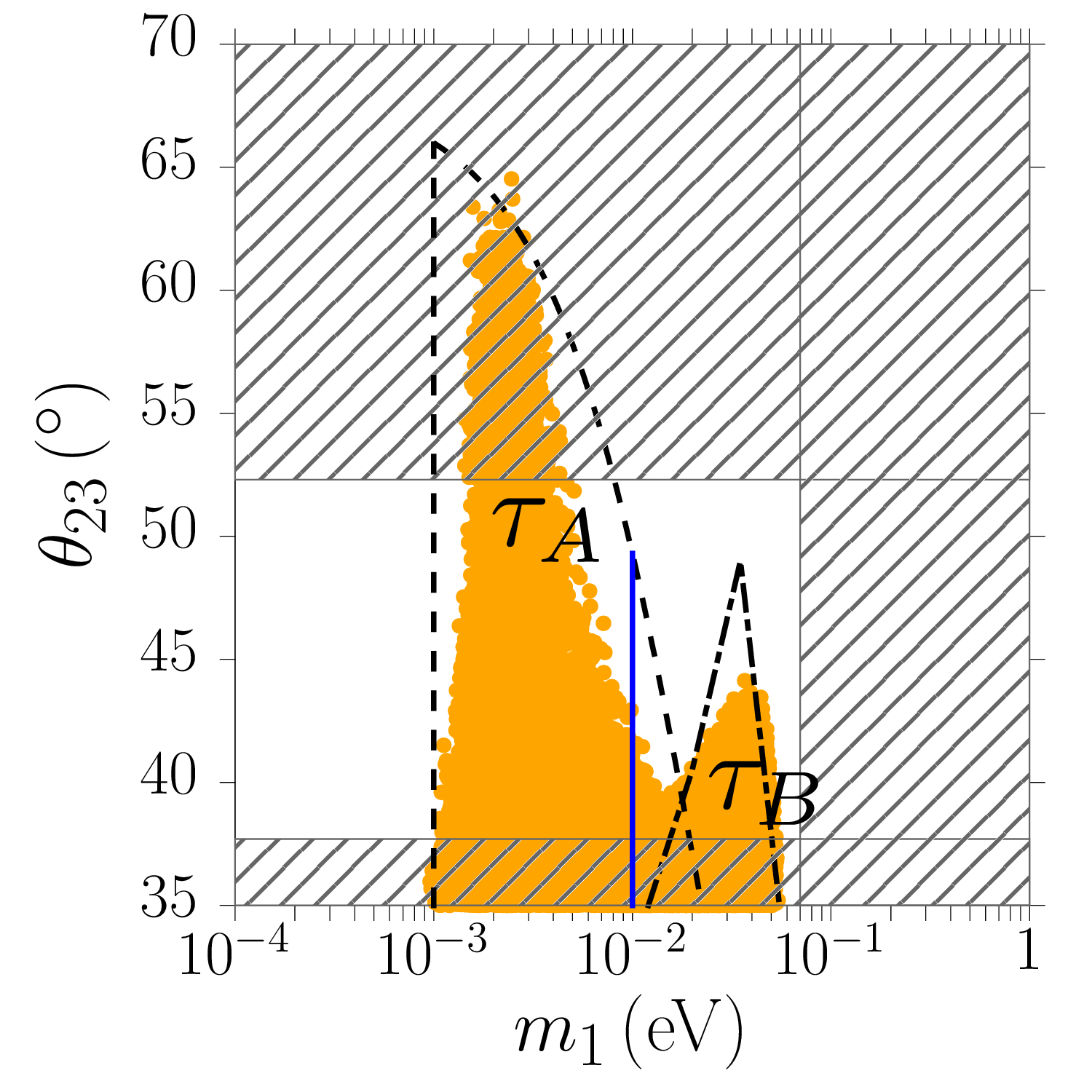,height=62mm,width=72mm}
\end{center}
\vspace{-11mm}
\caption{Scatter plot in the plane $m_1-\theta_{23}$ obtained imposing successful $SO(10)$-inspired leptogenesis with the asymmetry calculated from the analytic eq.~(\ref{NBmLflow}) as in the previous
figure but with enlarged $\theta_{23}$ range. The dashed  lines indicate the lower bound on $m_1$ 
eq.~(\ref{lbm1}) and the upper bound on $\theta_{23}$ at low $m_1$ eq.~(\ref{ubt23lm1}).
The dot-dashed lines indicate the upper bound on $m_1$ eq.~(\ref{ubm1}) 
and the upper bound on $\theta_{23}$ at high $m_1$.
As one can see, the two regions at low and high values of $m_1$, the type A and type B
solutions respectively, overlap around $m_1\sim 10\,{\rm meV}$.
The solid line is the lower bound from the ST condition eq.~(\ref{lbm1ST}) 
for $N^{\rm p,i}_{B-L}=10^{-3}$.}
\label{constrNO}
\end{figure}
In the top-left panel of Fig.~8 a scatter plot of $K_{2\t}$ vs. $m_1$ (orange points)
 confirms how for $m_1 \gtrsim 10\,{\rm meV}$ the value of $K_{2\t}$
becomes smaller and smaller for growing $m_1$  in order to minimise the wash-out at the production that  suppresses the asymmetry  $\propto m_{\star} / m_1$. The upper bound on $m_1$ is saturated for
an analytical minimum value of $K_{2\tau} \simeq 2.5$ well in agreement with the numerical result.  
\begin{figure}
%\vspace*{-35mm}
\hspace*{5mm}
\psfig{file=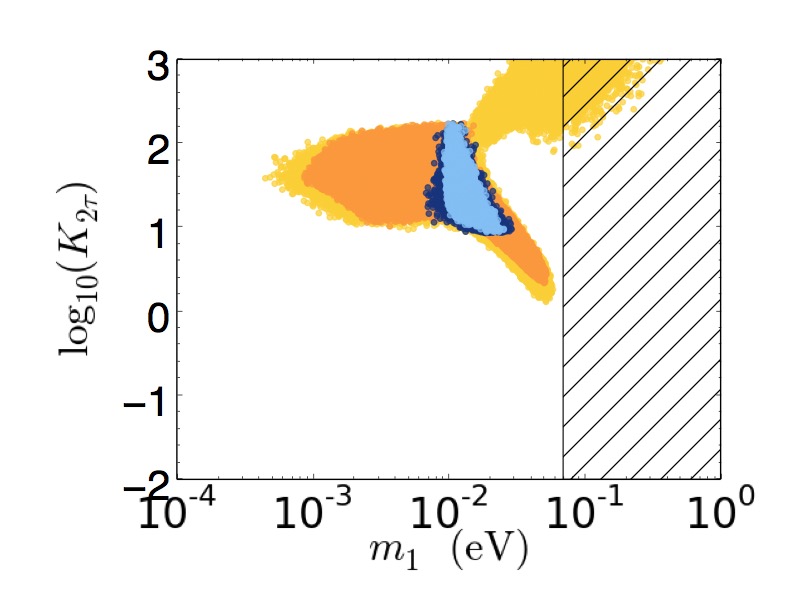,height=48mm,width=75mm}
\hspace{-3mm}
\psfig{file=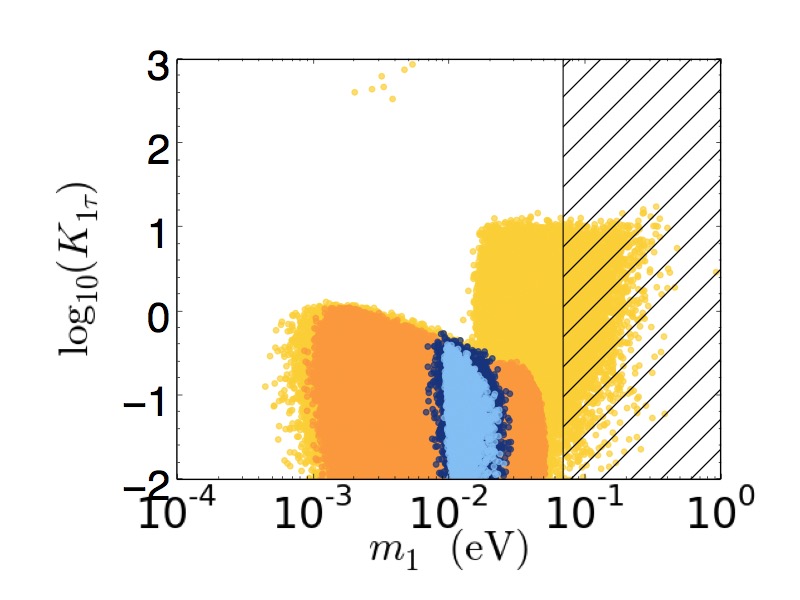,height=48mm,width=75mm}
\\
\vspace*{-4mm}
\hspace*{5mm}
\psfig{file=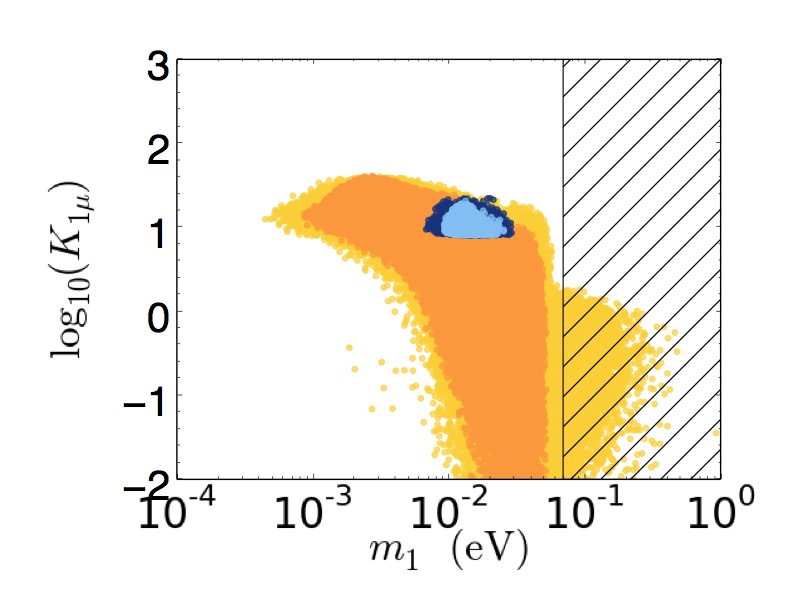,height=48mm,width=75mm} 
\hspace{-3mm}
\psfig{file=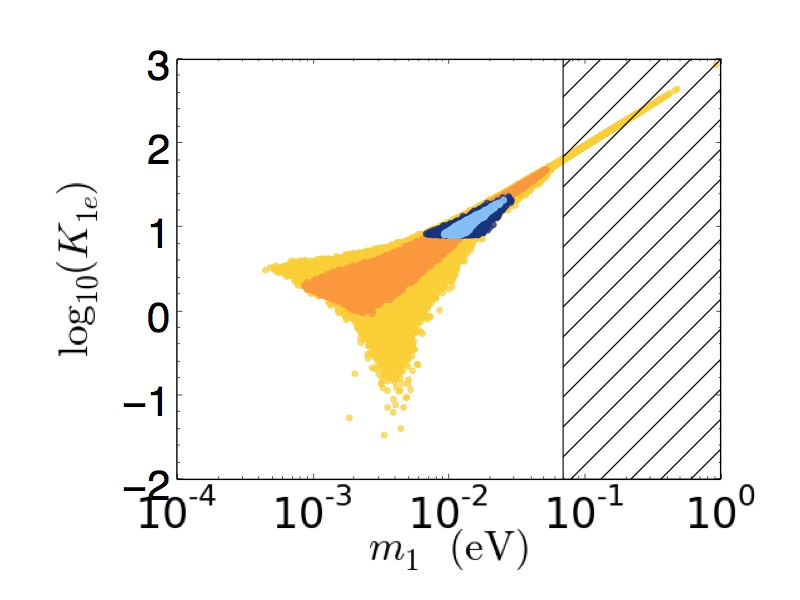,height=48mm,width=75mm}
\caption{Scatter plots for the four flavoured decay parameters 
$K_{2\t}$, $K_{1\t}$, $K_{1e}$, $K_{1\m}$ vs. $m_1$. The colour code is the
same as in Fig.~5.}
\end{figure}

\subsection{Type A solution ($m_1 \lesssim m_{\rm sol}$)}

We now describe what happens for values of $m_1$ between the lower and the upper bound. 
From this point of view, as we will see, the value of $m_{\rm sol} \simeq 10\,{\rm meV}$
represents a kind of border between two different solutions, the $\t_A$ and the $\t_B$
solutions, though the border is not sharp and the two solutions overlap somehow around
$m_1 \simeq 10\,{\rm meV}$. This distinction will be useful when we will discuss the ST solution
in the next Section. Let us start from values $m_1 \lesssim m_{\rm sol}$. 

\subsubsection{Upper bound on $\theta_{23}$}

An important feature of $SO(10)$-inspired leptogenesis is that, for NO as we are
considering, it places an upper bound on $\theta_{23}$.  In the case of low values of 
$m_1\lesssim m_{\rm sol}$, from the eq.~(\ref{K1taulowm1}),
imposing $K_{1\t} \lesssim 1$ and taking into account the dominant term 
$\propto m_1\,e^{2\,i\,\rho}$ in $K_{1\t}$, and approximating $\rho \simeq \pi/2$, one finds the upper bound
\be\label{ubt23lm1}
\theta_{23} \lesssim \arctan \left[{m_{\rm atm} - m_{\rm sol}\,s^2_{12}\over m_{\rm sol}+m_1}\,{s_{13}
\over c_{12}\,s_{12}} \right] \lesssim 65^{\circ} \,  ,
\ee
where the maximum value on the right-hand side is obtained clearly in the limit $m_1/m_{\rm sol}\ra 0$.
In Fig.~7 we show the results of a specific  scatter plot obtained starting from the analytic eq.~(\ref{NBmLflow}), holding for
$V_L \simeq I$, in the plane $m_1-\theta_{23}$, as for Fig.~5 but this time with $\theta_{23}$ in the
range $35^{\circ} \lesssim \theta_{23} \lesssim 70^{\circ}$. It can be seen how the 
analytical upper bound eq.~(\ref{ubt23lm1}) well reproduces the numerical result.

\subsubsection{Sign of the asymmetry and low energy phases}

Here we want to show how the sign of the asymmetry influences the values of the
phases. Looking at the $K_{i\a}$ one could indeed think that constraints on 
$\rho$ and $\sigma$ should exhibit a $\pi/2$-periodicity while constraints on $\d$
a $\pi$-periodicity. This is because the $K_{i\a}$ are defined in absolute values
and an overall change of sign of the argument leaves them unchanged.
However, it can be seen from the Figs. 5 and 6 
that actually $\s$ and $\rho$ constraints have a $\pi$-periodicity and constraints 
on $\d$ have a $2\,\pi$-periodicity. This is an effect of the sign of the asymmetry that,
 looking at the eq.~(\ref{NBmLflow}), is clearly given by 
\be
{\rm sign}(\eta_B) = {\rm sign}(\alpha_L) \, .
\ee
For the $\t_A$ solution, for $m_1 \ll m_{\rm sol}$,  one has  $\a_L \simeq 2\,(\rho - \s)$ and, 
therefore, one has that $\rho - \s \simeq \pi/4$ would maximise the asymmetry. We have already 
discussed how in the limit of lowest $m_1$ the value of $\rho \simeq 0.35\,\pi$ while indeed the value of 
$\s \simeq 0.1\,\pi$. This is because values $\rho=\pi/2$ would maximise the amplitude of the $C\!P$
asymmetry and at the same time minimise $K_{1\t}$. Values $\rho \simeq 0.35\,\pi$ are, therefore, a compromise that maximise the total final asymmetry.  At the same time the value $2\s-\d \simeq 0$ in order to minimise $K_{1\t}$, while $\s-\d \simeq 0$ in order to maximise $m_{ee}$. 

When $m_1/m_{\rm sol}$ increases, the values of the phases can be understood from the eq.~(\ref{K1tauexplicit})
for $K_{1\t}$.  The first term in the numerator $\propto m_1\,e^{2\,i\,\rho}$ becomes non-negligible. Since 
$\rho$ is slightly different from $\pi/2$, this term has some non-vanishing imaginary part that has also to be cancelled in order not to have $K_{1\t} \gg 1$. Since, as we have seen, at low $m_1$ necessarily 
$s_{13} \neq 0$, this is cancelled by the term 
$\propto m_3\,e^{i\,(2\s - \d)}$ with $2\s-\d < 0$. For increasing values of $m_1$, the value of $\rho$
has necessarily to tend to $\rho = n\,\pi$ in order to have $m_1 e^{2\,i\,\rho}- m_2 \simeq 0$. 
In this case there are two possibilities: either $\rho > \pi/2$ and in this case $2\s-\d > 0$
or $\rho < \pi/2$ and in this case $2\s-\d < 0$. The latter is clearly the dominant case, since, as we said, 
at lowest values of $m_1$ one has necessarily $\rho \simeq 0.35\pi < \pi/2$. For this case it is possible at the same time to have maximal phase and small value of $K_{1\t}$. It is important to realise that the dominance of this case is driven by the {\em positive sign of the asymmetry}. 

In order to make these visible from the scatter plots in a clear way, we have produced new scatter plots
constraining the reactor mixing angle in the current $3\s$ experimental range (cf. eq,~(\ref{expranges})).
The results are shown in Fig.~9. 
\begin{figure}
\vspace*{-25mm}
\hspace*{15mm}
\psfig{file=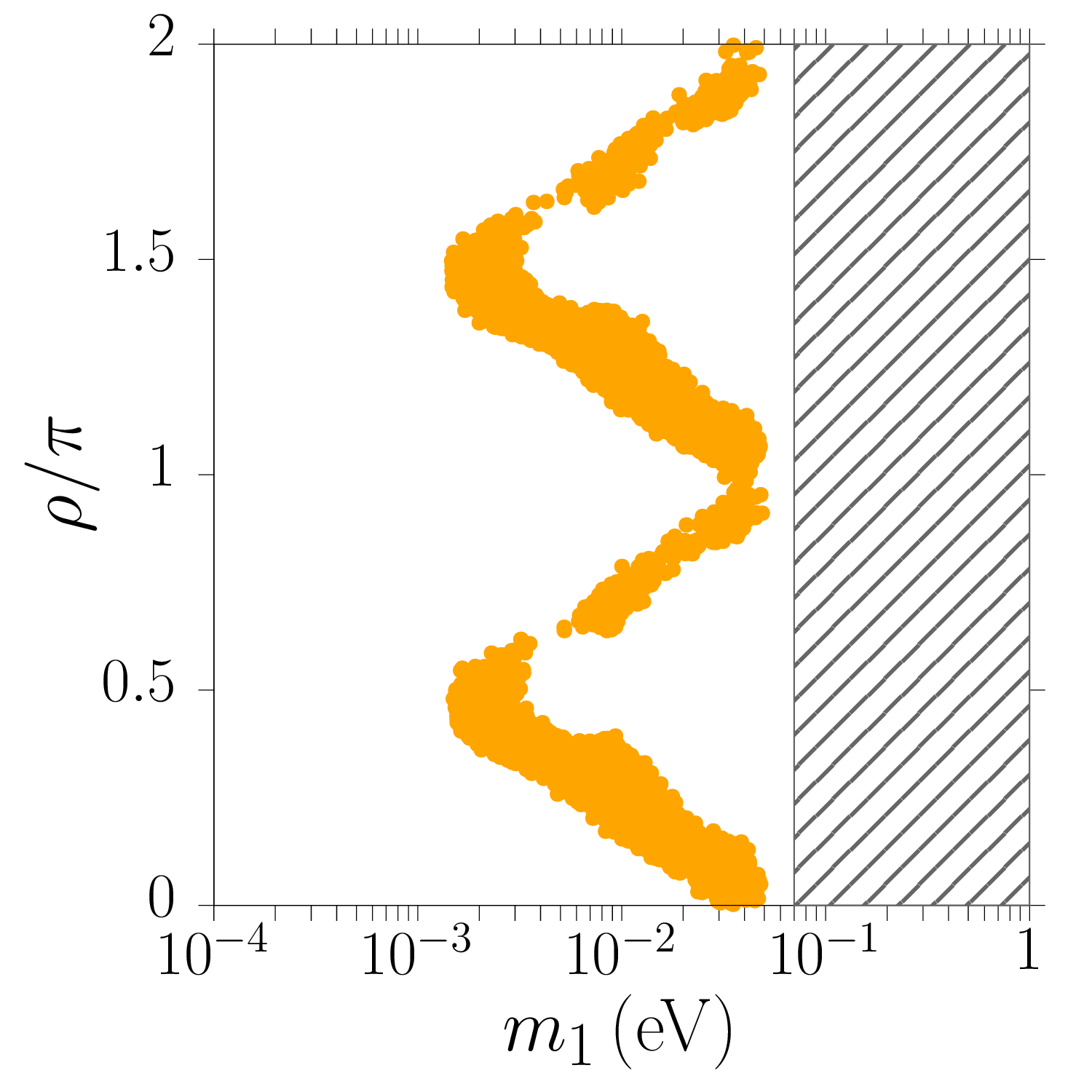,height=49mm,width=62mm}
\hspace{5mm}
\psfig{file=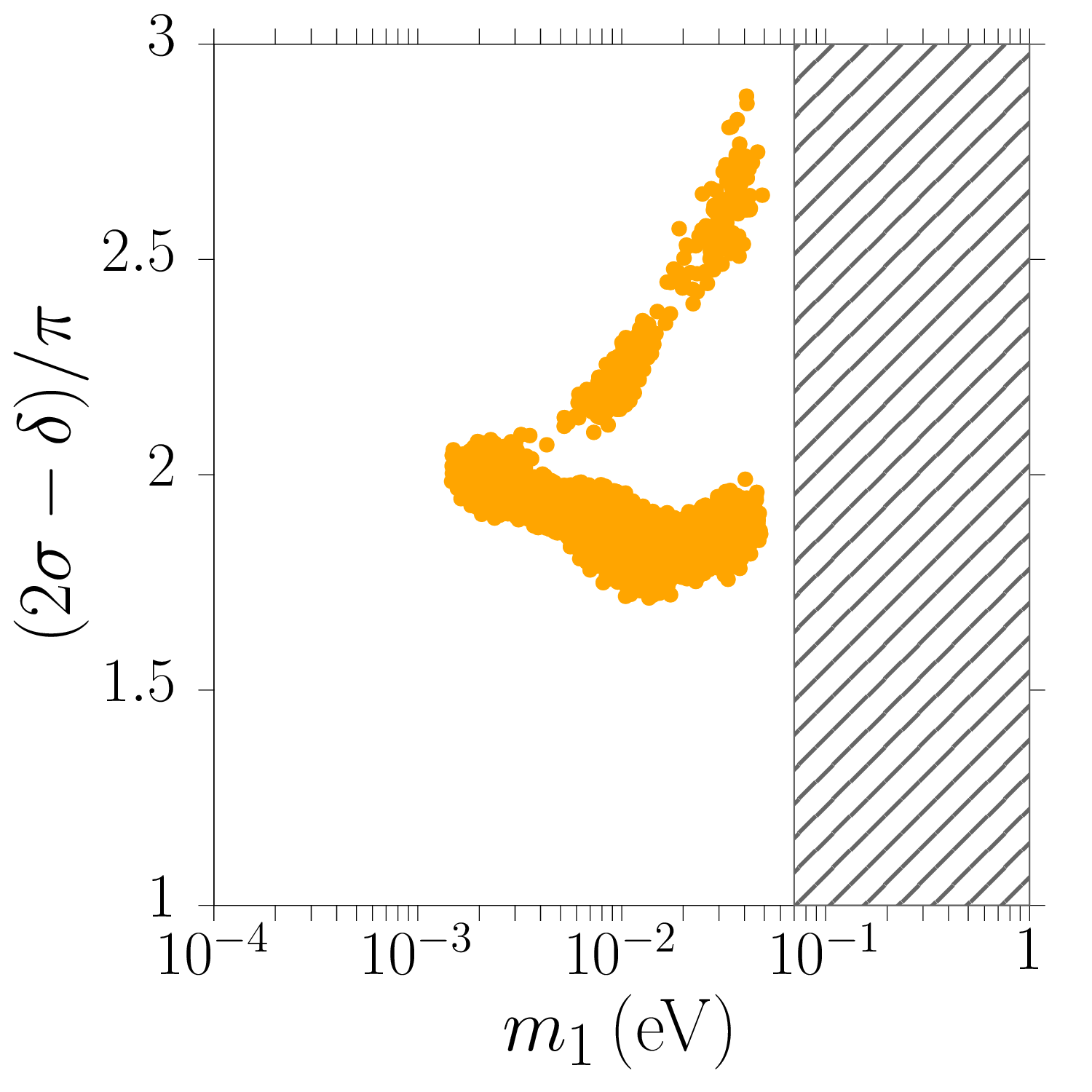,height=49mm,width=62mm}
\\
\hspace*{-3mm}
\psfig{file=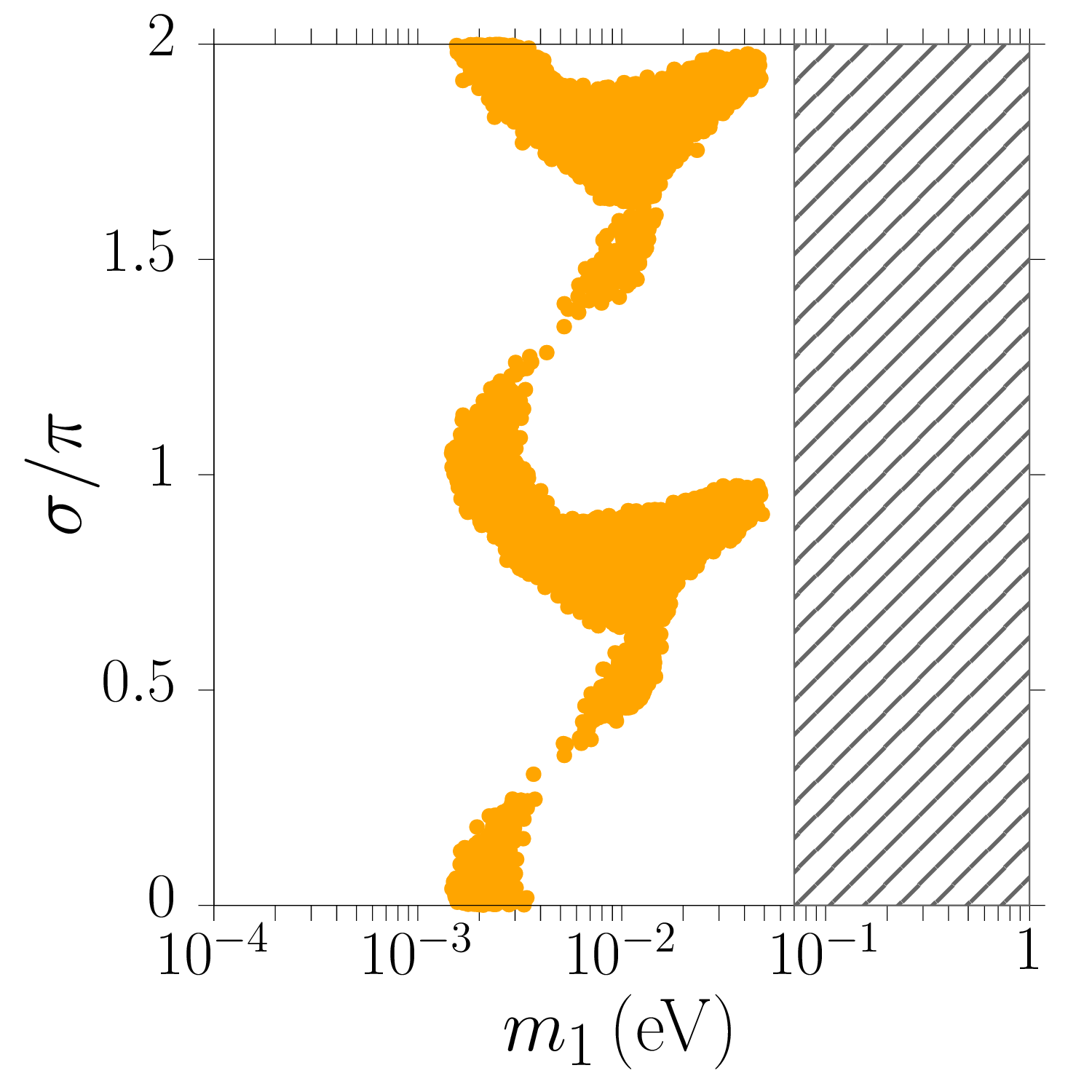,height=49mm,width=52mm} 
\hspace{-3mm}
\psfig{file=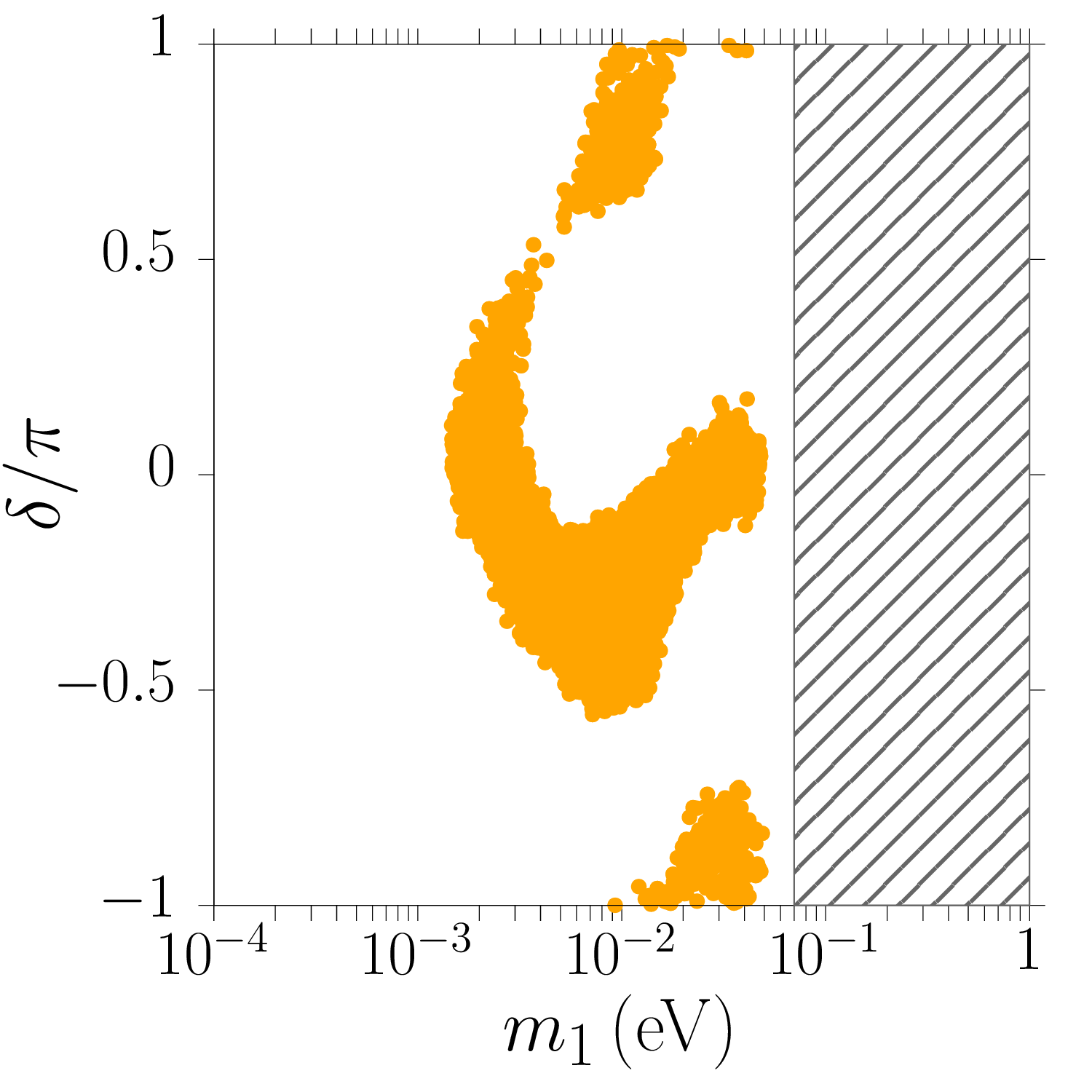,height=49mm,width=52mm}
\hspace{-3mm}
\psfig{file=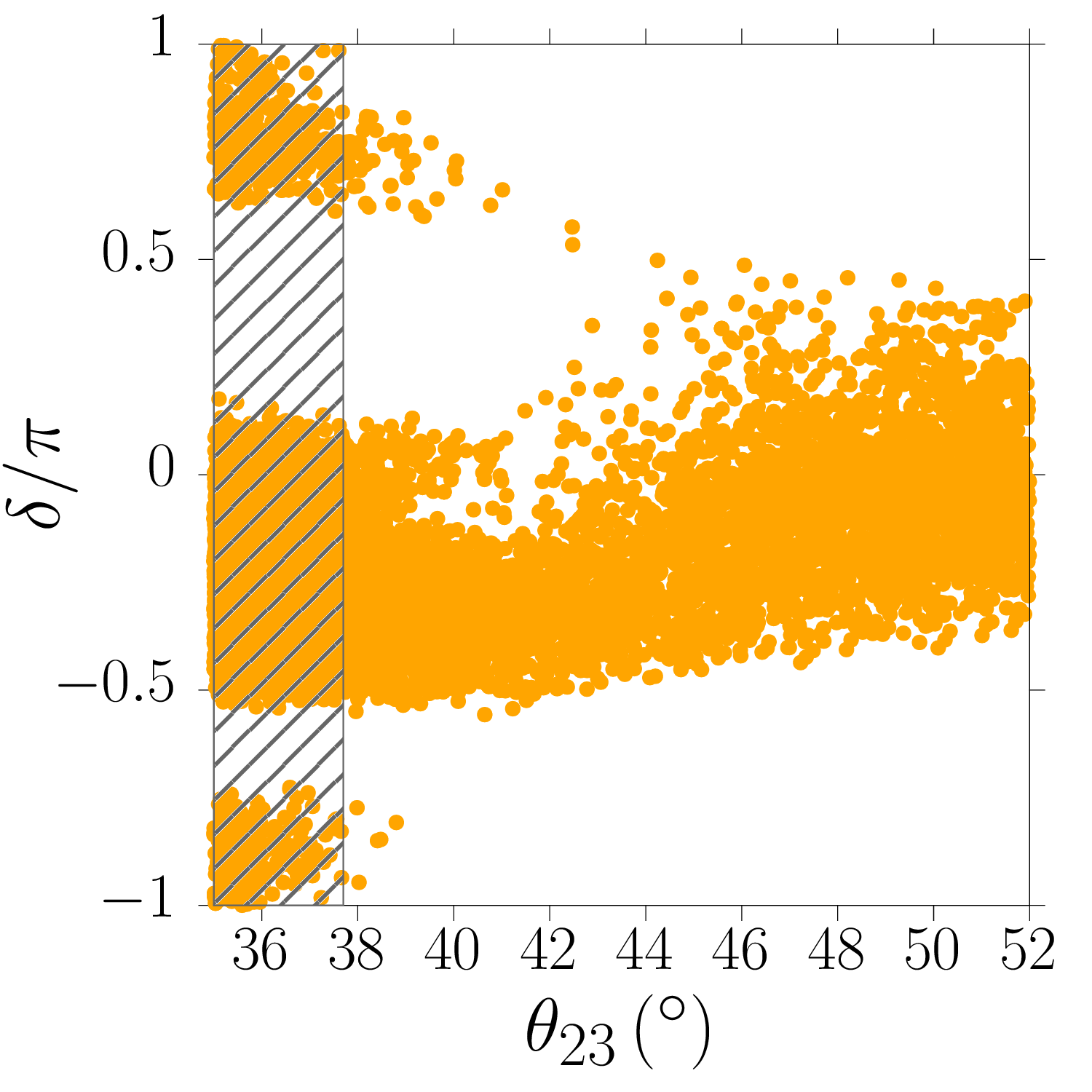,height=49mm,width=52mm}
\caption{Scatter plots of points satisfying successful leptogenesis 
generated using the analytical expression eq.~(\ref{NBmLflow}) for the
final asymmetry in different planes.  The mixing angle $\theta_{13}$ values are uniformly
randomly generated within the $3\,\s$ allowed experimental range in eq.~(\ref{expranges}).
Panels should be compared with the corresponding ones in Fig.~6, in particular
the last one for $\delta$ vs. $\theta_{23}$. }
\end{figure}
 In the first (top-left) panel we show the $\rho$ vs. $m_1$ scatter plot. One can first of all see that because of the much more restricted $\theta_{13}$ range, many points  disappeared compared to the corresponding plot in Fig. 5 and the behaviour is much cleaner. At the lower bound $m_1\simeq 1\,{\rm meV}$ one can see how indeed 
 $\rho \simeq 0.35\,\pi$.  For increasing values of $m_1$ there are two branches for $\rho$:
  in a first `high' branch the value of $\rho$ increases to $\pi$ and in a second `low' branch it decreases to $0$, where the two branches actually merge because of the $\pi$ periodicity. It is clearly noticeable how the
  low branch dominates, since it corresponds to values of $\rho$ that produce the correct sign of
  the asymmetry and to maximal leptogenesis phase ($\a_L$) values already at minimum $m_1$-values, while the high-$\rho$ branch is suppressed since it corresponds to non-maximal $\alpha_L$ values.
  
 In the second (top-right) panel we show the $2\s -\d$ scatter plot. This clearly shows how the 
 `low-$\rho$' branch corresponds to (dominant) $2\s -\d$ values below $2\,n\pi$, while the `high-$\r$' branch corresponds to (sub-dominant)  $2\s-\d$ values greater than $2\,n\pi$. 
 
 The next step is to understand what are the corresponding  values of $\s$.  In the third panel
 of Fig.~9 we show the $\s$ vs. $m_1$ scatter plot. As it could be expected by the fact that
 $\a_L \ra 2(\rho-\s)$ for $m_1 \ll m_{\rm sol}$, the (subdominant) high $\rho$ values branch 
 corresponds to high sub-dominant $\s$ values branch ($\s \gtrsim n\,\pi$) while the dominant
 low $\rho$ values branch corresponds to a low (dominant) $\s$ values branch. 
  
  Finally, combining these results on $\s$ with the results shown on $2\s - \d$, we can deduce
  the behaviour of $\d$. For the dominant low-$\rho$ values branch, corresponding to a low-$\s$ values branch
  and  values of $2\s-\d \lesssim 2\,n\,\pi$ one concludes that $\d$ shifts toward negative values. Vice-versa the sub-dominant high $\rho$ values branch, corresponding to $2\s -\d > 2n\,\pi$ values and high $\s > m\,\pi$ values,
 one has positive $\d$ values. The results are shown in the last (bottom-right) panel of Fig.~9. One can see the clear dominance of values of $\d$ in the fourth quadrant. This conclusion is strengthened even more by a scatter plot of $\d$ vs. $\theta_{23}$ showing that actually positive values of $\d$ are even more constrained if one 
 imposes the current $3\s$ lower bound  $\theta_{23} \gtrsim 38^{\circ}$. This result 
 should be mainly regarded as a proof that within $SO(10)$-inspired leptogenesis the sign of the asymmetry
 yields asymmetric constraints between positive and negative $\sin\d$ values. However, it would be certainly 
 interesting to see how these constraints relax going beyond the $V_L= I$ approximation since this could
also provide quite an effective way to test $SO(10)$-inspired leptogenesis with future experimental results. 
   
\subsection{Type B solution ($m_1 \gtrsim m_{\rm sol}$)}

Because of the upper bound $m_1 \lesssim m_{\rm atm}$, 
we can approximate $m_1 \simeq m_2$ and $m_3 \simeq m_{\rm atm}$.
In this case the general expression for $K_{1\t}$ eq.~(\ref{K1tauexplicit})
can be written as,
\be
K_{1\t} \simeq {|c_{13}\,c_{12}\,s_{12}\,s_{23}\,m_1\,(e^{2\,i\,\rho}-1) 
+ s_{13}\,c_{13}\,c_{23}\,e^{-i\,\d}\,[m_{\rm atm}\,e^{i\,2\,(\s-\d)} - m_1\,(s^2_{12} + \,c^2_{12}\,e^{2\,i\,\,\rho})|]^2\over m_{\star}\,|m_1\,\,c^2_{13}(c^2_{12}\,e^{2\,\,i\,\rho}+ s^2_{12}) + m_{\rm atm}\,s^2_{13}\,e^{2\,i\,(\s-\d)}|} \, .
\ee
There are two possibilities to minimise $K_{1\t}$. In the case $s_{13}\ra 0$ one can simply have
$\rho =n\,\pi$ and this immediately produces $K_{1\t}=0$, showing that it is quite easy to find a way for
the tauon asymmetry to escape the lightest RH neutrino wash-out. On the other hand for the measured values
$s_{13}\simeq 0.15$ a non-vanishing value of the first term in the numerator is necessary in order to cancel the second term. The exact value of $\rho$ depends on the value of $m_1$. The value of $\d$ is in this case able
to cancel the imaginary part of $e^{2\,i\,\rho}$ but at the same time has to be such to keep $\s-\d \simeq n\,\pi$
in order to maximise the value of $m_{ee}$.  Moreover since $\alpha_L\simeq -4\,\s$  has to be negative, this also leads to negative values of $\d$ and favours positive values of $\rho$. This is confirmed by the first panel of Fig.~9 showing a scatter plot of $\rho$ vs. $m_1$ for $\theta_{13}$ in the $3\s$ range eq.~(\ref{expranges}). It can be seen how this time, compared to the analogous plot of Fig.~5 where $0\leq \theta_{13} \leq 11.54^{\circ}$, one has $\rho=n\,\pi$ only when $m_1$ saturates its upper bound.  

We can maximise $K_{1\t}$ taking in both cases 
$\rho=n\,\pi$ even for $m_1 \simeq m_{\rm sol} \ll m_{\rm atm}$ and taking $\s-\d = m\,\pi$
in order to maximise $m_{ee}$, and in this case one finds 
\be
K_{1\t} \lesssim { s^2_{13}\,c^2_{23}\,\,(m_{\rm atm} - m_1)^2\over 
m_{\star}\,(m_{\rm atm} + s^2_{13}\,m_{\rm atm})} \simeq  2  \, ,
\ee
showing that indeed the lightest RH neutrino wash-out can be avoided in any case. 

Let us now consider the $C\!P$ asymmetry $\ve_{2\t}$ and the  wash-out at the production 
described by $K_{2\t}$. In the expressions one can still approximate $m_{ee}\simeq m_1$,
as for the quasi-degenerate case. However, this time one has
\be
|(m_{\nu}^{-1})_{\t\t}| \simeq {1\over m_1} \, \left|s^2_{23} + {m_1\over m_3}\,c^2_{23}\right| \, 
\hspace{5mm} \mbox{\rm and} \hspace{5mm}
|(m_{\nu}^{-1})_{\m\t}| \simeq {s_{23}\,c_{23}\over m_1} \, \left|1-  {m_1\over m_3}\right|  \,  , 
\ee
where we have approximated $\s \simeq n\,\p$. These two expressions produce a dependence
$\eta_B \propto s_{23}^{-4}$ that strongly suppresses the asymmetry for increasing $s_{23}$
and produces a tight upper bound on $\theta_{23}$.
On the other hand, however, now one also has $\eta_B \propto m_1/m_3$ and this makes
in a way that the upper bound gets relaxed at higher $m_1$ reaching a maximum toward
$m_1 \simeq 0.035\,{\rm meV}$. This is because for higher $m_1$ the term 
$|(m_{\nu}^{-1})_{\m\t}|\propto 1-m_1/m_3$ suppresses the asymmetry.

Finally one also has 
\be\label{K2tauhighm1}
K_{2\t}\simeq {m_3 \over m_{\star}} \, {s^2_{23}\,c^2_{23}\,(1-m_1/m_3)^2\over (s^2_{23}+ c^2_{23}\,m_1/m_3)}\, .
\ee
Combining together all results, one finds an implicit form  for the upper bound of $s_{23}$ vs. $m_1$.
In Fig.~7 we have plotted with the dot-dashed line the result. As one can see it somehow overestimates
the allowed region. This is a consequence of the crude approximations used for the phases. In any case
these results well explain the existence of an upper bound on $\theta_{23}$ also for values
$m_1 \gtrsim m_{\rm sol}$ and how this gets relaxed for increasing values of $m_1$ upto a peak value
that is reached for $m_1 \simeq 35\,{\rm meV}$. For values $m_1 \gtrsim 35\,{\rm meV}$
the upper bound on $\theta_23$ vs $m_1$ becomes more stringent and $\theta_{23}^{\rm max} \ra 0$
when $m_1 \ra m_1^{\rm max}$, where $m_1^{\rm max}\simeq 52\,{\rm meV}$ 
is the upper bound eq.~(\ref{ubm1}) found in the quasi-degenerate limit. 
It should be noticed how the regions for the $\tau_A$
and for the $\t_B$ solutions overlap to some extent for $m_1 \simeq 10\,{\rm meV}$. 
This is not contradictory since they are realised for
different values of the phases, in particular in the case of the $\tau_A$ solution the phase $\rho \simeq \pi/2$
for $m_1 \ra 0$, while for the $\t_B$ solution one has $\rho \simeq \pi$ for $m_1 \simeq m_1^{\rm max}$.
 Around $m_1 \simeq 10\,{\rm meV}$ the two solutions meet but, as we will discuss in the next Section, 
 the $\t_B$ solution  is incompatible with the ST leptogenesis condition. 

\section{Strong thermal leptogenesis condition}

In this section we finally over-impose the ST condition in addition to the successful leptogenesis and
$SO(10)$-inspired conditions, deriving all the features of the 
ST-$SO(10)$-inspired solution \cite{strongSO10solution}.
All $SO(10)$-inspired solutions, for $V_L = I$, are already tauon-type solutions satisfying 
$K_{1\t} \lesssim1$. Therefore, we need to impose, in addition, the conditions eqs.~(\ref{STcond}).

For pre-existing initial asymmetries $N_{\D_{\a}}^{\rm p,i}$ in the different flavours $\a=e,\m,\t$, 
one has to require \cite{strongthermal}
\be\label{Kstrong}
K_{2\t}\, , K_{1e} \,  ,  K_{1\m}  \gtrsim  
 {8\over 3\pi }\,\left[\ln \left({0.1 \over \eta_B^{\rm CMB}}\right) 
+ \ln \left|{N_{\D_{\a}}^{\rm p,i}}\right|\right] \simeq 8 + 0.85\,\ln 
\left|{N_{\D_{\a}}^{\rm p,i}\over 1.5\times 10^{-4}}\right| \,   ,
\ee
in order for these components to give a negligible contribution to the final asymmetry. Because
of geometric factors in general a total pre-existing asymmetry $N^{\rm p,i}_{B-L}$
corresponds to values of the  electron and muonic pre-existing  asymmetries
at the lightest RH neutrino wash-out of about $N_{\D_{\a}}^{\rm p,i} \simeq N^{\rm p,i}_{B-L}/6$
\cite{strongthermal}. 

\subsubsection*{The $\tau_B$ solution cannot realise ST leptogenesis}

First of all it is easy to show that the $\tau_B$ solutions, characterised 
by $m_1\gtrsim 10\,{\rm meV}$ and $\rho \simeq n\,\pi$ cannot satisfy the
ST condition. 

If one goes back to the expression (\ref{K2tauhighm1})
for $K_{2\t}$ in the $\tau_B$ case, one can immediately check that for $m_1\simeq m_{\rm sol}$
one has $K_{2\t}\simeq 13$, that would be still sufficient to wash-out a pre-existing asymmetry 
as large as $\sim 10^{-2}$.  The wash-out a pre-existing
electronic asymmetric is also not a problem. Indeed, starting from the expression eq.~(\ref{Kiaapp})
for the $K_{i\a}$ and using $M_1 = m^2_{D1}/m_{ee}$ (cf. eq.~(\ref{M1})) and that $U_{R 11}\simeq 1$,
one immediately obtains, in general and therefore also for $\t_B$ solutions, 
that $K_{1e} \simeq m_{ee}/m_{\star} \simeq m_1/m_{\star}$. This is  
sufficient to wash-out electronic pre-existing asymmetries as large as $10^{-3}$ 
for $m_1 \gtrsim 10\,{\rm meV}$ and even larger if $m_1$ increases (as we will see soon
this is indeed the origin of the lower bound on $m_1$). The real intrinsic problem for 
$\tau_B$ solutions is the wash-out of a pre-existing muon component, since one can easily see that
\be
\left. K_{1\m} \right|_{\tau_B} \simeq {m_{\rm atm}^2 \over m_{\star}}
\,{s^2_{13}\,s^2_{23}\over |m_1 + s^2_{13}\,m_{\rm atm}|} \lesssim 4 \,  ,
\ee
confirming in a general analytical way  the numerical examples shown in \cite{SO10lep1,SO10lep2,strongSO10solution}.
Therefore, we conclude that $\tau_B$-type solutions cannot realise ST leptogenesis
because they cannot wash-out a large pre-existing muon asymmetry. 
We can then now focus on $\tau_A$ solutions in the following discussion.

\subsubsection*{Lower bounds on $m_{ee}$ and on $m_1$}

As we have just seen, one finds easily from the general expression eq.~(\ref{Kialpha})
$K_{1e} = m_{ee}/m_{\star}$, interestingly showing how in $SO(10)$-inspired models a not too low neutrino-less
double beta decay effective neutrino mass  is required for the wash-out of the pre-existing 
electronic asymmetry. Indeed, from the eq.~(\ref{Kstrong}) we can immediately place the 
lower bound
\be\label{meelb}
m_{ee} \gtrsim 8\,{\rm meV}\,\left(1 + 0.095\,\ln 
\left|{N_{\D_{e}}^{\rm p,i}\over 1.5 \times 10^{-4}}\right| \right) \,  ,
\ee
that is quite interesting since it predicts that, despite neutrino masses are NO,
next generation $0\nu\b\b$ experiments should find a signal.
This lower bound translates into a lower bound on $m_1$.  From the explicit general expression for $m_{ee}$,
\bea
m_{ee} & = & |m_1\,U^2_{e1}+m_2\,U^2_{e2}+m_3\,U^2_{e3}| \\ \nonumber
            & = & |m_1\,c_{12}^2\,c_{13}^2\,e^{2\,i\,\rho} + 
            m_2\,s^2_{12}\,c^2_{13}+
            m_3\,s^2_{13}\,e^{2\,i\,(\s -\d)}| \\   \nonumber
            & \simeq & m_1\,|c^2_{12}\,e^{2\,i\,\rho}+s^2_{12}|\,  ,
\eea
where we approximated $m_1\simeq m_2$ and neglected the term $\propto m_3\,s^2_{13}$.
Considering that for the $\tau_A$ type solutions one has $2\rho \simeq \pm \pi/2$, 
one arrives to $m_{ee}/m_1 \simeq \sqrt{c^4_{12}+s^4_{12}} \simeq 0.75$, in very good agreement
with the numerical results. From this result combined with the lower bound eq.~(\ref{meelb}), 
one then obtains the lower bound 
\be\label{lbm1ST}
m_1 \gtrsim 10\,{\rm meV}\,\left(1 + 0.095\,\ln 
\left|{N_{\D_{e}}^{\rm p,i}\over 1.5\times 10^{-4}}\right| \right) \,  .
\ee
This result is a specific example of what happens more generally, beyond $SO(10)$-inspired models, 
for NO: the wash-out of the electronic pre-existing asymmetry implies a lower 
bound on $m_1$ \cite{strongthermal}.  In the case of $SO(10)$-inspired models this lower bound
is particularly stringent and implies $\sum m_i \gtrsim 75\,{\rm meV}$, a prediction that 
will be tested by future cosmological observations. 

\subsubsection*{Atmospheric mixing angle in the first octant 
and upper bound on $m_1$ and $m_{ee}$}

Plugging the lower bound on $m_1$ eq.~(\ref{lbm1ST})
in the eq.~(\ref{ubt23lm1}) giving the upper bound on $\theta_{23}$,
one finds, for $N_{\D_e}^{\rm p,i}=10^{-3}$, the upper bound $\theta_{23}\lesssim 40^{\circ}$,
quite in well agreement with the scatter plots in Fig.~5 (light blue points).

At the same time upper bound on $m_1$ is found simply by the value of $m_1$
corresponding to the minimum value of $\theta_{23}$ in the eq.~(\ref{lbm1ST}).
For $\theta_{23}=35^{\circ}$, as in the scatter plots in Fig.~5, one finds $m_1 \lesssim 20\,{\rm meV}$.
From this upper bound one can then straightforwardly write $m_{ee}\lesssim 0.8\,m_1 \lesssim 16\,{\rm meV}$, 
in fair agreement with the result from the scatter plots shown in Fig.~5 that give $m_1\lesssim 23\,{\rm meV}$.
This upper bound gets relaxed to $m_1 \lesssim 30\,{\rm meV}$ going beyond the approximation $V_L \simeq I$
and corresponds to $\sum m_i \lesssim 125\,{\rm meV}$.

\subsubsection*{Lower bound on $\theta_{13}$} 

From the general expression eq.~(\ref{Kiaapp}) for the $K_{i\a}$,
one obtains the following expression for $K_{1\m}$ 
for $\rho\simeq \pm \pi/2$ and $m_1 \simeq m_2$,
\be
K_{1\m} \simeq {c^2_{13}\,|s_{12}\,c_{12}\,c_{23}\,m_1\,(1\pm i)+m_3\,s_{13}\,s_{23}|^2 \over m_{\star}\,|m_1 + m_3\,s^2_{13}|}  \,  .
\ee 
It is easy to see that, for $s^2_{13}=0$, the condition $K_{1\m}\gtrsim 10$
(for $N^{\rm p}_{B/3-L_\m} \simeq 10^{-3}$) would imply $m_1 \gtrsim 30\,{\rm meV}$,
clearly incompatible with the upper bound $m_1 \lesssim 20\,{\rm meV}$ just obtained.
However, for $s^2_{13}\gtrsim 0.1$, corresponding to $\theta_{13} \gtrsim 5^{\circ}$,
one can simultaneously satisfy $K_{1\m} \gtrsim 10$ and $m_1 \lesssim 20\,{\rm meV}$.
This is an interesting feature of the ST solution, since it predicts
a non-vanishing reactor mixing angle \cite{talks}  as now firmly established 
by the experimental results. 

It should be noticed that this lower bound on $\theta_{13}$ strengthens the lower bound 
eq.~(\ref{t13lb}) derived from the condition $K_{1\t}\gtrsim 1$. 

\subsubsection*{Dirac phase in the fourth quadrant} 

As we discussed in 5.4.2, for $\tau_A$ solutions 
the Dirac phase is driven toward negative values because of 
its link with the phase $\s$ inside $K_{1\t}$ that requires,
for non-vanishing $\theta_{13}$, $2\s-\d \simeq 0$.
In addition there is a subdominant solution for $\d\simeq \pi$ 
When the ST condition is imposed this conclusion is strengthened even more
by the more stringent lower bound on $\theta_{13}$.  The last panel of Fig.~9,
where we plotted $\d$ vs. $\theta_{23}$ for the experimental allowed $3\s$ range
for $\theta_{13}$, clearly shows this situation. It also shows how $\d$ is basically constrained
in the fourth quadrant for $\theta_{23}$ in the first octant and $\theta_{23}\gtrsim 38^{\circ}$.
We have just seen how the ST condition necessarily requires $\theta_{23}$ in the first octant
as a result of the lower bound on $m_1$. Combining this result with the $3\s$ lower bound
$\theta_{23}\gtrsim 38^{\circ}$, we conclude that the only way for the ST condition to be satisfied
for such high values of $\theta_{23}$ is to have $\d$ in the fourth quadrant ($-\pi/2 \lesssim \d \lesssim 0$).
This is an interesting result in light of the experimental hint for $\sin\d <0$. The ST condition more definitely 
requires  also $\cos\d>0$. From the Fig.~5 panel showing $\d$ vs. $\theta_{23}$, 
it can be seen how, for $N_{B-L}^{\rm p,i} =10^{-3}$, the highest value of $\theta_{23}$
is obtained for $\d\simeq -60^{\circ}$ and is given by $\theta_{23} \simeq 41^{\circ}$
(light blue points for $V_L = I$). This upper bound relaxes to $\theta_{23} \lesssim 43^{\circ}$
going beyond the approximation $V_L \simeq I$.

\section{Inverted Ordering}

In this section we finally extend the discussion to the case of IO. 
The expressions (\ref{M1}), (\ref{M2}) and (\ref{M3})
still apply while the asymptotic limits for $m_1 \ra 0$ (cf. eqs. (\ref{Milowm1})) become now
\bea\label{Milowm1}
M_1  & \simeq &   {m^2_{D1} \over m_{\rm atm}\,c^2_{12} |c^2_{13}\, e^{2\,i\,\rho}+s^2_{12}|}
 \approx   \, {m^2_{D1} \over m_{\rm atm}\,c^2_{12}\,c^2_{13}} \, ,  \\ \nonumber
M_2 & \simeq & {m^2_{D2}\,|c^2_{13}\,e^{2\,i\,\rho}+s^2_{12}| \over m_{\rm atm} \,c^2_{23}\,c^2_{13}} 
  \approx  \, {m^2_{D2}\,c^2_{12} \over m_{\rm atm} \, c^2_{23}} \, ,  \\  \nonumber
M_3 & \simeq  & {m^2_{D3}  \over m_1} \,c^2_{23}\,c^2_{13}   \,  .
\eea
At the same time also the eq.~(\ref{URapp}) for the RH neutrino mixing matrix, 
the eq.~(\ref{CPanalytic}) for the $\ve_{2\a}$, the eq.~ (\ref{Kiaapp}) for the $K_{i\a}$ and, finally,
the eq.~(\ref{NBmLflow}) for the final asymmetry are also still valid. 
All the differences arise only when the neutrino masses and the leptonic mixing entries
are specified, since these are different from the NO case. 

\subsection{Successful leptogenesis}

As done for NO, we have verified that the
eq.~(\ref{NBmLflow}) is able to reproduce the numerical results of \cite{SO10lep2} for $V_L = I$ 
when the condition of successful leptogenesis is imposed. 
In particular it is confirmed that IO is only marginally allowed, requiring quite a restricted range of values 
$m_1$ between $20$ and $40\,{\rm meV}$ and the existence
this time  of a lower bound on the atmospheric neutrino mixing angle $\theta_{23} \gtrsim 48^{\circ}$,
that therefore has to lie in the second octant.  Moreover the lower bound on $\a_2$ is very stringent,
and values $\a_2 \lesssim 4.5$ are not allowed.  

It is interesting to show analytically the origin of some of the differences between IO and NO. 
The usual starting point is the calculation of $K_{1\t}$ that this time is particularly simple since
we can use the approximation $m_2 \simeq m_3$. In this way it is easy to show that 
$K_{1\t}$ is minimised for $\rho=n\,\pi$ and in this case one has
\be 
K_{1\t} \gtrsim {m^2_1 \, s^2_{13}\,c^2_{23} \over m_{\star}\, m_2} \,  ,
\ee
so that the condition $K_{1\t}\lesssim 1$ simply leads to the upper 
bound 
\be\label{upperboundm1IO}
m_1 \lesssim 0.1\,{\rm eV}\,{0.01\over s^2_{13}\,c^2_{23}} \,  .
\ee
As we will notice, however, the exact condition $\rho=n\,\pi$ would 
imply $\sin\a_L \simeq 0$ and, therefore, a small (positive) displacement from
$\rho=n\,\pi$ is necessary for the asymmetry not to vanish (see central plot in Fig.~10) and clearly this implies that the upper bound is more stringent. In the scatter plot (see left panel in Fig.~10) 
it is indeed found $m_1 \lesssim 50\,{\rm meV}$.  
\begin{figure}
\begin{center}
\psfig{file=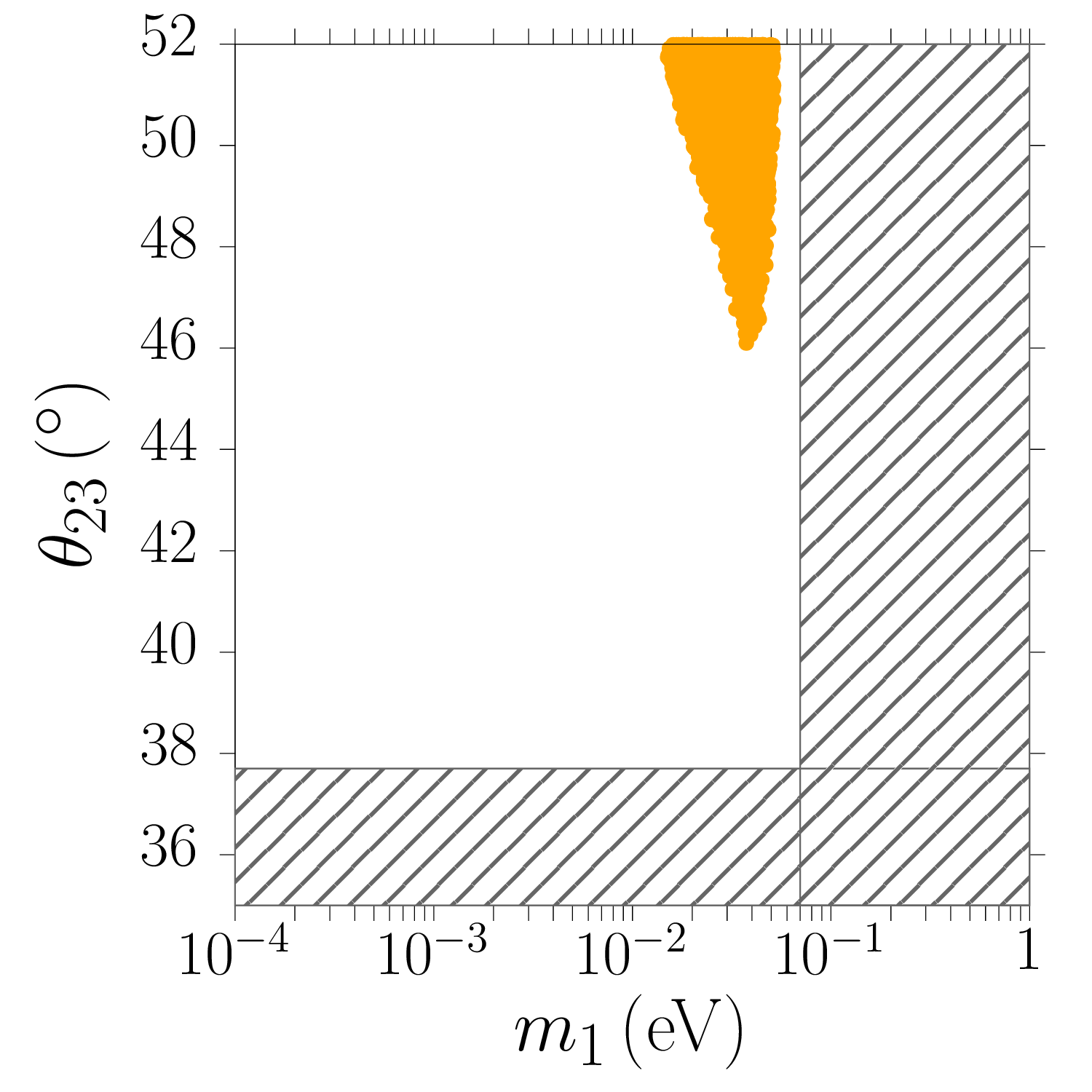,height=39mm,width=46mm}
\hspace{5mm}
\psfig{file=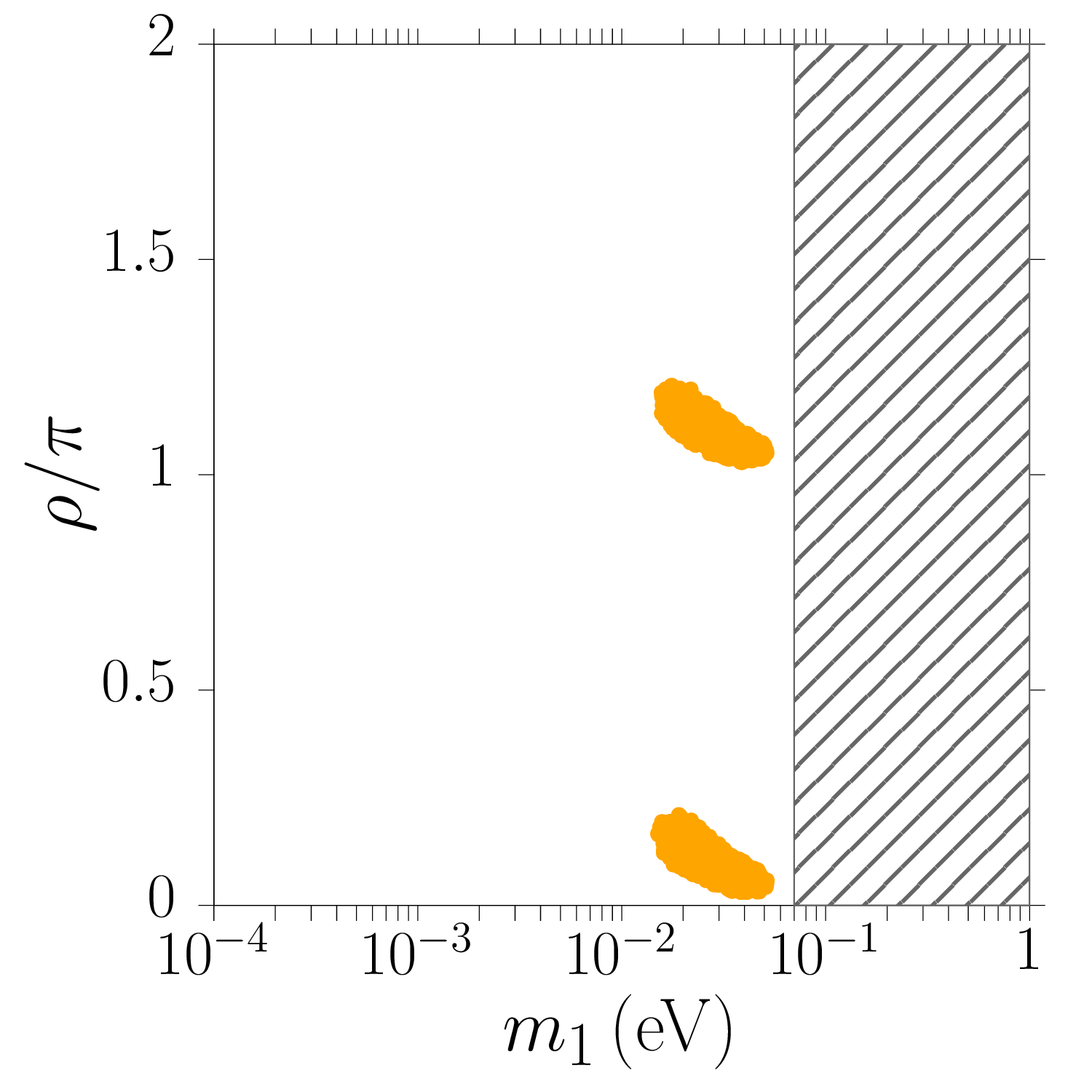,height=39mm,width=46mm} 
\hspace{5mm}
\psfig{file=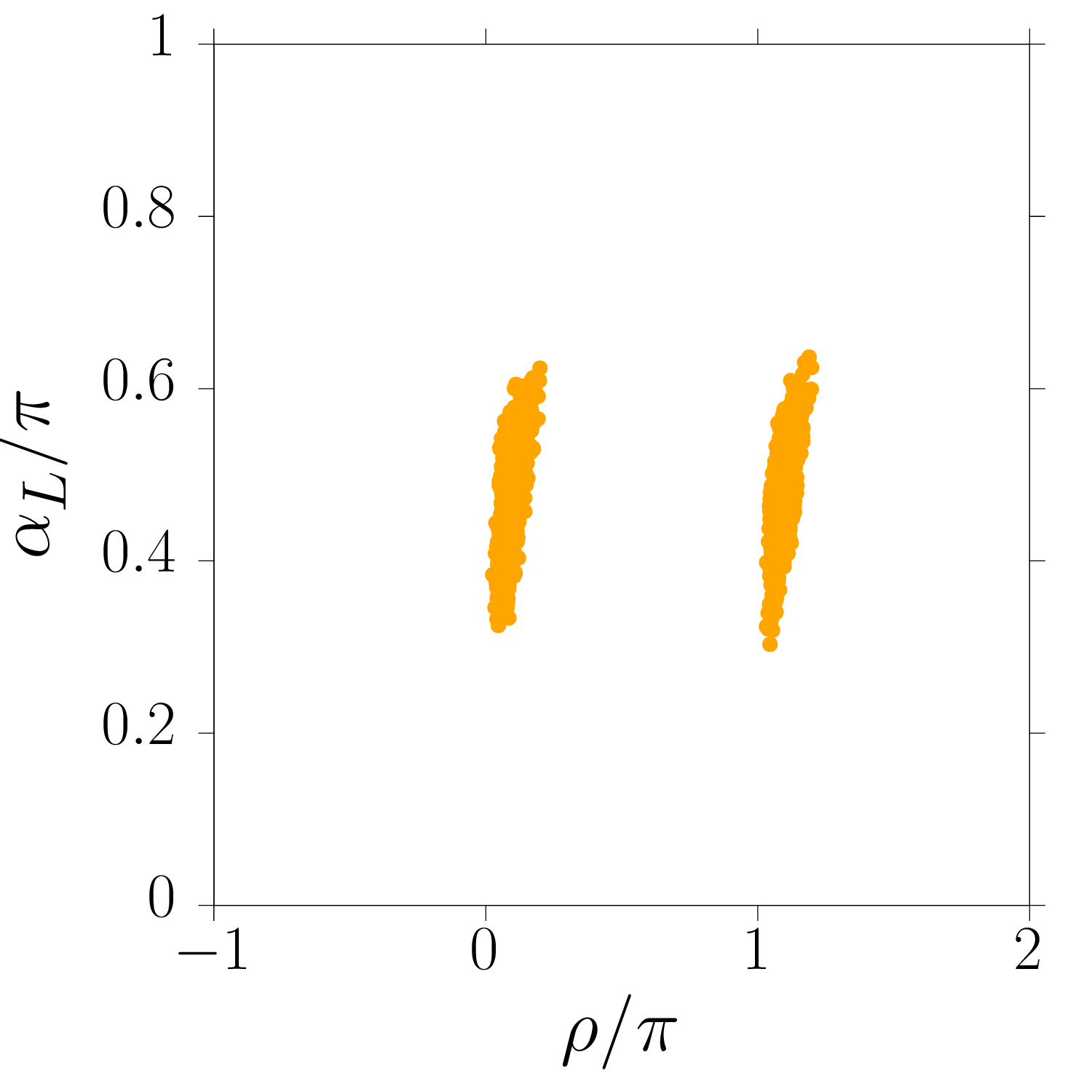,height=39mm,width=46mm} 
\end{center}
\vspace{-5mm}
\caption{IO case. Scatter plots for $\theta_{23}$ vs. $m_1$ (left), $\rho$ vs. $m_1$ (centre)
and $\alpha_L$ vs. $\rho$ (right).}
\end{figure}
We can also easily estimate the wash-out at the production, calculating
\be
K_{2\t} \simeq {m_3\over m_{\star}}\,s^2_{23} \,   ,
\ee
entering the efficiency factor $\kappa(K_{2\t})\simeq 0.5/K_{2\t}^{1.2}$.
We can then calculate the different terms entering the $C\!P$ asymmetry $\ve_{2\t}$
in the approximation $m_2\simeq m_{\rm atm}$, $m_1 \ll m_{\rm atm}$ 
and $\rho=n\,\pi$ finding
\be
m_{ee} \simeq m_2 \,  , \hspace{4mm}
|(m^{-1}_{\nu})_{\t\t}| \simeq {c^2_{23}\over m_1} \,  , \hspace{4mm}
|(m^{-1}_{\nu})_{\m\t}| \simeq {s_{23}\,c_{23}\over m_1}  \,  .
\ee
In this way we arrive to
\be
\ve_{2\t}\simeq {3 \over 16\,\pi}\,{\a_2^2\,m_c^2 \over v^2}\,
{s^2_{23}\over c^4_{23}}\, {m_1\over m_{\rm atm}}\,\sin\a_L  \,  .
\ee
Finally for the effective leptogenesis phase we find
\be
\sin\a_L \simeq \sin (2\rho - {\rm Arg[c^2_{12}\,e^{2\,i\,\rho}+s^2_{12}]})  \,  ,
\ee
showing that $\rho$ phase needs to deviate from $n\,\pi$ for the asymmetry not to vanish
and for this reason the upper bound eq.~(\ref{upperboundm1IO}) becomes more
stringent. 

Combining all terms together, and imposing the successful leptogenesis condition one
arrives to a lower bound on $m_1$ depending on $\theta_{23}$,
\be
m_1 \gtrsim 32\,\pi\,10^{-8}\,{m_{\rm atm}\,v^2 \over \a^2\,m_c^2}\,{c^4_{23}\over s^2_{23}}\,
[\k(K_{2\t})]^{-1}\,\sin\a_L^{-1} \,  .
\ee
When this lower bound is combined with the upper bound eq.~(\ref{upperboundm1IO}) 
one finds a lower bound  $\theta_{23}\gtrsim 45^{\circ}$ 
for $\sin\a_L\simeq 0.5$ (the phase cannot be maximal otherwise the condition 
$K_{1\t} \lesssim 1$ would be hardly violated: see right panel in Fig.~10)  
fairly reproducing the results from the  scatter plots (see Fig.~10). 

\subsection{Strong thermal leptogenesis}

It is quite straightforward to understand why the ST condition cannot be satisfied 
in the IO case.  Even though the $N_1$ wash-out of the electron pre-existing component 
is strong since again one has $K_{1e} = m_{ee}/m_{\star} \simeq m_2/m_{\star} \gtrsim 50$,
on the other hand the wash-out of the muonic component is very weak since one has
\be
K_{1\m} \simeq {|(m_{\nu})_{e\m}|^2 \over m_{\star}\, m_{ee}} \simeq 
{|m_1 \, s_{13} \,s_{23} \, e^{i\,(2\s-\d)} + c_{12}\,s_{12} \, c_{23}\,(m_3 -m_2)|^2 
\over m_{\star}\,m_2} 
\simeq
{m_1^2 \over m_{\star}\,m_2}\,s^2_{13}\,s^2_{23} 
\simeq 0.5 \,  .
\ee
This happens because for IO one has $m_2 \simeq m_3$ in a way that there is an almost exact 
cancellation of the two terms $\propto m_2, m_3$ (those not suppressed by $s^2_{13}$). This is again, as for NO,  a particular example of what happens more generally, beyond $SO(10)$-inspired models, 
for IO: the wash-out of a muonic pre-existing asymmetry is weakened 
by this cancellation and produces a lower bound on $m_1$ \cite{strongthermal}.
In the case of $SO(10)$-inspired models the cancellation is basically almost exact in a way 
that the lower bound becomes incompatible with the CMB upper bound and one can conclude that 
ST $SO(10)$-inspired leptogenesis is simply not viable for IO neutrino masses. 

\section{Theoretical approximations and uncertainties}

The eq.~(\ref{NBmLflow}) constrains all low energy neutrino parameters to lie on
a hyper-surface in the low energy neutrino parameter space. However, there are different
effects of different nature to be considered that make in a way that this hypersurface is 
actually a layer with some thickness:
the experimental errors on the low energy neutrino parameters; the theoretical uncertainties 
in the calculation; accounting for deviations from $V_L = I$ introduces a dependence on the parameters of the $V_L$ that 
since are not measured lead to an intrinsic indetermination within $SO(10)$-inspired models. Of course within
a specific model one is able to specify the parameters in the $V_L$ and in this case one can expect and calculate specifically the deviation from the hypersurface described by the eq.~(\ref{NBmLflow}).  

\subsubsection*{Beyond the approximation $V_L \simeq I$}

In Fig.~5 we have included the results of a scatter plot, for $\alpha=5$ and NO, of points respecting successful leptogenesis
for $I\leq V_L \leq V_{CKM}$ (yellow points), confirming once more previous results \cite{SO10lep2,strongthermal}. 
When these are compared with the results obtained for $V_L = I$ (the orange points), one can see that there are constraints that
do not get strongly modified (e.g. the lower bound on $m_1$) and constraints that are more greatly modified (e.g. the upper bound on $m_1$). The most remarkable difference can be noticed  in the panel $\theta_{23}$ vs. $m_1$ where a complete new region
at large values $m_1 \gtrsim m_{\rm sol}$ appears. This is now mostly excluded by the CMB upper bound on $m_1$. This region is due to the appearance of a muon-type solution that is possible since when deviations of $V_L$ from unity are taken into account the strong hierarchy in the $C\!P$ asymmetries (cf. eq.~(\ref{CPhierarchy})) gets much milder and a muonic solution becomes possible \cite{SO10lep2}. 

Another clear difference is that for type $\tau_B$ solution the upper bound on $\theta_{23}$ is much more relaxed.
On the other hand the constraints for the type $\tau_A$ solution do not change dramatically, except for the
well known effect that now there is no lower bound on $\theta_{13}$. 

We have also compared the results obtained in the approximation $V_L =I$ (light blue points) and for 
 $I\leq V_L \leq V_{CKM}$ (dark blue points)  in the case of ST. One can see how the constraints get 
 moderately relaxed. The lower bound on $\theta_{13}$ gets relaxed from $\theta_{13}\gtrsim 5^{\circ}$
 to $\theta_{13} \gtrsim 2^{\circ}$.  The upper bound on $\theta_{23}$ gets relaxed from 
 $\theta_{23}\lesssim 41.5^{\circ}$ to $\theta_{23}\lesssim 43^{\circ}$, probably the most important effect
 in light of the current experimental constraints on $\theta_{23}$ that tend to favour $\theta_{23}\gtrsim 40^{\circ}$ at least at $2\s$.  One can see how the lower bound on $m_{ee}$ gets greatly relaxed. 
 
 In Fig.~4 we have plotted of $\eta_B$ vs. $m_1$ for three examples
 where $V_L \neq I$ comparing them to the three examples for $V_L = I$. 
 In all three cases the low energy neutrino parameters are unchanged. 
 One can see how turning on angles and phases in the $V_L$ can significantly enhance the asymmetry,
 though not dramatically (approximately up to a factor $2$).
 
 \subsubsection*{Theoretical uncertainties}
 
 A detailed discussion of the theoretical uncertainties can be found in \cite{strongthermal}. 
 Here we just remind that  the main sources of corrections to our results should come by inclusion of 
 flavour coupling and  (in the case of the muonic solution)  of phantom terms \cite{fuller};  
 account of decoherence for $M_2 \gtrsim 10^{11}\,{\rm GeV}$ within a density matrix
 formalism \cite{zeno,riottodesimone,beneke,densitymatrix,garbrecht}. Minor effects should come from the running of neutrino parameters \cite{running}
and a  full relativistic calculation of the wash-out rates \cite{bodeker}.

\section{Summary}

We have seen how $SO(10)$-inspired models  motivate an interesting scenario of high energy scale 
leptogenesis with hierarchical RH neutrinos able to reproduce the correct 
asymmetry when the production from the decays of the next-to-lightest RH neutrinos
is taken into account.
This scenario implies constraints on the low energy neutrino parameters that will be
partially testable with low energy neutrino experiments.
In the approximation of negligible misalignment between the Yukawa basis and the charged lepton basis ($V_L=I$),
we found a very accurate analytical expression that links all low energy neutrino parameters in quite a
non-trivial way.  Constraints on each individual parameter depend on the experimental information 
on the other parameters and interesting predictions can gradually emerge with more experimental information. For example,
we have seen how the discovery of a non-vanishing $\theta_{13}$ seems to produce combined constraints on 
the Dirac phase, the atmospheric mixing angle and the absolute neutrino mass scale.  
This potential interesting feature should, however, be confirmed  relaxing the approximation $V_L = I$.

In addition, quite interestingly, for a subset of the successful leptogenesis solutions, $SO(10)$-inspired leptogenesis realises the ST condition, in a way that a large 
pre-existing asymmetry is efficiently washed-out  and the final asymmetry is independent of the initial conditions. 
This produces very tight constraints on the low energy neutrino parameters characterising 
the ST $SO(10)$-inspired solution whose predictions will be  (almost) completely tested during next years. 
For example, the discovery of a IO neutrino mass spectrum would certainly rule out the solution like also a value of the atmospheric angle in the second octant. Vice-versa a discovery of NO and atmospheric neutrino mixing angle in the first octant should certainly be regarded as a strong support to the solution. In this respect it is interesting that some of the current 
global analyses \cite{foglilisi2013,nufit14} mildly support either IO and $\theta_{23}$ in the second octant or NO and 
$\theta_{23}$ in the first octant: in the first case
the solution would be undoubtedly ruled out, while the second case should be regarded as a very successful test. 

It is also interesting  that the experimental value 
of the reactor mixing angle falls just within quite a narrow range of values allowed by the solution. 
Interestingly the solution, within the allowed current range for the atmospheric mixing angle, is also potentially able to explain the emerging hint for a negative value of $\sin\d$, and therefore of $J_{CP}$, with the additional prediction $\cos\d>0$, with $\d \sim -\pi/4$. We managed to provide a full analytical description of these constraints.  In particular we showed
why non-vanishing values of the reactor mixing angle are required.
Very importantly the ST leptogenesis condition also forces $m_1$ to lie within a
narrow range about $m_1 \simeq 20\,{\rm meV}$.  This narrow range necessarily implies
the atmospheric angle $\theta_{23}$ in the first octant and this explains indirectly why 
$\d$ has to lie in the fourth quadrant.  

%%%%%%%

The NO$\nu$A long baseline experiment \cite{NOVA} is currently taking data and, combined with the results from 
the other neutrino oscillation experiments, in particular T2K constraints on $\d$ \cite{T2K}
and $\theta_{13}$ determination \cite{T2Ktheta13,*dayabay,*reno}, 
will allow in the next years to test in quite a significant way the predictions on the leptonic mixing matrix unknowns  from the ST $SO(10)$-inspired solution. At the same time cosmological observations are starting to corner quasi-degenerate neutrino and might in a close future start to test the narrow window, $75\,{\rm meV} \lesssim \sum m_i \lesssim 125\,{\rm meV}$, required by the solution corresponding to semi-hierarchical NO neutrino masses. 
In longer terms neutrinoless double beta decay experiments  should also be able to test
the range of values predicted by the solution for $m_{ee}$ 
centred about $m_{ee}\simeq 15\,{\rm meV}$.  In this way one would just miss a further experimental
complementary constraint on the Majorana phases for a complete test of the solution (explaining why the
solution can `almost' completely be tested).
It is then quite exciting that with $SO(10)$-inspired leptogenesis  one has  a well motivated 
opportunity to probe at the same time not only leptogenesis and the see-saw mechanism, but also the
$SO(10)$-inspired conditions, shedding light on the model for the origin of neutrino masses and mixing.

\subsection*{Acknowledgments}
We wish to thank Ferruccio Feruglio and Steve King for many useful discussions. 
PDB acknowledges financial support  from the NExT/SEPnet Institute, 
from the STFC Rolling Grant ST/L000296/1 and from the  
EU FP7  ITN INVISIBLES  (Marie Curie Actions, PITN- GA-2011- 289442).
PDB also wishes to thank the CERN theory group, NORDITA, CP3-Odense and MITP Institutes
for the hospitality during  the preparation of the work.  MRF acknowledges financial support from the STAG Institute.  LM  acknowledges the European Social Fund for supporting his work under the grant MJD387.

\section*{Appendix A}
\appendix

\renewcommand{\thesection}{\Alph{section}}
\renewcommand{\thesubsection}{\Alph{section}.\arabic{subsection}}
\def\theequation{\Alph{section}.\arabic{equation}}
\renewcommand{\thetable}{\arabic{table}}
\renewcommand{\thefigure}{\arabic{figure}}
\setcounter{section}{1}
\setcounter{equation}{0}

In this Appendix we provide details on the derivation of the approximate 
expression (eq.~(\ref{URapp})) for the off-diagonal entries of the 
RH neutrino mixing matrix $U_R$. More precisely here we are 
calculating what we called $\widetilde{U}_R$ while the matrix $D_{\Phi}$ can be calculated
using the equation given in the body text though we will omit the tilde to simplify the notation. 
From the unitarity conditions, $U_{Rik}\,U^{\star}_{Rjk}=\d_{ij}$,
one finds, for $(i,j)=(1,2)$ and $(i,j)=(2,3)$ respectively,
\be\label{unitarity}
U_{R12}\simeq -U^{\star}_{R21}  \;\;\; \mbox{\rm and} \;\;\;
U_{R32} \simeq -U^{\star}_{R23}  \,  ,
\ee 
having neglected respectively
$U_{R13}\,U^{\star}_{R23}$ and $U_{R21}\,U^{\star}_{R31}$.
On the other hand one has 
\be\label{unitarity13}
U^{\star}_{R31} \simeq -U_{R13} - U_{R12}\,U^{\star}_{R32} \,  ,
\ee
since the second term in the RH side is also $\propto m_{D1}/m_{D3}$
and cannot be neglected in this case. 

From the eq.~(\ref{invMtakagi}) we can write 
$D_M^{-1} \simeq - U^{\dagger}_R\,D_{m_D}^{-1}\,m_{\nu}\,D_{m_D}^{-1}\,U_R^{\star}$
and, therefore, for the matrix elements we can write
\be\label{invMelements}
{\delta_{ij}\over M_i} \simeq -U^{\star}_{Rki}\,
(D_{m_D}^{-1}\,m_{\nu}\,D_{m_D}^{-1})_{kl} \, U^{\star}_{R l j} \,  .
\ee
For $(i,j)=(1,2)$ it is quite straightforward to find 
\be
U_{R21}\simeq {m_{D1}\over m_{D2}} \,  {m_{\nu e \mu }\over m_{\nu ee}} \,  ,
\ee 
that plugged into the eq.~(\ref{invMelements}) for $(i,j)=(3,1)$ gives, with the help of 
eq.~(\ref{unitarity13}) and the second  eq.~(\ref{unitarity}),
\be
U_{R 31} \simeq {m_{D1}\over m_{D3}}\,{m_{\nu e\t }\over m_{\nu ee}} \,  .
\ee
From the eq.~(\ref{invMelements}) for $(i,j)=2,3$ and using eq.~(\ref{unitarity})
to write $U^{\star}_{R 23}$ in terms of $U^{\star}_{R 13}$ 
and $U_{R 3 1}$ one finds
\be
U_{R 13} \simeq {m_{D1}\over m_{D3}}\,
{(m_{\nu}^{-1})^{\star}_{e\t}\over (m_{\nu}^{-1})^{\star}_{\t\t}}\, .
\ee 
Finally, from the second eq. in (\ref{unitarity}) and eq.~(\ref{unitarity13}), one can write
\be
U_{R 23} \simeq {U_{R 13}+U^{\star}_{R 31}\over U_{R 12}} 
\simeq  {m_{D2}\over m_{D3}}\,
{(m_{\nu}^{-1})^{\star}_{\m\t}\over (m_{\nu}^{-1})^{\star}_{\t\t}} \,  ,
\ee 
leading to the eq.~(\ref{URapp}). In writing these equations we have made use of the
fact that the entries of the inverse neutrino mass matrix are related to the entries
of the neutrino mass matrix by
\be
m_{\nu}^{-1} = {1\over {\rm det} (m_{\nu})}\,
\left( \begin{array}{ccc}
m_{\nu\m\m}\,m_{\n\t\t} - (m_{\n \m\t})^2 & 
m_{\nu\m\t}\,m_{\n e \t} - m_{\n e \m}\,m_{\n\t\t} &   
m_{\nu e\m}\,m_{\n \m \t} - m_{\n \m\m}\,m_{\n e\t} \\
m_{\nu\m\t}\,m_{\n e \t} - m_{\n e \m}\,m_{\n\t\t} &  
m_{\nu ee}\,m_{\n \t\t} - (m_{\n e \t})^2 & 
m_{\nu e\t}\,m_{\n e \m} - m_{\n e e}\,m_{\n\m\t}  \\
m_{\nu e\m}\,m_{\n \m \t} - m_{\n \m \m}\,m_{\n e\t}  & 
m_{\nu e\m}\,m_{\n e \t} - m_{\n e e}\,m_{\n \m\t}  &
m_{\nu e e}\,m_{\n \m \m} - (m_{\n e \m})^2  
\end{array}\right)   \,  .
\ee
The orthogonal matrix entries are given by the eq.~(\ref{Oij}). Using
the eq.~(\ref{URapp}) and that
$(m_{\nu}\,U^{\star})_{ei}= - (U\,D_m)_{ei}$ one arrives 
to the eq~(\ref{Omegaapp}).

%\newpage
%=============================================================================
\bibliographystyle{./apsrev}
\bibliography{./mybib}
%=============================================================================

\end{document}